\documentclass[10pt]{article}
\usepackage[a4paper,margin=1.8cm]{geometry}
\usepackage[T1]{fontenc}
\usepackage[utf8]{inputenc}
\usepackage{lmodern}
\usepackage{amsmath, amssymb, amsfonts, bm}
\usepackage{mathtools}
\usepackage{physics}
\usepackage{graphicx}
\usepackage{float}
\usepackage{svg}
\usepackage{caption}
\usepackage{subcaption}
\usepackage[hidelinks]{hyperref}
\usepackage{xcolor}
\usepackage[normalem]{ulem}
\usepackage{cancel}
\usepackage{placeins}
\usepackage{pdfcomment}
\usepackage{subcaption}
\usepackage[percent]{overpic}
\usepackage[numbers,sort&compress]{natbib}
\setcitestyle{super}  
\newcommand{\mobius}{M\"obius~}

\begin{document}
	\begin{center}
		{\Large\bfseries Topological Quenching of Noise in a Free-Running \mobius Microcomb \par}
		\vspace{1.4em}
		Debayan Das$^{1}$, Antonio Cutrona$^{1}$, Andrew C. Cooper$^{1}$, 
		Luana Olivieri$^{1}$, Alexander G. Balanov$^{1}$,\\
		Sai Tak Chu$^{2}$, Brent E. Little$^{3}$, Roberto Morandotti$^{4}$,
		David J. Moss$^{5}$,\\
		Juan Sebastian Totero Gongora$^{1}$,
		Marco Peccianti$^{1}$,
		Gian-Luca Oppo$^{6}$,
		and Alessia Pasquazi$^{1,*}$\\		
		\vspace{1.2em}
		{\small
			$^{1}$Emergent Photonics Research Centre (EPicX), Dept. of Physics, Loughborough University, UK \\
			$^{2}$Dept. of Physics, City University of Hong Kong, China SAR \\
			$^{3}$QXP Technologies Inc., Xi’an, China \\
			$^{4}$INRS-EMT, Varennes, Québec, Canada \\
			$^{5}$Optical Sciences Centre, Swinburne University of Technology, Australia \\
			$^{6}$SUPA, University of Strathclyde, Glasgow, UK \\
			$^{*}$Correspondence: a.pasquazi@lboro.ac.uk
		}
	\end{center}	
	\vspace{1.5em}


	\noindent
	\textbf{
		Microcombs require ultralow-noise repetition rates to enable next-generation applications in metrology, high-speed communications, microwave photonics, and sensing, where spectral purity is a central performance metric.
	Best-performing sources operate actively locked at “quiet points’’ in parameter space, fixed by device and material properties. Creating broad, low-noise operating regions with relaxed constraints—especially in simplified free-running architectures that avoid electronics-heavy control—remains an open challenge.
	Here, we demonstrate a symmetry-protected topological \mobius soliton molecule that enables intrinsically low phase noise in a fully free-running microcomb, operating without any external referencing or control. Using a microresonator-filtered laser, we implement a \mobius geometry via interleaved microcavity modes. Upon the formation of a topological \mobius soliton molecule, the free-running laser exhibits over 15\,dB of phase-noise suppression across 10\,Hz–10\,kHz at a 100\,GHz repetition rate, yielding $-63$\,dBc/Hz phase noise at 1\,kHz and an Allan deviation of $4\times10^{-10}$ at 10\,s average time—without any external control. We show that the \mobius structure brings dynamic robustness to the comb, and we demonstrate a symmetry-protected topological regime that enables long-term drift-invariant operation. Our results establish a route to intrinsically noise-quenched microcombs operating in a fully free-running configuration, governed by internal physical principles and suitable for field-deployable, low-noise photonic systems.
	}

	\vspace{1.0em}
Optical solitons are nonlinear waves that preserve their structure during
propagation by balancing dispersion or diffraction with nonlinearity~\cite{Kivshar2003Optical},
and, in dissipative systems, by achieving equilibrium between loss and
gain~\cite{Lugiato2015Nonlinear,Grelu2016Nonlinear}.
For microcombs, or optical frequency combs in microresonators,~\cite{Kippenberg2018Dissipative,Pasquazi2018Micro} solitons
have represented a major advancement,~\cite{Herr2014Temporal,Brasch2016Photonic,Huang2016broadband,Obrzud2017Temporal,Stern2018Battery,Pavlov2018Narrow,Bao2019Laser,Huang2019Temporal,Piccardo2020Frequency,Tikan2021Emergent,Helgason2021Dissipative}
enabling key breakthroughs in metrology, optical communications, and microwave
distribution.~\cite{Liang2015High,Marin2017Microresonator,Suh2018Soliton,Spencer2018optical,Hu2018Single,Fulop2018High,Corcoran2020Ultra,Xu202111,Feldmann2021Parallel,Riemensberger2020Massively,Liu2020Monolithic,Meng2020Mid,Boggio2022Efficient,Wu2025Vernier,Zhao2024All,Jang2018Synchronization,Kim2021Synchronization}
For many applications, such as metrology, reducing their intrinsic noise is crucial, particularly when considering the repetition rate.
Critical progress has been made in synchronising microcombs to ultra-stable
references, establishing them as a promising platform for high-purity microwave
sources, using both all-optical approaches~\cite{Jang2018Synchronization,Kim2021Synchronization,Moille2023Kerr,Zhao2024All,Lei2024Self}
and electronic-based locking,~\cite{Wu2023Vernier,Kudelin2024Photonic,Sun2024Integrated,Wildi2024Phase,sun_microcavity_2025,jin_microresonator-referenced_2025}
which can improve free-running repetition-rate noise by over 30–60 dB once the
system is actively locked.
Moreover, operation at \emph{quiet points}~\cite{Yi2017Single}
effectively reduces intrinsic noise arising from thermal fluctuations.~\cite{Lei2022Optical}
Approaches that lock the continuous-wave source in externally driven microcombs
to these operating points underpin a critical innovation for metrological microcombs~\cite{Yi2017Single,Lei2022Optical,Liu2020XKBandMicrocombs,Lucas2020Ultralow,Yang2021DispersiveNoise,Triscari2023Quiet}.

In contrast, significantly less progress has been made in achieving low-noise
repetition rates in standalone free-running systems. Recent results on
self-starting microcombs,~\cite{Shen2020Integrated,Rowley2022Self,Nie2022Dissipative,Cutrona2023Nonlocal,Opacak2024NozakiBekki,Ulanov2024Synthetic,MkrtchyanMicroring}
based on self-consistent laser–microcavity architectures, have shown that
microcombs can operate without high-end control electronics. These devices
maintain a stable nonlinear state even when cavity parameters (length, gain, loss)
vary, producing a robust self-locked regime with relaxed physical constraints. 
Metrologically, however, one would ideally require the comb frequencies—and especially the repetition rate—to remain invariant under adiabatic, environmental drifts and fluctuations of the parameters.
These self-locked sources would therefore benefit from a method capable of
providing a \emph{quiet region} of operation, rather than a \emph{quiet point.}

In parallel, topological photonics—including nonlinear, active, and
non-Hermitian implementations—has significantly advanced optical resilience to
perturbations~\cite{Lu2014Topological,Ozawa2019Topological,Smirnova2020Nonlinear,Nasari2023NonHermitian,Khanikaev2024Beyond,Wong2025Nonlinear, SzameitRechtsman2024NonlinearReview}.
A topological state retains its properties under smooth parameter
changes unless a transition is reached—typically marked by a spontaneous symmetry breaking, where the system’s state no longer remains invariant under a symmetry of the governing equations.
In periodic structures~\cite{HasanKane2010TopIns,Wang2009TopoEM,Rechtsman2013FloquetTI} topological states are naturally
described through symmetry-defined operators and transitions, where bandgaps,
symmetry protection, and geometrical phases play a central role~\cite{Zak1989,Berry1984,Resta1994}.
In nonlinear optics, topological states are associated with phase defects in the nonlinear states—
vortices, dislocations, and phase singularities~\cite{Trillo1997Stable,Oppo1999From,Englebert2024Topological} and with localisation in lattices,~\cite{aceves_optical_2000,Lumer2013SelfLocalized,LeykamChong2016EdgeSolitons,MukherjeeRechtsman2020FloquetSolitons,Pernet2022GapSolitons, Bongiovanni2021DynamicalTopo,Hu2021HOTIBIC,Wang2023SubSymmetryTopo}.
In the microcomb context, topological solitons have been explored in self-organised temporal
lattices\cite{Fan2022Topological} and extended to coupled-resonator
geometries~\cite{Mittal2021Topological,Flower2024Observation,Coen2024Nonlinear}.

----------------

\begin{figure*}[htbp]
	\centering
	\includegraphics[width=\textwidth]{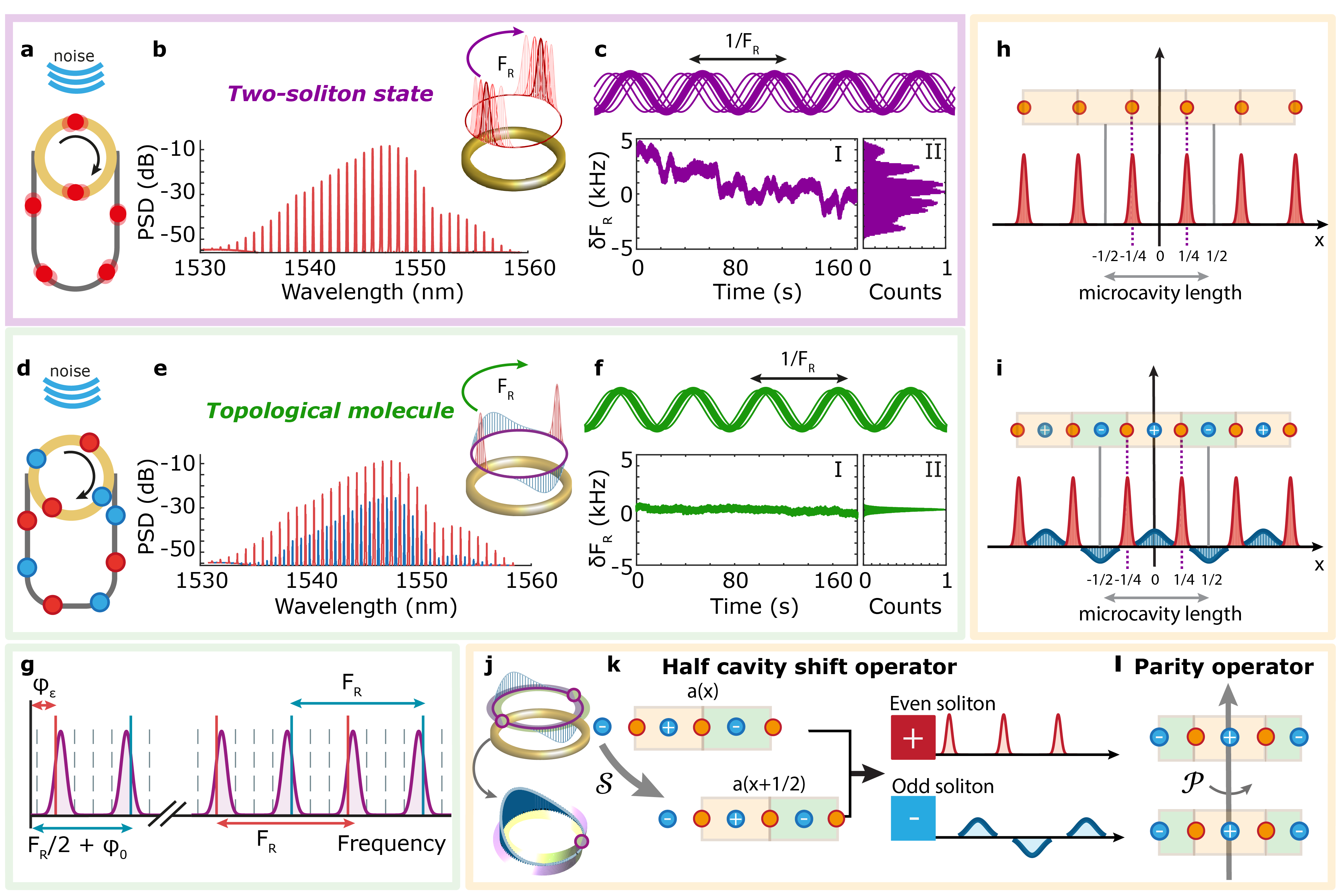}
	\caption*{\footnotesize\textbf{Figure 1. {Topological \mobius soliton molecule} and intrinsic noise quenching.}		
		\textbf{a--g} Microcomb pulsed laser operation: comparison between (a–c) a standard two-soliton state and (d–f) a topological \mobius molecule formed by two independent soliton pairs (red and blue) bound into a single molecular state and circulating with a shared repetition rate $F_R$.
		\textbf{a,d} Schematic of the dual-cavity laser. A nonlinear microring (yellow, cavity $a$, round-trip $T_a$) is nested inside an amplifying cavity (grey, cavity $b$, round-trip $T_b$) and subject to noise. In (a), the amplifier-cavity round trip is an integer multiple of $T_a$; in (d), it is an odd multiple of $T_a/2$.
		\textbf{b,e} Measured optical spectra (free-spectral range, $\mathrm{FSR}=1/T_a=50\,\mathrm{GHz}$). Standard two-soliton (b) and \mobius molecule (e) combs. Interleaved even (red) and odd (blue) comb lines are highlighted.
		\textbf{c,f} Sketches of the pulse circulation under external noise, operating at $F_R=2/T_a=100\,\mathrm{GHz}$, twice the ring round-trip frequency, producing a repetition-rate oscillation with period $1/F_R$ (arrow). 
		I: measured instantaneous repetition-rate variation $\delta F_R(t)$ and II: measured histogram, for standard (c) and \mobius molecule (f) states.
		\textbf{g} Mode structure of the \mobius molecule. 
		Purple solid lines mark the microring resonances; red and blue lines indicate the even and odd lasing modes. An odd number of amplifier-cavity modes (grey dashed lines, frequency spaced by $1/T_b$) lie between two consecutive microring resonances. The even and odd combs acquire different carrier--envelope--offset frequencies, $\varphi_E$ and $\varphi_O + F_R/2$.
		\textbf{h–l}~Lattices and operators. The microcavity coordinate $x$ is normalised to the microring round-trip. The fundamental lattice element spans $|x|<1/2$, several replicas are shown for clarity.
		\textbf{h,i} Trivial and non-trivial soliton lattices; 
		\textbf{h} Trivial even-soliton train preserving two independent symmetries: 
		(i) spatial parity $a(x)\!\to\! a(-x)$ and
		(ii) half-cavity shift $a(x)\!\to\! a(x{+}1/2)$.	
		\textbf{i}~Non-trivial even–odd soliton train (\mobius molecule).
		This state preserves only spatial parity $a(x)\!\to\! a(-x)$.
		\textbf{j,k}~\mobius geometry and half-cavity shift operator. 
		\textbf{j} \mobius geometry: a half-period translation combined with a sign flip produces a characteristic \mobius structure in which the waveform reverses sign every half cavity length. The odd mode effectively displays an antiperiodic relationship over half-period length, $a(x{+}1/2)=-a(x)$. 
		\textbf{k}~\emph{The half-cavity shift operator}. It decomposes any field into two orthogonal subspaces: an even sector $a(x)+a(x{+}1/2)$ and an odd sector $a(x)-a(x{+}1/2)$. Applied to the molecule, this operator separates the even and odd solitons.
		\textbf{l}~\emph{Parity operator.}
		Spatial parity $a(x)\!\to\! a(-x)$ reflects a state at $x=0$. A molecule can be invariant under this operator, which underpins the protection of the topological state.}
	\label{fig:Fig1Combined}
\end{figure*}

A key open question is whether a topological microcomb can exhibit enhanced
stability in its characteristic frequencies—particularly its microwave-range
repetition rate—under free-running operation, where cavity parameters vary
slowly. Importantly, in self-locked systems, the nonlinear states can persist
under such drifts~\cite{Shen2020Integrated,Rowley2022Self,Nie2022Dissipative,Cutrona2023Nonlocal,
	Opacak2024NozakiBekki,Ulanov2024Synthetic,MkrtchyanMicroring}, yet their
frequencies are still affected.

Figure~1a–c illustrates a representative free-running operation of a
self-starting microresonator-filtered laser, consisting of a nonlinear (Kerr)
microcavity embedded in a laser loop~\cite{Bao2019Laser} (Fig.~1a).
The soliton pulses—whose characteristic spectrum (Fig.~1b) is consistent with  
Refs.~\cite{Rowley2022Self,Cutrona2023Nonlocal}—remain structurally intact under
external noise, yet their repetition-rate frequency undergoes slow drift
(Fig.~1c), as typical of free-running systems.

Here, we introduce a soliton molecule in a microresonator-filtered laser, featuring a \mobius geometry implemented across interleaved microcavity modes by simply adjusting the laser-cavity length.
We show that this molecule exhibits a
self-organised, symmetry-protected topological regime in which long-term drifts
of the repetition rate are intrinsically suppressed in fully standalone, free-running
operation (Fig.~1d–f). Beyond this symmetry-protected regime, the \mobius
structure preserves the robustness of the state and maintains excellent
phase-noise performance. The topological \mobius soliton molecule exhibits over
15\,dB of phase-noise suppression across 10\,Hz–10\,kHz at a 100\,GHz repetition
rate, yielding $-63$\,dBc/Hz at 1\,kHz and an Allan deviation of
$4\times10^{-10}$ at 10\,s average time—without any external control.
A M\"obius optical cavity emulates the well-known M\"obius strip—consisting of a one-sided
surface formed by a half-twist in a closed loop—by imposing an antiperiodic
	relationship on the optical field \(a(x)\):
\begin{equation}
	a\!\left(x+\tfrac{1}{2}\right) = -\,a(x),
	\label{boundaryM}
\end{equation}
so that the field reproduces itself only after two cavity round trips.
The example in Eq.~(\ref{boundaryM}) uses a normalised cavity length of \(1/2\),
leading to a period equal to one. Such cavities are typically implemented with an element imposing
a \(\pi\)-phase shift and have been realised in both fibre and integrated
platforms~\cite{Wang2023Experimental,Maitland2020Stationary,
	Chen2023Topological,Coen2024Nonlinear}. Remarkably, a M\"obius cavity can
preserve the integrity of the optical state even across nonlinear bifurcations~\cite{Coen2024Nonlinear}.
We now extend this concept to a standard cavity, whose period is normalised to one, realising a \emph{relative \mobius relationship}. 
In a simple Fourier picture, a generic intracavity field
\( a(x)=\sum_{m} a_m\, e^{2\pi i m x} \)
populates the resonance labelled by the integer index \(m\) with amplitude \(a_m\).
We then separate the modes \(a_m\) into two interleaved families, which we denote
as \emph{even} ($2m$) and \emph{odd} ($2m+1$) .

These families are populated by the fields
\( a_e(x)=\sum_{m} a_{2m}\, e^{2\pi i (2m)x} \)
and
\( a_o(x)=\sum_{m} a_{2m+1}\, e^{2\pi i (2m+1)x} \),
respectively.
Each has an effective period of \(1/2\) and satisfies the
\emph{relative} periodic/antiperiodic relations
\begin{equation}
a_e\!\left(x+\tfrac{1}{2}\right)=a_e(x), \qquad
a_o\!\left(x+\tfrac{1}{2}\right)=-\,a_o(x).
\label{boundary}
\end{equation}

We leverage these relations to implement an effective M\"obius structure—
mimicking a cavity of period \(1/2\) with a \(\pi\) defect, by enforcing that the fields \(a_e(x)\) and \(a_o(x)\) remain distinct, thereby preventing the two mode families from merging into a single period-one state.

In our implementation, such condition is realised by a \emph{soliton molecule}~\cite{Kivshar2003Optical,
	Grelu2016Nonlinear}, a bound configuration of nonlinearly interacting solitary states. Here, the molecule consists of two combs (Fig.~1d–f), each with
its own carrier--envelope-offset (CEO) frequency~\(\varphi\) (Fig.~1g), yet
sharing the same repetition rate~\(F_R\). Each comb line is given by
\begin{equation}
	f_m = m F_R + \varphi.
	\label{comb}
\end{equation}

A simple way to impose this constraint in a microresonator-filtered laser is to set the ratio between the FSRs of the microcavity
(\(F_a \approx 50\,\mathrm{GHz}\)) and the main cavity (\(F_b \approx
75\,\mathrm{MHz}\)) to a half-integer (Fig.~1g). In this way, the
interleaved main-cavity resonances acquire a fixed offset \(F_b/2\) while
keeping a common spacing \(2F_a\), so that \(F_R = 2F_a\). The resulting combs
satisfy
\begin{equation}
	\begin{split}
		f_m^{(\mathrm{e})} &= 2m F_a - F_b\,\Delta, \\
		f_m^{(\mathrm{o})} &= (2m+1) F_a - F_b\,\Delta + \frac{F_b}{2},
	\end{split}
	\label{eoComb}
\end{equation}
where \(\Delta\) is the detuning normalised to \(F_b\). The fixed offset \(F_b/2\) is small compared to the microcavity FSR \(F_a\).
Here we separate the action of the two contributions. The CEO separation is
dominated by \(F_a\), which is half of the repetition rate; this means that
the two combs are effectively interleaved and obey the periodic relations in
Eq.~\eqref{boundary}. The small term \(F_b/2\), importantly, lifts the
degeneracy between the two families and preserves their independence
(Supplementary~S1).

Despite its simplicity, the frequency relationship expressed by Eq.~\eqref{eoComb} imposes non-trivial constraints on the system symmetry, which we analyse using symmetry operators.  By definition, a symmetry operator maps a solution into another solution of the governing equations.~\cite{NonlinearDynamics2007} When a symmetry is \emph{discrete}, i.e. associated with a specific finite transformation, it often allows splitting the solution space into complete and orthogonal subspaces.~\cite{Resta1994,NonlinearDynamics2007} Here, the critical transformation is a shift by \emph{half a cavity period}.
We therefore consider the translation operator \(\mathcal S\) (\(x \mapsto x+\tfrac{1}{2}\)), which we term the half-cavity shift, as it translates a generic waveform \(a(x)\) by half a period.
The frequency relationship in Eqs.~\eqref{eoComb} modifies the system (Fig.~1h--i) such that \(\mathcal S\) becomes a discrete symmetry of the underlying equations (Methods, Supplementary~S1, S2), defining two complete and orthogonal subspaces
by the symmetric and antisymmetric decomposition of a waveform \(a(x)\) under \(\mathcal S\) (Fig.~1j--k):
\begin{equation}
	a_e(x)=\tfrac{1}{2}\bigl[a(x)+a(x+\tfrac{1}{2})\bigr], \qquad
	a_o(x)=\tfrac{1}{2}\bigl[a(x)-a(x+\tfrac{1}{2})\bigr].
	\label{prS}
\end{equation}
	These relationships map precisely the even and odd families, which experience different effective detunings in Eq.~\eqref{eoComb} and therefore have different lasing thresholds.

Since the system can self-start into solitons~\cite{Rowley2022Self,
	Cutrona2023Nonlocal}, it is natural to begin from the simplest configuration, in
which only one family of resonances initially supports a soliton state—two
equidistant pulses per round trip. We associate these solitons with the even
{family} for simplicity (Fig.~1h,Supplementary~S3).

When the odd modes rise above their effective lasing threshold, a new pulse forms between
the two even solitons, emerging from a destabilisation of the cavity background
(Fig.~1i, Supplementary~S4, S5). This growth mechanism is reminiscent of potential
mode-locking~\cite{Heckelmann2023Quantum,Bourgon2025Mode}.
The resulting state comprises an even soliton and an odd pulse with necessarily
different CEO frequencies fixed by Eqs.~\eqref{eoComb}, yet nonlinearly
synchronising at twice the microcavity FSR and bonding into a \emph{molecule}, which we refer to as a "\mobius soliton molecule".

\begin{figure*}[htbp]
	\centering
	\includegraphics[width=\textwidth]{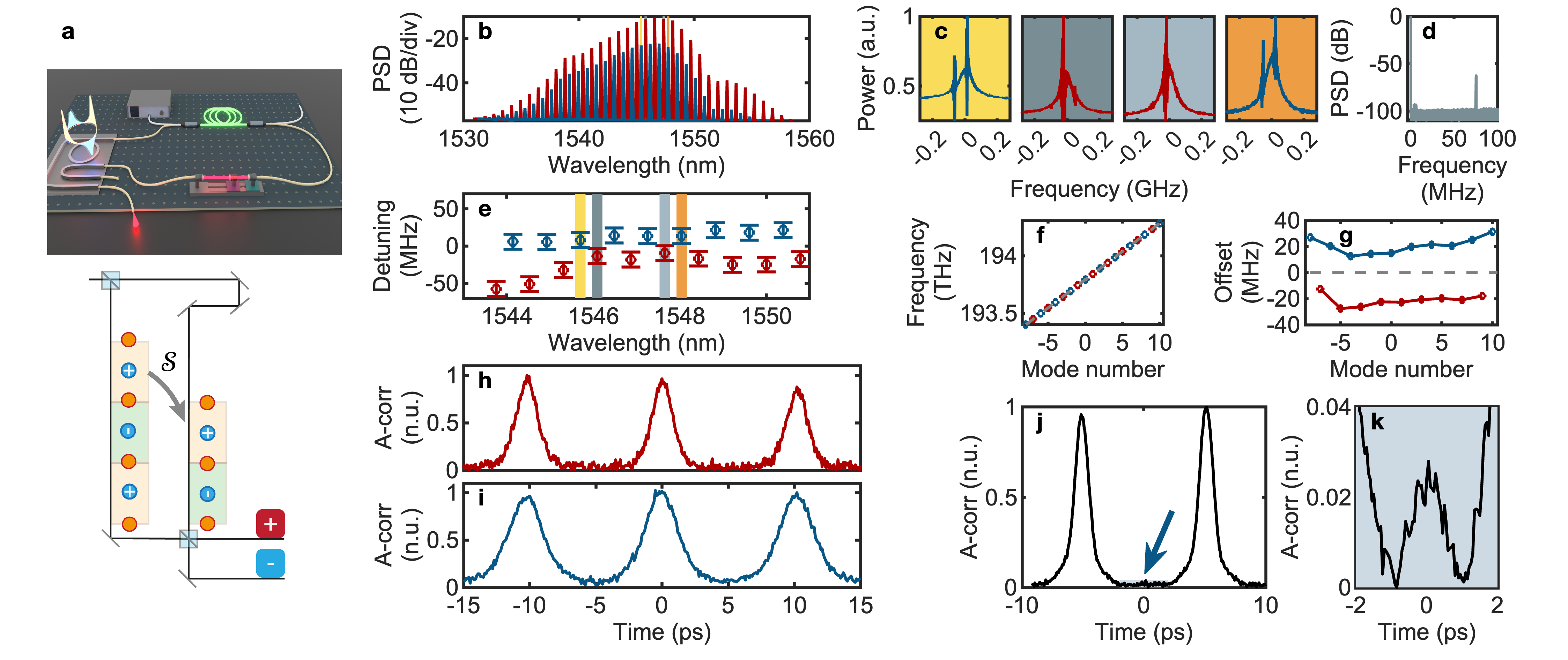}
	\caption*{\footnotesize \textbf{Figure 2 Experimental characterisation of the M\"obius soliton molecule.}
		\textbf{a} Microresonator-filtered setup. 
		A $\sim $50~GHz-FSR microring resonator is nested within a larger $\sim $75 MHz-FSR laser cavity, where a tunable delay line controls the effective cavity length.	
		\textbf{b} Optical spectrum of the M\"obius soliton molecule. 
		The trace is a single measured spectrum, with even and odd lines in red and blue, respectively.
		\textbf{c} Radio-frequency spectrum showing a narrow tone at the 
		\emph{FWM CEO-mismatch frequency} $f_{\mathrm{CM}} \sim 75$~MHz.
		\textbf{d} Laser-scanning spectroscopy of selected comb lines, highlighting the lasing modes within the microcavity resonances for the two soliton families (red, blue). 
		A FWM tone at $f_{\mathrm{CM}}$ appears mainly in the odd family.
		\textbf{e--g} Analysis of optical frequencies extracted from LSS as in (d), for even (red) and odd (blue) resonances. 
		(e) Detunings from the microcavity resonance centres, revealing a systematic red and blue shift for even and odd modes, respectively. 
		(f) Absolute optical comb frequencies versus microcavity resonance index, showing that the two families share the same repetition rate. 
		(g) Frequency offsets obtained by subtracting the linear fit in (f). 
		The two soliton families differ by $\sim 38$~MHz in carrier–envelope–offset frequency—about half the main-cavity intermode spacing.
		\textbf{h--k} Time-domain characterisation via second-harmonic autocorrelation. 
		\textbf{h--i} Autocorrelation for the even (h) and odd (i) soliton trains, separated using an unbalanced Mach–Zehnder interferometer (sketched on the left). 
		\textbf{j} Autocorrelation of the full train. 
		\textbf{k} Detail around $0$~s showing the odd soliton between the two even soliton peaks.
	}
	\label{fig:Fig2}
\end{figure*}
Experimentally, we verify the existence of a molecule implementing the
M\"obius geometry associated with the action of the operator~$\mathcal S$.
Our microresonator-filtered laser (Fig.~2a, Methods), based on a $\sim 75$~MHz
Erbium-doped fibre amplifier (EDFA) cavity and a $\sim 50$~GHz, high-index
doped-silica microring is arranged in the interleaved
configuration of Fig.~1g using a simple tunable delay line.

Figure~2b displays a representative molecule spectrum, where we highlighted the
interleaved even (red) and odd (blue) components for presentation purposes. To observe the lasing modes
within the microcavity, we use laser-scanning spectroscopy (LSS, Methods);
examples for selected wavelengths are shown in Fig.~2d. These modes consistently
lase on opposite sides of the microcavity resonance: the even (red) family on
the red-detuned slope and the odd (blue) family on the blue-detuned slope.
These detunings—defined as the difference between the peak lasing frequency and
the centre of the microcavity resonance—remain systematically opposite across
the entire comb, as summarised in Fig.~2e, confirming the presence of the interleaved structure
introduced in Fig.~1g.

When plotted versus mode number (Fig.~2f), the lasing frequencies of both
soliton families exhibit the same linear slope, demonstrating that they share a
common repetition rate $F_R \approx 2F_a \approx 100$~GHz. Subtracting this
linear trend reveals a constant offset of $\Delta\varphi \approx F_b/2
\approx 38$~MHz between consecutive comb lines of the two families (Fig.~2g),
confirming that they possess distinct CEO frequencies. An independent CEO
measurement via electrical down-conversion~\cite{Cutrona2023Stability} is
reported in the Methods.

Additionally, the odd-family traces in Fig.~2d exhibit a secondary, weaker
resonance displaced by $2Delta\varphi$ from the main lasing peak. This
feature arises from four-wave mixing (FWM) and produces the narrow RF
tone at $f_{\mathrm{CM}} = 2\Delta\varphi \approx F_b \approx 75~\mathrm{MHz}$
visible in Fig.~2c (Supplementary S4). This RF tone constitutes a third, independent signature of
the CEO independence of the even and odd combs; in what follows, we refer to
it as the \emph{FWM CEO-mismatch frequency} to emphasise its origin.
To probe the temporal structure of each component, we use an unbalanced
Mach–Zehnder interferometer (Fig.~2h,i), which implements exactly the
projectors of Eq.~\eqref{prS} illustrated in Fig.~1j. The isolated even and odd
fields are then characterised via second-harmonic autocorrelation, confirming
their pulsed nature at $\sim 10$~ps spacing and their distinct pulse durations.
Finally, we built a second-harmonic autocorrelator with a high dynamic range, which clearly identifies the odd pulses in the temporal region
between successive even pulses, as reported in Fig.~2k (Methods).

Together, these measurements provide clear evidence for the generation of
the M\"obius soliton molecule introduced above and enabled by the
operator~$\mathcal S$.
\begin{figure*}[htbp]
	\centering
	\includegraphics[width=\textwidth]{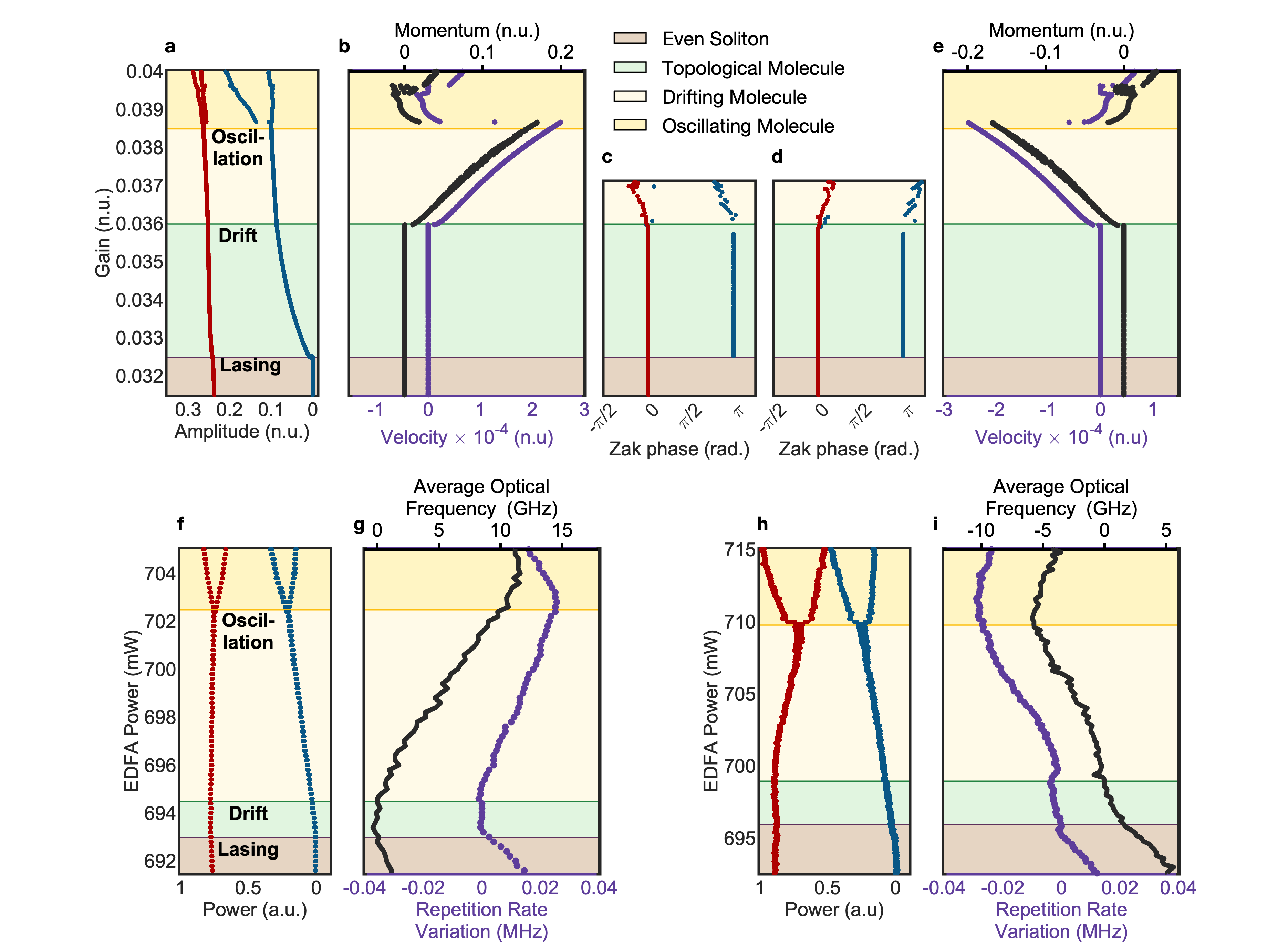}
	\caption*{
		\footnotesize
		\textbf{Figure 3 \, \mobius molecule transitions and topological regime.}
		\textbf{a--e} \emph{Theoretical bifurcation diagram as a function of the gain}
		(Methods: ``Theoretical Model''; full parameter set in Extended Fig.~E2).
		Starting from $g = 0.031$, the gain is increased adiabatically and the
		intracavity state is recorded after transient dynamics.
		\textbf{a} Average intracavity power of the even (red) and odd (blue)
		components. For slow oscillatory solutions, maxima and minima are plotted.
		Two transitions appear:
		(i) formation of the molecule at the effective lasing threshold of the odd
		mode ($g \approx 0.0325$);
		(ii) onset of oscillations ($g \approx 0.0385$).
		\textbf{b} Group velocity (purple; bottom axis) and average optical
		frequency (``momentum'', black; top axis).
		A transition near $g \simeq 0.036$ marks the onset of drift, indicating
		the breaking of spatial parity~$\mathcal{P}$ (Extended Fig.~E1).
		\textbf{c} Zak phase of the even (red) and odd (blue) sectors.
		A quantised $\pi$ phase signals a topological state.
		The odd sector acquires a Zak phase of $\pi$ upon its formation in the
		parity-symmetric (resting) regime; after drift onset, the Zak phase evolves
		following the momentum.
		\textbf{d,e} Same analysis as in (c) and (b) for a different realisation.
		The opposite drift directions correspond to the two symmetry-related
		outcomes.
		\textbf{f--i} \emph{Experimental bifurcation diagram as a function of EDFA
			power}, showing two realisations in the same parameter range and reproducing
		the two symmetry-breaking outcomes.
		The system is initialised in the even-soliton state below the odd-mode
		lasing threshold; EDFA power is increased in steps and the output
		recorded.
		\textbf{f,h} Average power of the even (red) and odd (blue) components,
		measured with narrowband filters (1547.32\,nm and 1548.51\,nm).
		For slow oscillatory states, maxima and minima are shown.
		Lasing and oscillatory thresholds occur at
		(f) $P \simeq 693$\,mW and $P \simeq 703$\,mW;
		(h) $P \simeq 694$\,mW and $P \simeq 710$\,mW.
		\textbf{g,i} Repetition-rate variation (purple; bottom axis) and average
		optical frequency (black; top axis).
		A transition near $P \simeq 695$\,mW (g) and $P \simeq 700$\,mW (i)
		marks the onset of drift, indicating the breaking of spatial parity in
		agreement with the theoretical prediction.
	}
	\label{fig:Fig3}
\end{figure*}

We now show that the birth of the odd pulse — and hence the onset of the molecule —
corresponds to a \emph{symmetry breaking} of the half-cavity shift operator
$\mathcal S$ (Section S5, S6). To introduce this concept, it is useful to
represent the intracavity field as a one-dimensional \emph{lattice}, in which
the field profile inside the cavity is taken as the basic element (Fig.~1h,i) and
repeated according to the periodic boundary conditions.

In the pure even-soliton train (Fig.~1h), the cavity contains two identical
subcells per round trip, which become \emph{distinct} once the odd pulse forms
(Fig.~1i). A one-dimensional lattice with two different sites is reminiscent of
the Su–Schrieffer–Heeger (SSH) model\cite{Resta1994,Lu2014Topological,
	Ozawa2019Topological,Smirnova2020Nonlinear,Nasari2023NonHermitian,
	Khanikaev2024Beyond,Wong2025Nonlinear}, and such a non-trivial configuration
can host topological states in the presence of a protecting symmetry.

In our system, the relevant protecting symmetry is the spatial parity inversion
$\mathcal P$ ($x \mapsto -x$) (Fig.~1l), which maps the lattice onto itself when mirrored
about the cavity centre. This symmetry is intrinsic to the even-soliton train and,
provided that the odd pulses appear symmetrically between the even ones, it remains preserved. 
The temporal reconstruction in Fig.~2k is consistent with a near-realisation of this condition (Methods).
A \mobius molecule retains spatial parity and can realise
\emph{a symmetry-protected topological state}.
To rigorously determine the extent of the parity-protected, topological regime,
we perform (Fig.~3) an \emph{adiabatic sweep of the gain}, under which the
properties of a topological state must remain invariant%
\cite{Resta1994,Lu2014Topological,Ozawa2019Topological,Smirnova2020Nonlinear,
	Nasari2023NonHermitian,Khanikaev2024Beyond,Wong2025Nonlinear}.

\begin{figure*}[htbp]
	\centering
	\includegraphics[width=\textwidth]{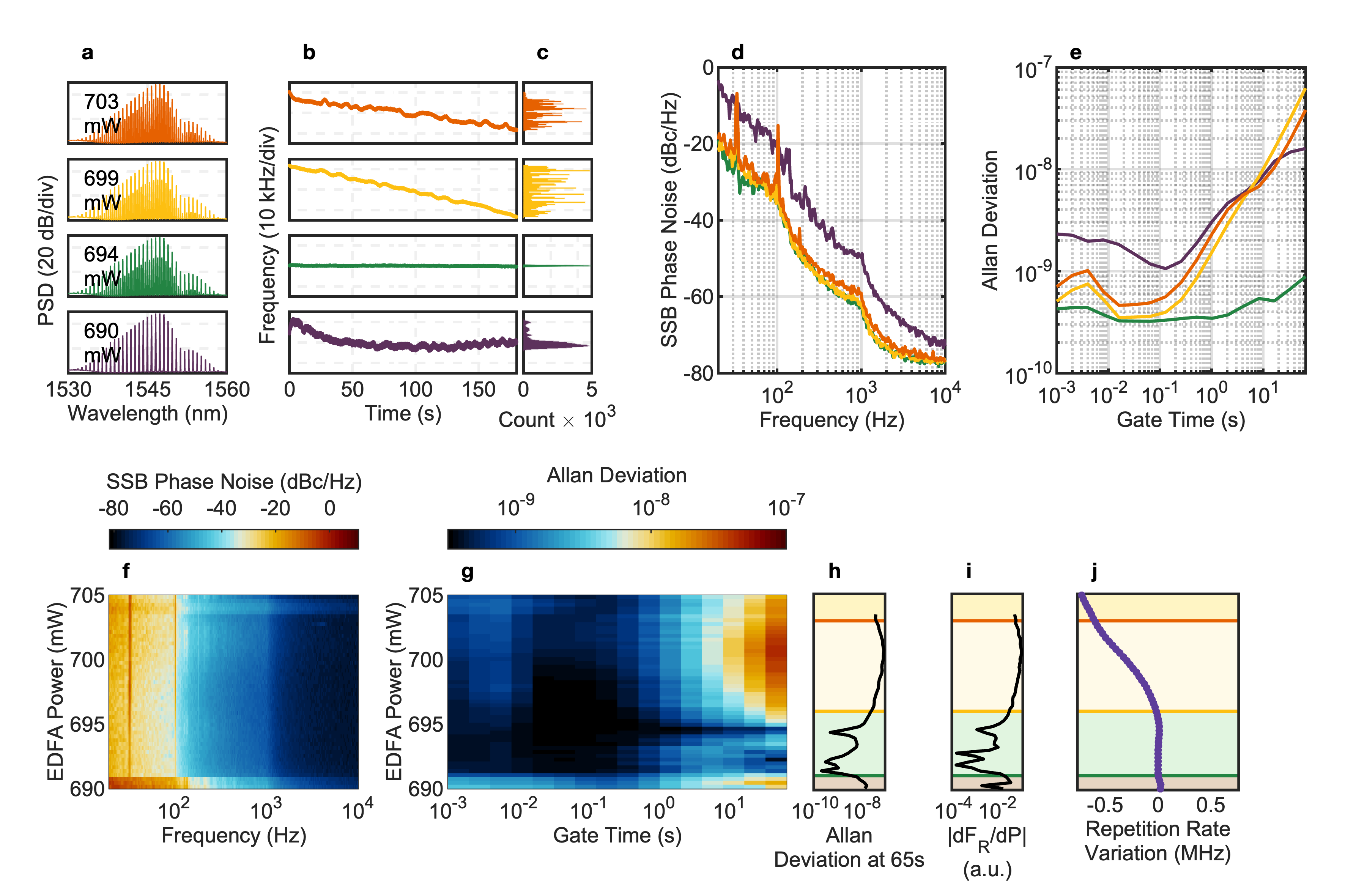}
	\caption*{\footnotesize 
		\textbf{Figure 4 \, Noise characterisation of the M\"obius molecule.}
		\textbf{a} Representative optical spectra at selected EDFA powers spanning the even-soliton regime (brown), the topological M\"obius molecule (green), the drifting molecule (orange), and the oscillating molecule (yellow).
		\textbf{b} Time evolution of the repetition rate over 180~s for the same operating points.
		\textbf{c} Corresponding histograms of the repetition-rate fluctuations.
		The state in the topological region displays the smallest long-term drift, consistent with the reduced sensitivity of the group velocity to pump-power variations.
		\textbf{d} Single-sideband phase-noise spectra (colour-coded as in panel~\textbf{c}), confirming that the M\"obius molecule maintains significantly lower phase noise than the pure even-soliton state across the measurable range.
		\textbf{e} Allan deviation of the repetition rate for the traces in \textbf{b}, showing that the M\"obius molecule preserves low noise across both short and long averaging times.
		\textbf{f} Single-sideband phase-noise spectra across EDFA powers.
		\textbf{g} Allan deviation of the repetition rate across EDFA powers and averaging gate times.
		\textbf{h} Allan deviation at fixed gate time of $65$~s extracted from panel~\textbf{g} (logarithmic scale).
		\textbf{i} Derivative of the repetition rate with respect to pump power, $|dF_R/dP|$, showing strong correlation with the long-term frequency drift in panel~\textbf{g} (logarithmic scale), as expected for adiabatic drifts.
		\textbf{j} Repetition-rate variation from its average value (98~GHz) versus EDFA power.
	}	\label{fig:Noise}
\end{figure*}

Starting from pump values for which only the even-soliton train is present
(\emph{Even Soliton}, brown region), the sweep reveals the symmetry-breaking threshold
of the half-cavity shift $\mathcal S$ that marks the onset of the molecule
(\emph{Topological Molecule}, green region). A subsequent breaking of spatial parity
$\mathcal P$ delineates the transition into a \emph{Drifting Molecule}
(yellow region), which eventually destabilises into an \emph{Oscillating Molecule}
(dark yellow region) through an Andronov--Hopf bifurcation%
\cite{Lugiato2015Nonlinear,NonlinearDynamics2007}, typical of soliton-breather
dynamics~\cite{Kivshar2003Optical,Lugiato2015Nonlinear}.

To characterise these transitions, we monitor—both theoretically (Fig.~3a–e)
and experimentally (Fig.~3f–i)—the average power (Fig.~3a,f,h), the
repetition rate (purple traces in Fig.~3b,e,g,i), and the average optical
frequency (black traces) for the even and odd modes. The latter acts as an effective \emph{momentum} of
the molecule: it reflects the mean spectral shift of the comb and is
dynamically linked to the molecule’s velocity, which corresponds to the
repetition rate (Methods, Fig.~E1).
 Numerical modelling further enables us to
associate a geometrical phase to the even and odd sectors (Fig.~3c,d).
This so-called Zak phase~\cite{Zak1989} is the Berry phase implementation for
one-dimensional lattices~\cite{Resta1994,Berry1984}. It takes quantised
values of $0$ or $\pi$ within the symmetry-protected region, serving as a
topological invariant (Supplementary~S5, S6), providing a consistent theoretical
marker for the transitions observed experimentally.
The onset of the \mobius molecule is marked by a sharp increase in the power
of the odd sector, rising from noise as expected at an effective lasing threshold
(labelled ``Lasing'', brown line in Fig.~3a,f,h). This distinct transition originates
from a destabilisation of the even-soliton train—whose average power remains
essentially constant—and corresponds to the symmetry breaking of the half-cavity shift operator $\mathcal S$.
At this transition, the Zak invariant of the odd sector acquires the quantised
value $\pi$ (Fig.~3c,d). Together with key properties of the molecule—
momentum and velocity (Fig.~3b,e), both experimentally accessible
(Fig.~3g,i)—these quantities \emph{remain unchanged} under adiabatic variations
of the gain/EDFA pump. This invariance is a hallmark of a
symmetry-protected topological state, here enabled by the preservation of the
spatial parity operator~$\mathcal P$ after the transition.

A clear signature of the parity-protected topological regime is its dissolution
via a second symmetry-breaking transition (``Drift'', green boundary between the
green and yellow regions). This transition is clearly
visible in Fig.~3b,e,g,i: in the theoretical panels (b,e), the molecule
velocity and momentum show a sharp change in their dependence on the gain
parameter. Similarly, in the experimental panels (g,i) the repetition rate and
average optical frequency display the same sharp transition as a
function of the EDFA pump power.

Importantly, the molecule
can change velocity, i.e., drift, in either direction, accompanied by a corresponding shift of the
average spectral frequency (momentum). This bidirectional behaviour is evident
in both the theoretical (Fig.~3b,e) and experimental (Fig.~3g,i) realisations.

Across this transition, the Zak phase evolves adiabatically by following the
momentum~\cite{Resta1994}, the latter being an experimentally accessible
quantity that we track with close agreement to the theoretical model.
The loss of the invariant regime under adiabatic changes, together with the appearance of a \emph{dynamical bifurcation} manifested by the occurrence of two possible drift directions, is a signature of symmetry breaking. Specifically, this transition manifests as a
dramatic rest-to-motion event (Extended Fig.~E1) and is the optical analogue of the
Ising--Bloch transition—originally identified in magnetism and later observed
in nonlinear optical domain-wall systems~%
\cite{Coullet1990Breaking,Michaelis2001Universal,De2002Domain,Esteban2005Controlled}.
It marks the spontaneous breaking of the protecting parity symmetry, here the
spatial parity $\mathcal P$, with concomitant breaking of continuous translational symmetry in a
non-equilibrium cavity (Methods, Supplementary~S4).

At higher pump powers, the system undergoes an oscillatory bifurcation,
evidenced by the appearance of power maxima and minima (Fig.~3a,f,h). This
transition corresponds to an Andronov–Hopf bifurcation%
\cite{Lugiato2015Nonlinear,NonlinearDynamics2007}. 
Remarkably, even in these highly dynamical regimes—including routes to
chaos (Methods)—the overall molecule structure remains intact.

Such robustness is connected to the M\"obius geometry itself (Supplementary S6), 
as further evidenced in Fig.~4, which reports a noise study across four examples 
corresponding to the main regimes identified above. 
The optical spectra, time evolution, 
and histograms of the repetition rate are shown in Fig.~4a–c, while Fig.~4d,e 
illustrate the single-sideband (SSB) phase-noise spectra and Allan deviations 
extracted from the traces in panel~\textbf{b}.
The Allan deviation (see Methods and Supplementary S7), is the standard metric in metrology for quantifying frequency 
stability across different timescales: unlike a simple variance, it measures how 
the average value of a signal changes between consecutive time intervals, called 
gate times. Hence, for long gate times, stabilisation of the time traces reflects 
suppression of long-term drift.

The most evident effect in the time traces is the suppression of drift in the topological 
regime, clearly visible in the Allan deviation of Fig.~4e.
This occurs because, in this region, the repetition rate becomes insensitive to parameter 
variations. For the topological molecule, this insensitivity is enforced by parity 
protection. We observe an almost drift-free Allan deviation ($4\times10^{-10}$ at 
10\,s averaging time), achieved without any active stabilisation
a critical feature for 
metrological applications\cite{Kudelin2024Photonic,Sun2024Integrated,Wu2023Vernier,Wildi2024Phase,Zhao2024All,Jang2018Synchronization,Kim2021Synchronization,Moille2023Kerr,	sun_microcavity_2025,jin_microresonator-referenced_2025}. 

Figure~4g shows that this performance is maintained across the 
entire topological region. This is particularly evident when comparing the long gate-time
Allan deviation (Fig.~4h) with the evolution of the repetition rate (Fig.~4j) and its 
derivative (Fig.~4i) with respect to EDFA power. The physical effect observed here is the 
\emph{invariance} of the repetition rate under adiabatic parameter drifts within the 
symmetry-protected topological region. 

An additional important property of the M\"obius geometry related to noise is its 
intrinsic robustness, which is a feature of the molecule itself—not only of the strictly 
parity-protected topological regime—and arises from the population of both parity sectors 
of the operator~$\mathcal S$. 
At short timescales, the Allan deviation shows noise quenching across all molecular states. 
This behaviour is confirmed by the single-sideband (SSB) phase-noise spectra in Fig.~4d,f. 
Notably, once the topological molecule forms, the phase noise is consistently reduced by 
more than 15\,dB across the full offset-frequency range, achieving $-63$\,dBc/Hz at 1\,kHz 
in free-running operation.

In conclusion, we demonstrate spontaneous and intrinsic noise suppression in a
free-running soliton microcomb by realising a \mobius molecule composed of even
and odd solitons. Under fully autonomous operation, the repetition-rate noise is
self-quenched by more than 15~dB for a 100~GHz comb. This noise reduction
originates from the \mobius geometry of the system.

We identify a symmetry-protected topological regime — preserving spatial parity —
in which the long-term drift of the repetition rate is minimised. Remarkably,
this regime emerges in a standalone laser system with no external control or
stabilisation, governed solely by intrinsic and easily accessible cavity
parameters.

This work represents a significant step toward intrinsically low-noise,
free-running microcombs. The topologically quenched noise performance enables
autonomous, low-noise microwave and optical sources with inherent long-term
stability. Moreover, it benefits externally referenced systems, where the
ultimate spectral purity benefits from a low-noise seed comb. Our results have
direct implications for chip-scale microwave generation, robust optical clocks,
and frequency-comb metrology, particularly in scenarios where complex
referencing architectures are impractical.

	\section*{Methods}

\subsection*{Theoretical Model}

To model our experimental observations, we extend the mean-field framework of
perfectly matched cavities~\cite{Bao2019Laser}—where the ratio between the free-spectral ranges of
the two resonators is an integer—to the interleaved configuration used
here, in which the ratio is a half-integer. A full derivation is provided in
Supplementary~S1 and leads to the set of coupled equations

\begin{align}
	\partial_t a (x,t) &=
	\frac{i \zeta_a}{2}\partial_x^{2} a
	+ i |a|^2 a
	- \kappa a
	+ \sqrt{\kappa}\sum_{q=-N}^{N} b_q ,
	\label{S2.1} \\
	\partial_t b_q (x,t) &=
	\frac{i \zeta_b}{2}\partial_x^{2} b_q
	+ 2\pi i\!\left(\Delta - q - \tfrac14\right)b_q
	+ \tfrac{2\pi i}{4}\, b_q(x+\tfrac12,t)
	+ g\, b_q
	+ \sigma\, \partial_x^{6} b_q
	- \sum_{p=-N}^{N} b_p
	+ \sqrt{\kappa}\, a .
	\label{S2.2}
\end{align}

Here, \(a\) and \(b_q\) denote the field envelopes in the microresonator and
amplifying cavities. Even and odd components are not explicitly present at this
stage; they are extracted from these fields by applying the relations in
Eqs.~\eqref{prS}. The spatial coordinate \(x \in [-\tfrac12,\tfrac12]\) is
normalised to one microcavity round trip with periodic boundary conditions.
The slow time \(t\) is normalised to the main-cavity round trip,
\(F_b^{-1}\). The indices \(q \in [-N,N]\) define a truncated supermode basis
of \(2N{+}1\) modes for the main cavity; we use \(N=5\) in the simulations
following Ref.~\cite{Bao2019Laser}.

The key effect of the interleaved structure of Eqs.~\eqref{eoComb} appears in the
spatial coupling term \(2\pi ib_q(x+\tfrac12)/4\). Because of this term, the
half-cavity shift operator \(\mathcal S\) acts as a discrete symmetry of the
system (Supplementary~S2).
As parameters, we use the normalised anomalous group-velocity dispersions
\(\zeta_a = 1.25\times10^{-4}\) and \(\zeta_b = 2.5\times10^{-4}\), the spectral
filtering coefficient \(\sigma = (1.5\times10^{-4})^{3}\), and the coupling
constant \(\kappa = 1.5\pi\). These values are fixed by the experimental setup
(Supplementary~S2.4) and are the same as those used in
Ref.~\cite{Rowley2022Self}. They remain unchanged across all simulations.
Group-velocity mismatch is neglected here and discussed in Supplementary~S8.

Conversely, the detuning \(\Delta\) and the normalised gain \(g\) (with
\(0<g<1\)) vary across the simulations, as they are actively modified in the
experiments.
To gain insight into the accessible operating regimes, we first classified into the
\((\Delta,g)\) plane all dynamical states that the system develops when
initialised with the even-soliton solutions, across their full existence domain
(Supplementary~S3, S4). All our experiments are consistently started from
even-soliton states; this classification therefore provides the appropriate
starting point for the analysis. The map in Fig.~E1a identifies the
even-soliton (brown), topological molecule (green), drifting molecule (yellow),
and oscillatory molecule (dark yellow) regimes. Outside these regions, the pink
and light-blue areas denote, respectively, non-molecule unstable regimes and
parameter values for which the even-soliton state does not exist
(Supplementary~S3, S4). The topological, drifting, and oscillatory molecule
states appear consistently in these domains, with representative examples shown
in Fig.~E1b–e.
In the \emph{parity-symmetric regime} (Fig.~E1b, \(g=0.0343\), \(\Delta=0.34\)),
the pulse propagates at rest in the co-moving frame. The odd soliton is
generated during propagation and exhibits a discrete \(\pi\)-phase jump on
top of the even-soliton peak, a clear topological signature
(Supplementary~S4).

In the \emph{drifting regime} (Fig.~E1c–d, \(g=0.0366\), \(\Delta=0.35\)),
the discrete \(\pi\)-jump disappears. Spatial parity \(\mathcal P\) is broken,
and the odd component shifts toward one of the even solitons. The system
spontaneously develops a drift in either spatial direction, via an Ising-Bloch-like transition (Supplementary S4).

A third \emph{oscillatory regime} (Fig.~E1e, \(g=0.0386\), \(\Delta=0.36\))
exhibits periodic oscillations of both position and intensity.

To interpret the experiment and the transition sequence in Fig.~3, it is
essential to clarify how the EDFA pump modifies the normalised gain
\(g\) and detuning \(\Delta\). The detuning \(\Delta\), in particular,
	acquires a dependence on the pump rate due to variations of the refractive index 
	originating in the amplifying medium.
 In our
system, soliton formation arises spontaneously through energy-dependent gain
and thermal nonlinearities~\cite{Rowley2022Self, Cutrona2023Nonlocal}, which
self-lock the system into the soliton state. Once formed, the soliton energy
remains essentially constant; further increases of the EDFA pump primarily raise
\(g\) and induce only small red shifts of \(\Delta\), as long as the soliton is
preserved~\cite{Rowley2022Self}. In this regime, the temporal dynamics of the slow nonlinearities can
be neglected for adiabatic parameter changes.

The same phenomenology holds for the \mobius{} system: once the even-soliton
state is established, measurements show that the energy remains nearly constant
across the entire operating region. Hence, Eqs.~\eqref{S2.1}--\eqref{S2.2} fully
capture the molecule phenomenology. The classification in Fig.~E1a further
highlights a key point: the molecule parity-symmetric, drifting and oscillating states appear consistently in well-defined
regions of the \((\Delta,g)\) plane. These domains are separated by sharp
transition lines corresponding to the breaking of \(\mathcal S\) at the molecule
onset, and to the Ising–Bloch and Andronov–Hopf bifurcations associated with
the drifting and oscillatory molecules (Supplementary~S5, S6).
\begin{figure}[htbp]
	\centering
	\includegraphics[width=\textwidth]{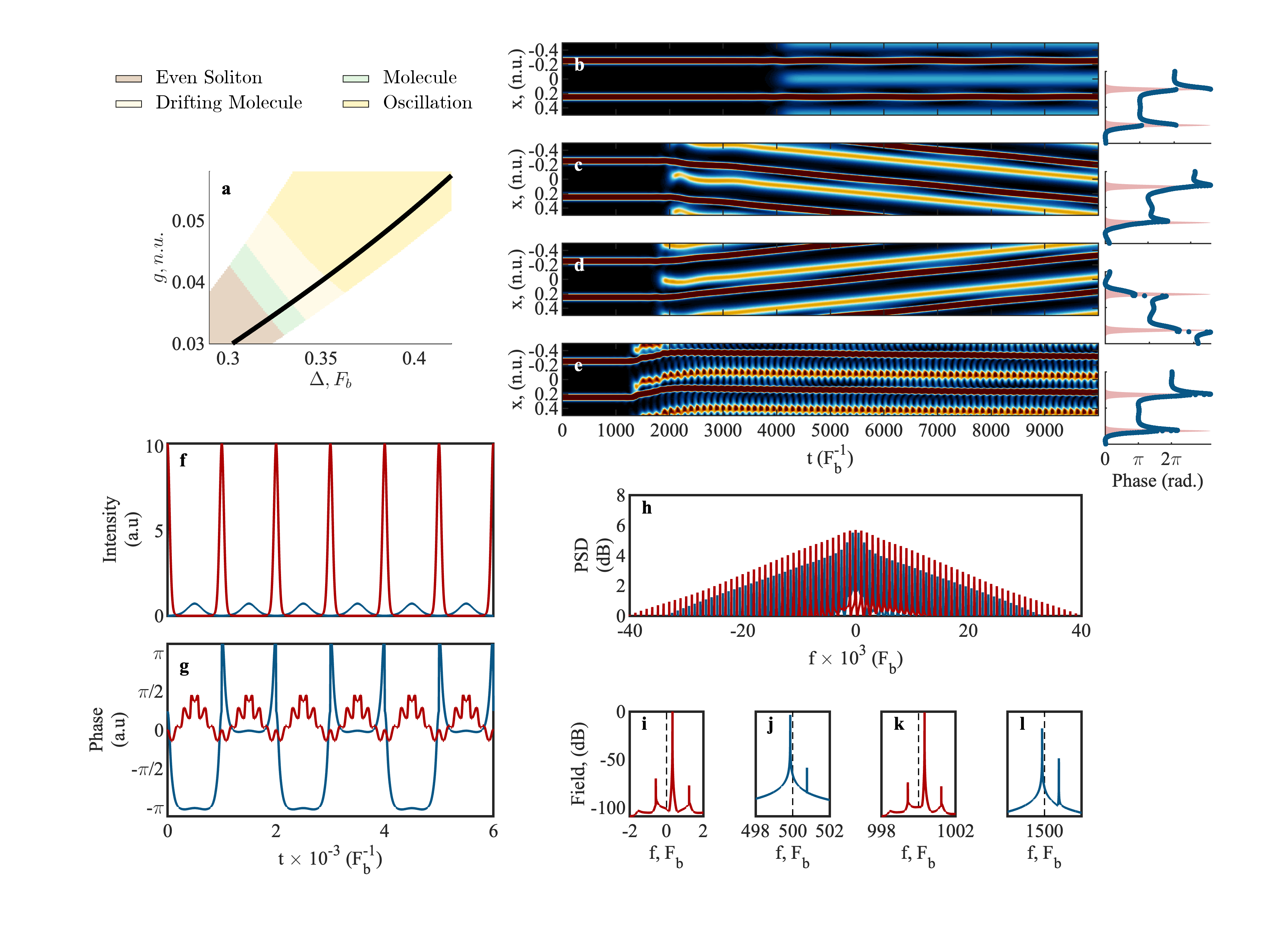}
	\caption*{\footnotesize 
		\textbf{Figure E1 \, Theoretical generation and dynamical regimes of the M\"obius soliton molecule.} 
		Microcavity field \(a(x,t)\) as a function of space \(x\) (normalised to one microcavity round trip) 
		and slow time \(t\) (in units of \(F_b^{-1}\)) obtained from Eqs.~\eqref{S2.1}--\eqref{S2.2} (Methods). 
		\textbf{a} Map in the \((\Delta,g)\) plane summarising the dynamical regimes observed in simulations: 
		even-soliton (brown), topological molecule (green), drifting molecule (yellow), oscillating molecule (dark yellow). 
		The reference path (black) intersects these regions sequentially and forms the numerical basis for Fig.~3 and Fig.~E2.
		\textbf{b--e} Dynamical evolution starting from two even solitons with small random spatial noise, 
		leading to the spontaneous lasing of the odd soliton. 
		Left panels: intensity evolution of \(a(x,t)\). 
		Right panels: phase of the odd component at \(t = 10{,}000\), with the red band marking the even-soliton positions. 
		\textbf{b} Parity-symmetric (resting) state (\(g=0.0343\), \(\Delta=0.34\)). 
		\textbf{c,d} Drifting states obtained for identical parameters (\(g=0.0366\), \(\Delta=0.35\)), 
		but with different initial noise seeds, manifesting opposite velocities due to spontaneous 
		breaking of spatial parity. 
		\textbf{e} Oscillatory state following an Andronov--Hopf bifurcation (\(g=0.0386\), \(\Delta=0.36\)).
		\textbf{f--g} Propagation along the full cavity map using Eq.~\eqref{map}, 
		with \(F_a = 500\,F_b\). Periodic temporal pulses are obtained by sampling the microcavity field at a fixed coordinate, \(a(x{=}0,t)\), 
		thus reproducing the laser output. 
		\textbf{f} Intensity of the even (red) and odd (blue) components. 
		\textbf{g} Corresponding phase profiles, showing a localised \(\pi\)-phase jump in the odd component at each even-soliton position. 
		\textbf{h} Comb spectrum.
		\textbf{i--l} Representative comb lines for mode indices \(m=0,1,2,3\), extracted from panel~\textbf{h} 
		around the frequencies \(f = m\times500\,F_b\). 
		Even and odd components exhibit distinct CEO frequencies, consistent with the interleaved structure.
	}
	\label{fig:FigE1}
\end{figure}

Critically, trajectories that cross the same bifurcation lines exhibit the same
qualitative transitions, provided the path intersects these lines in a fixed
order. A sequential increase of the EDFA pump, as done in the experiments,
traces an effectively monotonic path in \((\Delta,g)\), with increasing gain and
a small red detuning. Because the experiment always starts below the molecule
onset threshold, this path consistently crosses first the molecule-formation
threshold, then the drift onset, and finally the oscillatory instability—the
resulting regimes are therefore dynamically equivalent. \emph{This explains the
	repeatability of our results.}

Naturally, spatial parity \(\mathcal P\) must be preserved in the experiments.
Operatively, this requires matching the cavity lengths to the half-ratio.
A small mismatch introduces a group-velocity offset that biases, but does not
destroy, the transition sequence, as shown in Supplementary~S8. This is why we
are able to adjust our experimental parameters to reproduce the bifurcation
diagrams consistently.

For the numerical analysis, we focus on describing the sequence of transitions.
We therefore adopt a monotonic \emph{reference path} in \((\Delta,g)\) that
reproduces the observed ordering of bifurcations, and we use it consistently to
study the dynamics (Fig.~E1a, black line). The parameters in Fig.~E1 belong to
this set. For the bifurcation diagrams, the corresponding values of \(g\) and
\(\Delta\) are reported in Fig.~E2, which extends the parameter range compared
to the theoretical results of Fig.~3.

We report a representative optical spectrum for each regime in Fig.~E2a,f. To
the amplitude (Fig.~E2b,g) and repetition-rate (Fig.~E2e,i) traces, we add the
radio-frequency noise obtained by Fourier transforming the evolution of the
intracavity energy (Supplementary~S4, S6). Figures~E2c,h show the baseband
noise, revealing the onset of the Hopf frequency at the oscillatory
transition. Figures~E2d,i, conversely, show the radio-frequency noise around the
FWM CEO-mismatch frequency \(f_{\mathrm{CM}} \approx F_b\), which appears at the
molecule onset. The extended data in Fig.~E2 also illustrate the richer
dynamics that emerge beyond the Andronov–Hopf bifurcation, including a route to
chaos.

\begin{figure}[h!]
	\centering
	\includegraphics[width=\textwidth]{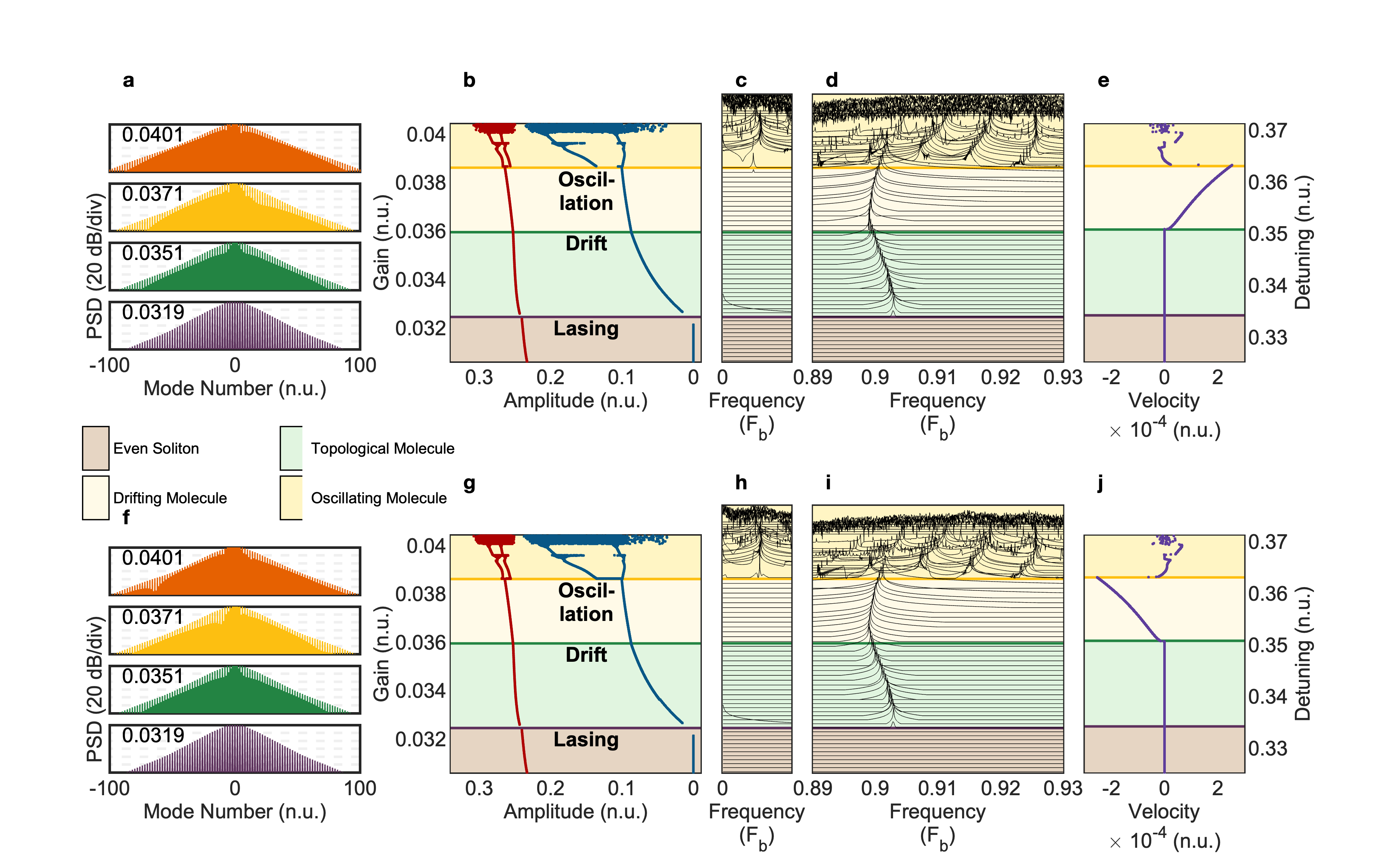}
\caption*{\footnotesize 
	\textbf{Figure E2 \, The \mobius molecule dynamics: extended theoretical bifurcation diagram.}
	Same parameter set as Fig.~3 of the main text, but extended here to show the full sequence
	of dynamical transitions, including the route to chaos.
	Gain $g$ and detuning $\Delta$ co-vary along the monotonic \emph{reference path}.
	The gain scale is shown on the left $y$-axis of panels \textbf{b},\textbf{g} and the detuning on the right
	$y$-axis of panels \textbf{e},\textbf{j}. 
	\textbf{a},\textbf{f} Comb spectra (PSD versus mode number) at four representative gain values
	along the reference path, corresponding to the even-soliton (brown), topological
	molecule (green), drifting molecule (yellow) and oscillatory molecule (dark yellow) regimes. Gain values are indicated in each plot.
	\textbf{b},\textbf{g} Average amplitudes of the high-energy (even, red) and low-energy (odd, blue)
	components versus gain. Amplitude oscillations identify stationary, drifting and oscillatory
	regimes.
	\textbf{c},\textbf{h} Baseband radio-frequency spectra obtained from the intracavity energy
	evolution, revealing the onset of the Hopf frequency at the oscillatory transition.
	\textbf{d},\textbf{i} Radio-frequency spectra around the CEO-mismatch frequency
	$f_{\mathrm{CM}} \approx F_b$ (insets: zooms on the sidebands).
	\textbf{e},\textbf{j} Molecule velocity versus gain (purple), with the corresponding
	detuning $\Delta$ reported on the right axis.
	The odd-mode lasing threshold occurs at $g \simeq 0.0325$; in the resulting topological
	(resting) regime the velocity remains zero (repetition rate invariant) under gain and
	detuning variations. At $g \simeq 0.036$ an Ising--Bloch-like rest-to-drift transition
	occurs, and at $g \simeq 0.038$ an Andronov--Hopf bifurcation generates symmetric RF
	sidebands. At higher gain the system undergoes a route to chaos, without compromising
	the structural robustness of the M\"obius molecule.
	Panels \textbf{f--j} show a second realisation with identical parameters, where the
	drifting regime appears with opposite sign, reflecting the spontaneous breaking of
	spatial parity $\mathcal{P}$.
}
	\label{fig:FigE2}
\end{figure}

The following describes the operative definitions of the velocity, momentum, and
Zak phase used in Fig.~3. The velocity of the molecule is obtained by tracking
the coordinate \(x_c(t)\) of one of the soliton peaks along the propagation and
computing its derivative \(v_c(t) = \partial_t x_c(t)\). As illustrated in
Fig.~E1b–e, once the molecule has settled after any transient,
\(v_c(t)\) approaches a constant value in both the topological and drifting
regimes (with \(v_c = 0\) in the topological case). In the oscillatory regime,
\(v_c(t)\) acquires a small modulation at the Hopf frequency, reminiscent of
breather-soliton dynamics~\cite{Kivshar2003Optical,Grelu2016Nonlinear}. We
extract this value by averaging \(v_c(t)\) over a sufficiently long propagation
interval after the transient.
The optical “momentum” is defined as
\begin{equation}
	\hat{L}[a(x,t)] = 
	\frac{
		\int_{-1/2}^{1/2} \mathrm{Im}\!\left[a^*(x,t)\,\partial_x a(x,t)\right]\,dx
	}{
		\int_{-1/2}^{1/2} |a(x,t)|^2\,dx
	}
	=\;
	\frac{\sum_m (2\pi m)\,|a_m|^2}{\sum_m |a_m|^2},
	\label{momentum}
\end{equation}
where the second identity follows directly from the Fourier expansion
\(a(x) = \sum_m a_m\, e^{2\pi i m x}\). This expression shows that
\(\hat{L}\) represents the average mode index (or equivalently, the average
optical frequency) of the waveform. It provides a sensitive indicator of
spectral asymmetries, as illustrated in Fig.~E2a,f.
Finally, to identify the regions of parameter space in which the nonlinear state
preserves its topological character, we compute a Zak invariant that can be
directly applied to the nonlinear solutions of our system. In his seminal work,
Zak introduced a method to evaluate the Berry phase in one-dimensional systems
by extending the parameter space through an auxiliary Bloch momentum
\cite{Zak1989,Resta1994}. This construction is standard in condensed matter physics
and has since been generalised to extract topological invariants of nonlinear
waves, including in nonlinear optical systems
\cite{Li2023Zak,Smirnova2020Nonlinear}.

Among more recent developments, the work of Fan \emph{et al.}~\cite{Fan2022Topological} on
topological soliton metacrystals in microcombs provides a directly relevant
methodology. We adopt this approach in Supplementary~S5–S6 to perform the
numerical evaluation of the invariant. This constitutes the most technical part
of our analysis, discussed extensively in those sections together with the
state–symmetry breaking transitions. Here, we summarise only the essential
elements required to convey the spirit of the calculation.

As introduced in Supplementary~S5, we compute the Zak phase by introducing an
extended Bloch momentum \(l \in [-\tfrac12,\tfrac12]\) and replacing the spatial
derivatives in Eqs.~\eqref{S2.1}–\eqref{S2.2} with the covariant derivatives
\[
D_{\pm l} = \partial_x \pm 2\pi i\, l.
\]
Starting from \(l=0\), we take a molecule state from the simulation set and vary
\(l\) adiabatically in small steps, extracting the fields
\(a(l,x)\) and \(b_q(l,x)\) after transient relaxation at a fixed slow time \(t\)
for each value of \(l\). The Zak phases \(\mathcal{Z}_{(e,o)}\) associated with
the even and odd components are then defined as
\[
\mathcal{Z}_{(e,o)}
= \mathrm{i}
\int_{-\tfrac12}^{\tfrac12}\!\!\int_{-\tfrac12}^{\tfrac12}
a_{(e,o)}(l,x)^*\,\partial_l a_{(e,o)}(l,x)\, dl\,dx
\;+\;
\mathrm{i}\!\sum_{q=-N}^N\!\int_{-\tfrac12}^{\tfrac12}\!\!\int_{-\tfrac12}^{\tfrac12}
b_{q,(e,o)}(l,x)^*\,\partial_l b_{q,(e,o)}(l,x)\, dl\,dx,
\quad \mathrm{mod}(2\pi).
\]

This procedure extracts the geometrical phase invariant, which
quantises to \(\pi\) for the topological states, as demonstrated in
Fig.~3c,d.
Before closing this section, we illustrate the simulation of a full temporal comb,
providing a representative example in which \(\mathcal P\) is preserved
(Fig.~E1f–l for \(g = 0.0343\), \(\Delta = 0.34\)). Specifically, we propagate
the field along the full cavity map rather than in the co-moving frame. This is
implemented by modifying the left-hand side of the equations to include the
Galilean correction
\begin{align}
	\partial_t a(x,t)\;\longrightarrow\;\partial_t a(x,t)
	-\frac{F_a}{F_b}\,\partial_x a(x,t),\qquad
	\partial_t b_q(x,t)\;\longrightarrow\;\partial_t b_q(x,t)
	-\frac{F_a}{F_b}\,\partial_x b_q(x,t),
	\label{map}
\end{align}
which accounts for the ratio of the free-spectral ranges of the two cavities
and correctly reproduces the physical circulation.

Sampling the field at a fixed coordinate, e.g.\ \(a(x{=}0,t)\), yields the
pulsed temporal behaviour shown in Fig.~E1f,g. Importantly, the odd component
displays the characteristic \(\pi\)-phase jumps, while the even component
remains flat-phased, consistent with a transform-limited soliton invariant
under propagation. Discontinuous \(\pm\pi\) slips are a well-known signature of
topological states in nonlinear optical cavities, appearing as domain walls in
parametric oscillators%
~\cite{Coullet1990Breaking,Michaelis2001Universal,De2002Domain,Esteban2005Controlled,
	Trillo1997Stable,Oppo1999From,Englebert2024Topological}
and in vectorial polarization resonators%
~\cite{Gallego2000Self,Garbin2021Dissipative,Lucas2024Polarization}.

The simulation also reproduces the experimental spectral features (Fig.~2d):
the even and odd comb lines lase on opposite sides of the microresonator
resonance, as visible in the interleaved detuning pattern of Fig.~E1i–l.

	\subsection*{Molecule Transitions}
\begin{figure}[t]
	\centering
	\includegraphics[width=\textwidth]{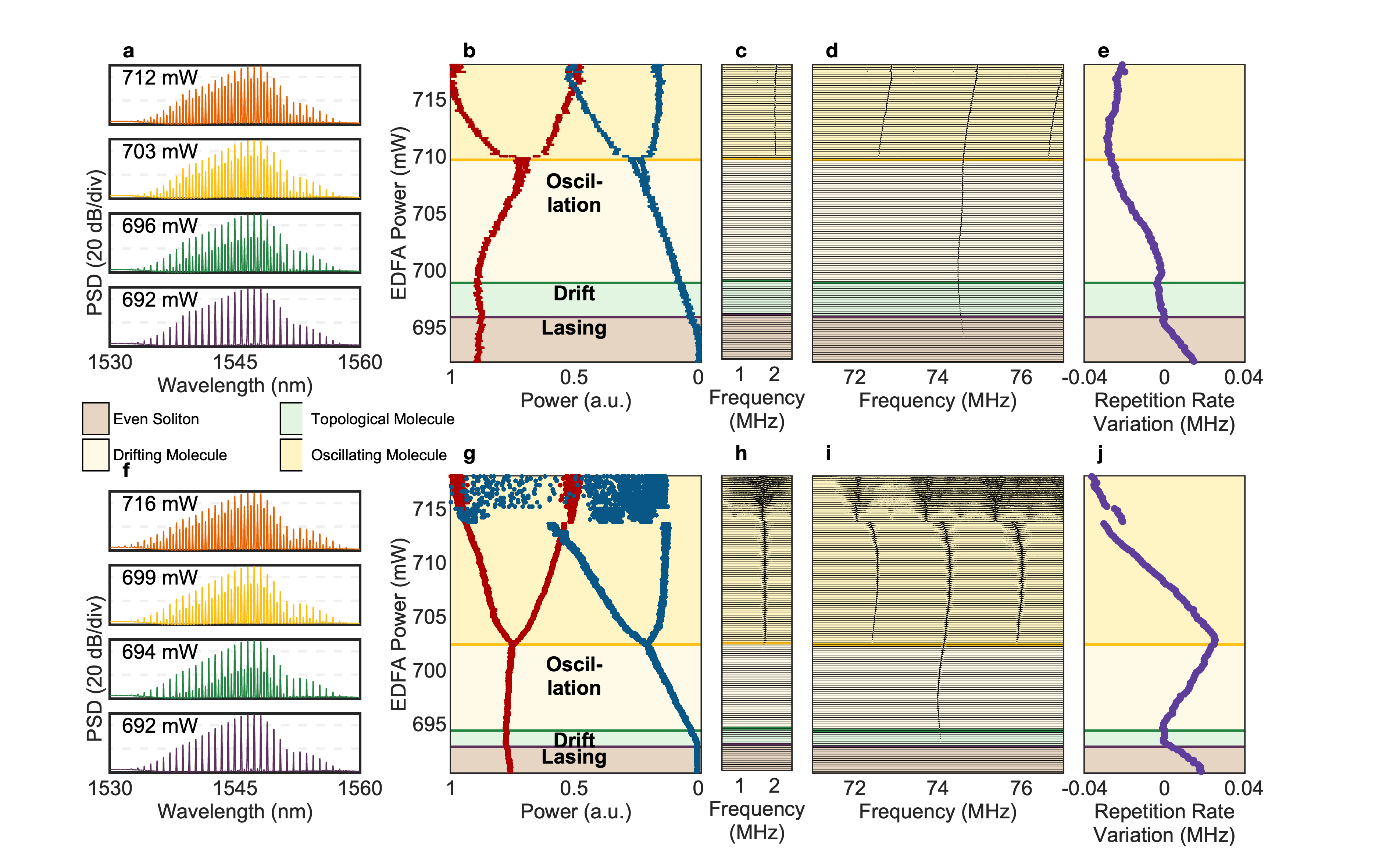}
\caption*{
	\textbf{Figure~E3 The \mobius molecule dynamics: extended experimental bifurcation diagram.}
	\textbf{a,f}, Optical spectra at representative EDFA pump powers, covering the 
	four dynamical regimes identified in the main text: 
	Even Soliton (brown), Topological Molecule (green), Drifting Molecule (yellow), 
	and Oscillating Molecule (orange). 
	Each spectrum shows the characteristic redistribution of power between even and odd 
	comb lines as the molecule forms and destabilises.
	\textbf{b,g}, Average power of the high-energy (red) and low-energy (blue) components 
	as a function of EDFA pump power. 
	The Lasing boundary marks the birth of the odd soliton; 
	the Drift boundary corresponds to the breaking of spatial parity~$\mathcal{P}$; 
	the Oscillatory boundary indicates an Andronov--Hopf bifurcation, followed at higher pump 
	powers by oscillations that evolve along a route to chaotic dynamics.
	\textbf{c,h}, Radio-frequency (RF) spectra around baseband showing the appearance of a 
	periodic oscillation tone at $\sim$2~MHz at the AH bifurcation.
	\textbf{d,i}, RF spectra around the cavity-mode frequency 
	$f_{\mathrm{CM}} \sim F_b \approx 75$~MHz. 
	The Hopf bifurcation produces symmetric RF sidebands whose amplitude grows with pump, 
	followed by spectral broadening indicative of dynamics approaching chaos.
	\textbf{e,j}, Repetition-rate evolution relative to the topological molecule. 
	In both independent realisations, the repetition rate remains nearly invariant within the 
	topological region, undergoes a rest-to-drift transition at the parity-breaking point, 
	and displays oscillatory (and, in the second dataset, near-chaotic) fluctuations beyond the Hopf line. 
	Despite experimental fluctuations, the same sequence of dynamical transitions 
	(Lasing $\rightarrow$ Topological Molecule $\rightarrow$ Drift $\rightarrow$ Oscillation) 
	is reproduced in both datasets, demonstrating the robustness and repeatability of the 
	underlying bifurcation structure.
}
\end{figure}

Figure E3 reports additional data from the set in Fig.~3, with Fig.~E3a,f showing representative spectra. 
We constructed an experimental bifurcation diagram (Fig.~E3b--g) by isolating a single comb line from each dataset 
(1547.32\,nm and 1548.51\,nm, respectively), allowing us to track their time evolution and report the peak 
dynamics of the two soliton components in Fig.~3a.

We verified that the spectral properties of the system were preserved across the full operating range, and 
extracted the RF noise both at baseband (Fig.~E3c,h) and around the characteristic CEO-mismatch frequency 
$f_{\mathrm{CM}}$ (Fig.~E3d,i) arising from the FWM process (Fig.~2c). 
The repetition rate of the molecule was down-converted 
(see \textit{Repetition Rate Measurement and Relative CEO Extraction}), and its dependence on the EDFA 
pump power was recorded (Fig.~E3e,j). 
For the second dataset, we report the full bifurcation sequence up to the onset of chaos.

The theoretical (Fig.~E2) and experimental (Fig.~E3) bifurcation 
diagrams reveal a consistent sequence of transitions. 
The effective lasing threshold of the odd state appears at EDFA powers $P = 692$\,~mW (Fig.~E3b) and 
$P = 694$\,~mW (Fig.~E3g), coinciding with the emergence of the RF tone at 
$f_{\mathrm{CM}} \approx 75$\,\text{MHz}. 
This marks the formation of a topological M\"obius molecule, characterised by an almost flat repetition 
rate across the accessible EDFA range. 
As in hybridised soliton states observed in semiconductor cavities~\cite{Letsou2025Hybridized}, the 
odd-resonance soliton would not be observable without the presence of the even-resonance soliton.

The ``resting'' configuration is disrupted at $P \approx 698$\,~mW in Fig.~E3b--e and $P = 694$\,~mW in 
Fig.~E3g--j, where the molecule begins to drift, marking the spontaneous breaking of continuous 
translational symmetry: the Ising--Bloch-like transition. 
Notably, no new spectral components appear in the RF spectrum at this transition, but the frequency 
$f_{\mathrm{CM}}$ shifts, indicating a change in the relative soliton phases, also visible in the theoretical analysis in Fig. E2,d,i.

The two experiments display opposite drift directions, a clear signature of symmetry breaking 
(see Supplementary Section~6 for the eigenvalue analysis). 
In both cases, the repetition-rate traces show a mild bias immediately before molecule formation, 
arising from the dependence of the cavity length on EDFA pump power—a behaviour also visible in the 
optimised two-soliton state under group-velocity-matched conditions reported in Fig.~E4a. 
Importantly, this initial bias (i.e.\ the slope of the repetition rate before molecule formation) is 
small and identical in both measurements, suggesting that the direction of motion after the drift 
transition is not externally imposed but instead emerges from spontaneous symmetry breaking.

At higher pump powers, the system undergoes an Andronov--Hopf bifurcation 
($P \approx 708$\,~mW and $P \approx 704$\,~mW), generating sidebands in the RF spectrum that evolve into 
nearly-chaotic oscillations. These oscillations appear in the dataset of Fig.~E3g--j above 
$P \approx 712$\,~mW. 
The molecular state persists across all regimes, even close to the onset of chaos, demonstrating that 
the molecule is highly resilient. The bifurcation sequence (even soliton $\rightarrow$ topological 
$\rightarrow$ drift $\rightarrow$ Hopf $\rightarrow$ route to chaos) is consistently observed across 
independent runs. 
These experimental observations closely match the theoretical modelling in Fig.~E2, confirming the 
M\"obius molecule dynamics.

For comparison, Fig.~E4 summarises the noise properties of an optimised two-soliton state under 
matched cavity conditions, at similar EDFA power values. 
A typical dependence of the repetition rate on EDFA pump power is shown in Fig.~E4a, consistent with the 
dynamics reported in Ref.~\cite{Rowley2022Self}. 
As discussed above, in our system variations in laser pumping act on multiple parameters (gain and phase) 
through slow, nonlocal nonlinearities (thermal and gain effects), which constitute the primary mechanism 
of self-induced locking and therefore sustain the soliton. 
Compared to this standard soliton case, the M\"obius molecule exhibits at least a 10\,dB reduction in 
 phase noise (cf. Fig.~4d). 
Moreover, the robustness of the group-velocity variation against parameter changes reduces the long-term 
drift, as also observed in the direct comparison reported in Fig.~1c,f.

\begin{figure}[h!]
	\centering
	\includegraphics[width=\textwidth]{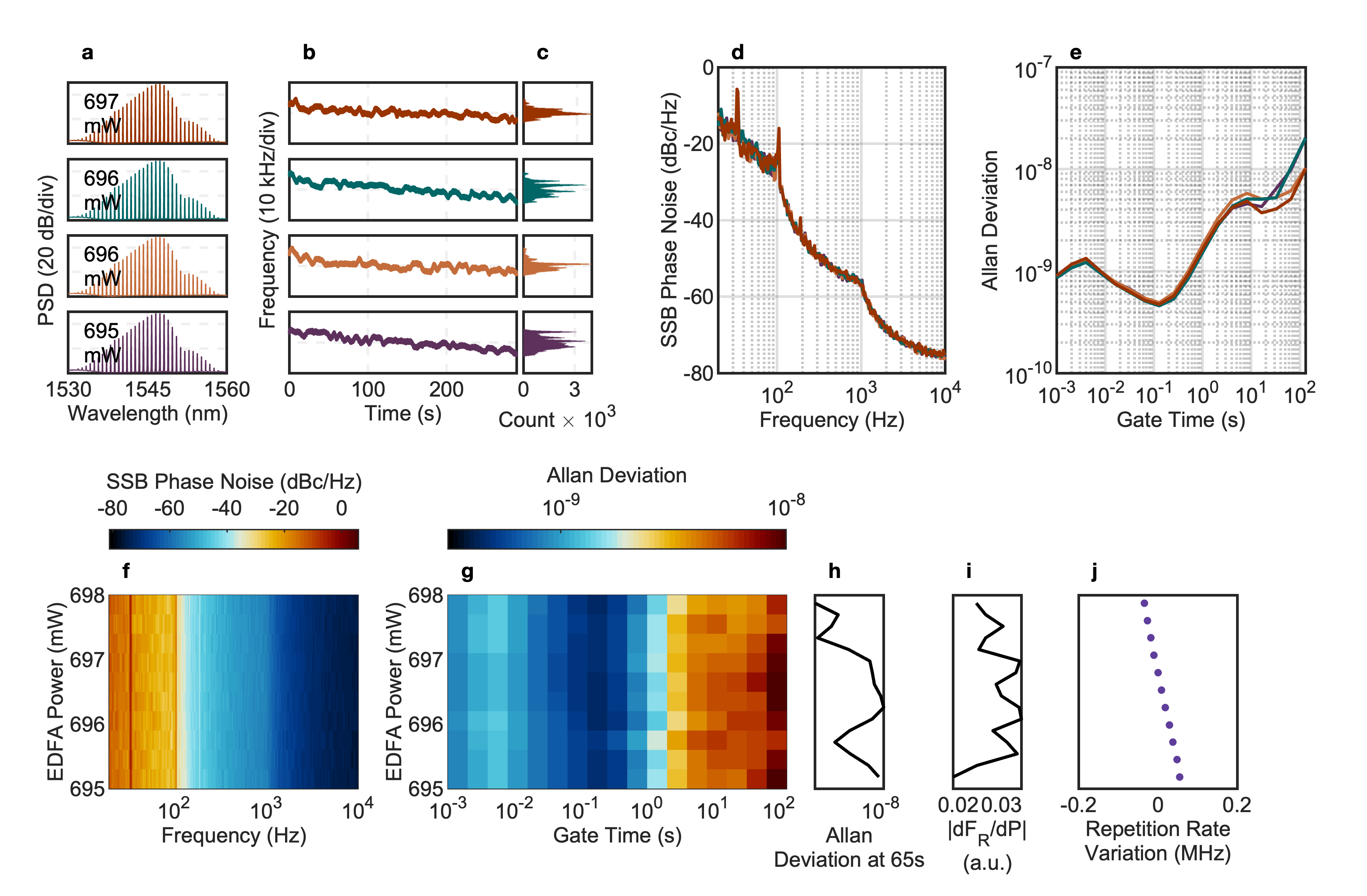}
\caption*{
	\footnotesize
	\textbf{Figure E4 Standard two-soliton state: noise study as a function of EDFA power.}
	Reference case obtained for group-velocity-matched main and ring cavities, where a standard 
	two-soliton state forms without M\"obius geometry.
	\textbf{a} Representative optical spectra at selected EDFA powers.
	\textbf{b} Time evolution of the repetition rate over 180~s for the same operating points.
	\textbf{c} Corresponding histograms of the repetition-rate fluctuations. 
	\textbf{d} Single-sideband phase-noise spectra (colours matching panels \textbf{c}). 
	\textbf{e} Allan deviation of the repetition rate for the traces in \textbf{c}.
	\textbf{f} Single-sideband phase-noise spectra across EDFA powers.
	\textbf{g} Allan deviation of the repetition rate across EDFA powers and averaging gate times.
	\textbf{h} Allan deviation at fixed gate time ($65$~s) extracted from panel~\textbf{f} (logarithmic scale).
	\textbf{i} Derivative of the repetition rate with respect to pump power ($|dF_R/dP|$), showing strong 
	correlation with the long-term frequency drift in panel~\textbf{g} (logarithmic scale).
	\textbf{j} Repetition-rate variation from its average value (98~GHz) versus EDFA power.
}
\label{fig:FigE4}
\end{figure}

\subsection*{Setup}

The microcomb source is implemented using a self-starting, microresonator-filtered laser system
\cite{Bao2019Laser,Rowley2022Self}. The experimental layout is shown in Fig.~2a.
The microcavity is a high-index doped-silica four-port on-chip microring with 
FSR $= 49.05$~GHz, linewidth $140$~MHz, and $Q$-factor of $1.4$ million, operating in the anomalous-dispersion regime. 
The microring completes the main cavity, with its input port connected to the amplifier output and the cavity closed at the drop port. 
The laser output is collected at the through port of the resonator.

The laser cavity is a unidirectional fibre loop (FSR $\approx 75$~MHz) comprising an EDFA, a polarisation controller, 
a tap coupler to extract the intracavity power (at the resonator drop port), and an optical isolator to enforce 
unidirectional propagation. 
A spectral bandpass filter centred at $1550$~nm (FWHM $=12$~nm) is inserted into the main cavity to flatten the EDFA’s 
asymmetric gain spectrum, thereby determining the laser gain bandwidth. 
The cavity length is fine-tuned using a motorised free-space delay line. 
By adjusting the delay line, the laser modes can be positioned between every two microcavity resonances, yielding the 
interleaved configuration required for the formation of \mobius soliton molecules. 
The EDFA pump power is controlled by adjusting the current of a temperature-stabilised $976$~nm pump diode 
(linewidth $15$~MHz).

To track the onset of the dynamical transitions, we monitor the temporal power evolution of individual soliton 
components. 
The through-port output is spectrally resolved using two narrowband WDM filters (bandwidth $\sim 50$~GHz), which isolate 
representative even and odd comb lines at $1547.32$~nm and $1548.51$~nm. 
The filtered signals are detected using amplified photodiodes, and their temporal oscillations are recorded on an 
oscilloscope (RIGOL MS05074). 
From these traces we extract the maximum and minimum power values, which are used to construct the bifurcation maps of 
Fig.~3 and Fig.~E3. 
An optical spectrum analyser (YOKOGAWA AQ6370D) is used to measure the optical spectra.

The positions of the main-cavity lasing lines (and therefore the absolute comb-line positions) relative to the microring 
resonances are obtained via laser-scanning spectroscopy using a tuneable laser (TUNICS T100S-HP), calibrated against a 
commercial frequency comb (MENLO system)~\cite{Bao2019Laser}.
We characterised the temporal structure of the even and odd soliton pulses using a custom-built, high-dynamic-range second-harmonic autocorrelator.
To isolate the two components, the through-port output was sent through a Mach--Zehnder interferometer (MZI) configured as a 100~GHz spectral filter, which separates the interleaved even and odd comb lines.
This yields two periodic pulse trains with average durations of 1.5~ps (even) and 3~ps (odd), consistent with their respective optical spectra, as measured with our autocorrelator and reported in Fig.~2h,i.
Extended Fig.~E5 reports the molecule spectrum at the MZI input and the spectra at the two output ports, confirming isolation of the even and odd components.
	\begin{figure}[htbp]
	\centering
	\includegraphics[width=0.8\textwidth]{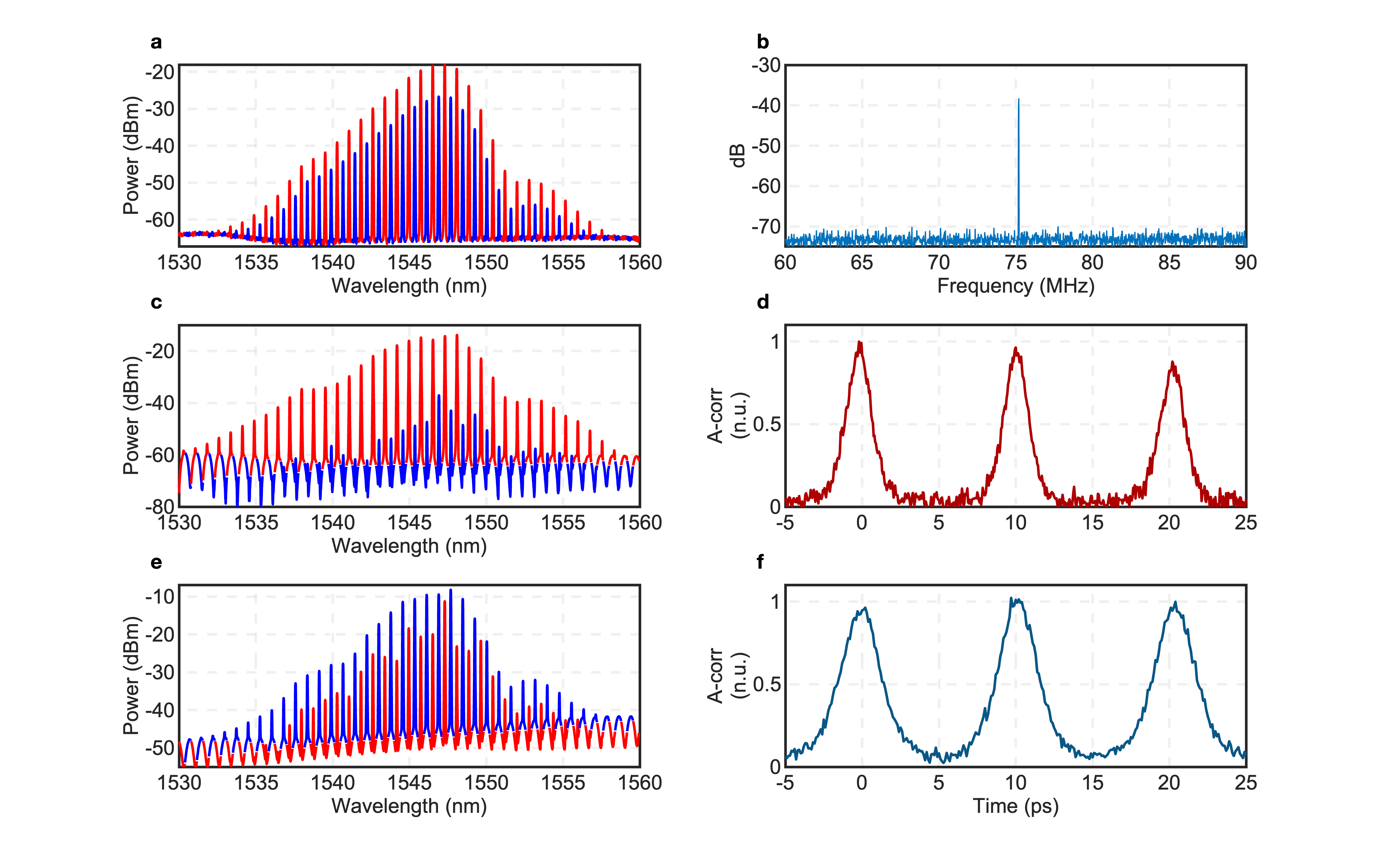}
	\caption*{
		\footnotesize
		\textbf{Figure E5. Additional experimental measurements related to Fig.~2h,i.}
		\textbf{a} Optical spectrum of the molecule; even and odd comb lines are plotted in red and blue for clarity.
		\textbf{b} Radio-frequency spectrum highlighting the four-wave-mixing CEO-mismatch tone. The presence of a single narrow tone indicates the absence of oscillatory sidebands (no Hopf modulation).
		\textbf{c--f} Spectra (\textbf{c,e}) and autocorrelations (\textbf{d,f}) at the two outputs of the Mach--Zehnder interferometer used to separate even and odd resonances.
		\textbf{c,d} Output port selecting the even resonances.
		\textbf{e,f} Output port selecting the odd resonances; the optical spectra show that the even soliton lines are strongly suppressed at this output port after the interferometer.
	}
	\label{fig:FigE5}
\end{figure}

Below the Hopf bifurcation, the odd (M\"obius-symmetric) soliton carries typically $\geq 10$~dB less total pulse energy than the even soliton, making full-train characterisation challenging.
At the repetition rates and average powers involved here (100~GHz, 5--10~mW), the pulses have peak powers on the order of  50~mW (even) and $\sim$1~mW (odd), consistent with the relative spectral amplitudes in Fig.~2h,i and Fig.~2k.
To resolve this contrast, we developed a femtosecond SHG autocorrelator providing $>30$~dB dynamic range at the autocorrelator input while resolving femtowatt-level second-harmonic cross-product power.
The autocorrelator uses a standard non-collinear SHG geometry in a BBO crystal; one arm was amplitude-modulated (`chopped'), and the SHG signal was detected with a photomultiplier tube followed by lock-in demodulation.
This architecture enables observation of both pulses in a single trace: because the even pulse energy is much larger than that of the M\"obius sector, the odd feature is dominated by the cross-correlation between the even and odd sectors.
Although the even-soliton tails still mask the full odd-pulse shape, a clear single peak appears at the midpoint between adjacent even-soliton autocorrelation maxima. Its temporal width is qualitatively consistent with the independently inferred odd-sector pulse duration from Fig.~2h,i, indicating that the odd soliton is interstitial in the time domain between two even solitons.
 Notably, because an autocorrelation trace is intrinsically symmetric
in the delay, a displacement of the odd pulse away from the midpoint would, in
principle, produce a double-peaked feature at the midpoint delay. In practice,
for the small displacements induced by symmetry breaking, this signature cannot
be separated from small symmetric distortions of the pulse envelope and the
associated broadening of the inferred odd-pulse autocorrelation. For this
reason, the rigorous demonstration of parity breaking relies not on the
autocorrelation, but on the bifurcation measurements reported in Fig.~3 and
Fig.~E2.

	\subsection*{Repetition Rate Measurement and Relative CEO Extraction}
	
	We start by recalling the frequencies of each soliton comb line for the $n$th mode:
	\begin{align*}
		f_n^{(\mathrm{e})} &= n F_R + \varphi_E, \\
		f_n^{(\mathrm{o})} &= n F_R + \frac{F_R}{2} +\varphi_O.
	\end{align*}
	where $\varphi_E$ and $\varphi_O$ are the CEO frequencies (in Hz here; a normalised definition is used in the Supplementary) of the even and odd combs. The four-wave mixing interaction 
	\[
	2f_n^{(\mathrm{e})} - f_{n-1}^{(\mathrm{o})} = n F_R + \frac{F_R}{2} +(2\varphi_E - \varphi_O)
	\]
	produces the additional lines in the resonances occupied by $f_n^{(\mathrm{o})}$, as visible in Fig.~2c. This leads to the four-wave mixing tone
	\[
	f_{\mathrm{CM}} = 2 |\varphi_E - \varphi_O|.
	\]
	
	The repetition rate is extracted using the down-conversion scheme~\cite{Cutrona2023Stability} described in Fig.~E6. The molecule is amplitude modulated through an electro-optic modulator (EOM) driven in saturation. In this configuration, a rich harmonic content is generated in the form of sidebands around each comb line (i.e., up to the third harmonic of the modulation frequency in our implementation). The interaction between high-order sidebands of adjacent comb lines produces easily accessible RF beat notes that encode the stability properties of the comb spacing (see Fig.~E6a).
	
	The down-conversion scheme creates three RF lines at frequencies $F_{\mathrm{mix}}$ and $F_{\mathrm{mix}}^{\pm}$. These are obtained by mixing the comb lines with harmonics of an electro-optic modulator at the frequency $F_{\mathrm{EO}} = 16.3174$~GHz ($< F_R/6$), driven with a GPS-referenced Keysight EXG N5173B at approximately 20~dBm.
	
	Specifically,
	\[
	F_{\mathrm{mix}} = (f_{n+1}^{(\mathrm{e,o})} - 3F_{\mathrm{EO}}) - (f_n^{(\mathrm{e,o})} + 3F_{\mathrm{EO}}) = F_R - 6F_{\mathrm{EO}}.
	\]
	Mixing two lines from different solitons yields:
	\[
	F_{\mathrm{mix}}^{(\pm)} = f_n^{(\mathrm{e,o})} - (f_n^{(\mathrm{o,e})} - 3F_{\mathrm{EO}}) = \frac{F_R}{2} - 3F_{\mathrm{EO}} \pm  |\varphi_E - \varphi_O| = \frac{F_{\mathrm{mix}}}{2} \pm \frac{f_{\mathrm{CM}}}{2}.
	\]
	Similar derivations apply to other harmonics producing the same frequency difference. These components are shown in Fig.~E6, with $F_{\mathrm{mix}}^{\pm}$ containing the CEO mismatch term $|\varphi_E - \varphi_O|$.
	
	It follows that
	\[
	F_{\mathrm{mix}}^+ - F_{\mathrm{mix}}^- = 2 |\varphi_E - \varphi_O| = f_{\mathrm{CM}}.
	\]
	
	The phase noise of the down-converted repetition rate beat note was measured using an electrical spectrum analyser (Anritsu MS2840A), and ten repeated frequency scans were performed to construct the phase noise measurements reported in the main text. The beat note frequency was tracked using a frequency counter (Keysight 53220A) with a gate time of 1~ms.
	
The Allan deviation $\sigma(\tau)$ was computed from the fractional-frequency record $y(t)$ using the (overlapping) Allan-variance estimator
\[
\sigma^2(\tau) = \frac{1}{2K} \sum_{i=1}^{K} \left( \bar{y}_{i+1} - \bar{y}_i \right)^2,
\]
where $\bar{y}_i$ is the average of $m$ consecutive samples of $y$ starting at index $i$, $\tau = m\tau_0$ is the averaging time for the sampling period $\tau_0$, and $K = N - 2m$ for a record of $N$ samples.

\begin{figure}[ht]
	\centering
	\includegraphics[width=\textwidth]{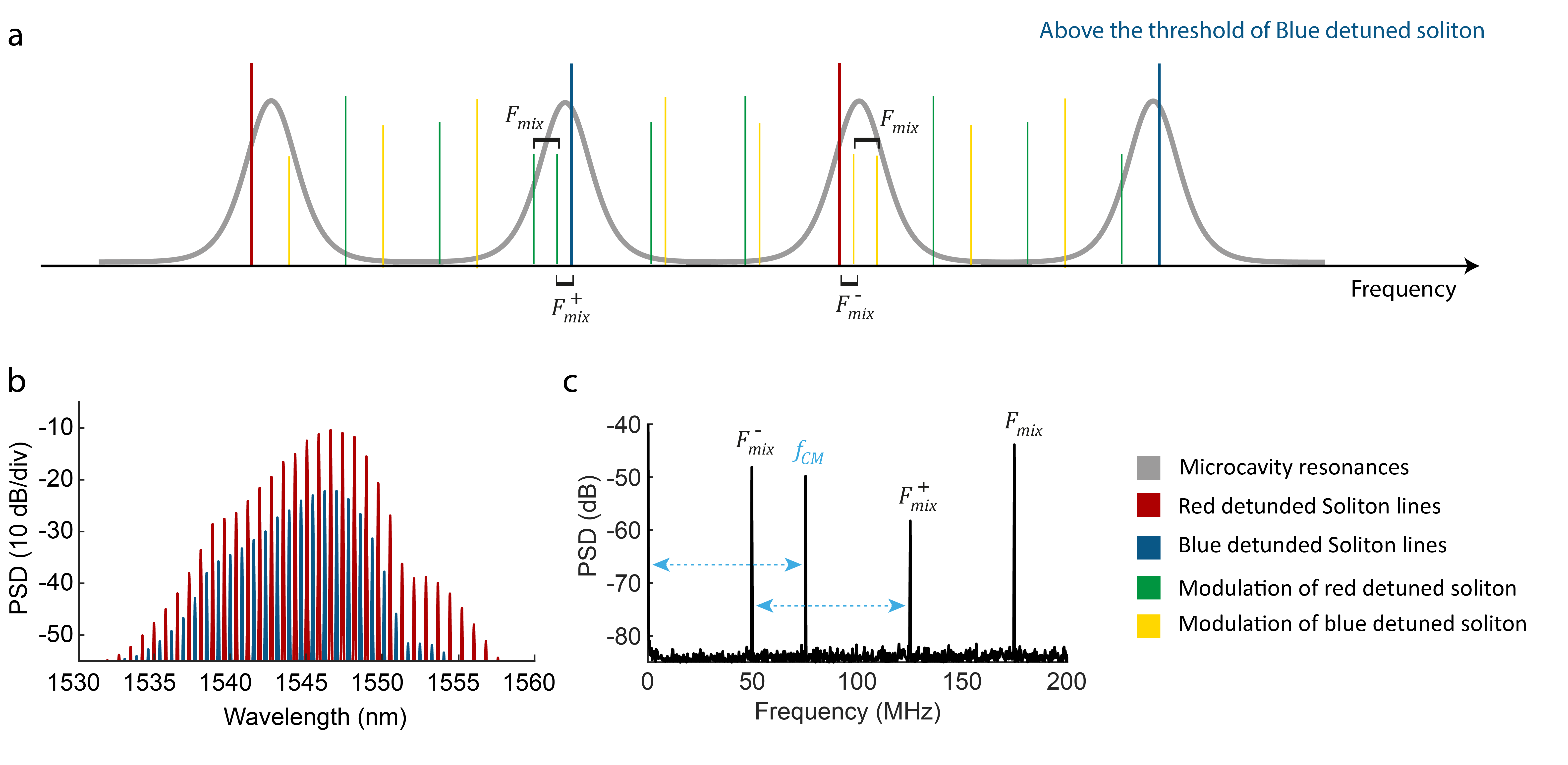}
	\caption*{
		\footnotesize
		\textbf{Figure E6. Down-conversion of the microcomb laser when the system is operated above the blue-detuned soliton threshold.}
		\textbf{a} Frequencies distribution. All the microcavity lines are occupied by blue- and red-detuned laser lines in an interleaved fashion. In addition to the beatnote at frequency $F_{\mathrm{mix}}$, the beatnotes $F_{\mathrm{mix}}^{-}$, $F_{\mathrm{mix}}^{+}$ are formed, due to the interaction of the third-order sidebands relevant to a certain comb with lines of the other comb. 
		\textbf{b} Optical spectrum of a red-detuned two-soliton state, obtained when the system is operated above blue-detuned soliton threshold. Note the interleaving between the red-detuned (in red) and blue-detuned (in blue) combs. 
		\textbf{c} Radio-frequency (RF) spectrum of the down-converted comb, for the case described in (a). Note the presence of beatnotes at frequencies $F_{\mathrm{mix}}^{-}$, $F_{\mathrm{mix}}$, $F_{\mathrm{mix}}^{+}$, and $f_\mathrm{CM}$.
	}
	\label{fig:FigE6}
\end{figure}

	\subsection*{Author contributions}
	D.D. and A.P., developed the original concept. D.D. performed the experiments. A.P. and G.-L.O. developed the theoretical analysis. A.Cu and M.P. supported the experimental investigation, A.Co, L.O., J.S.T.G. and A.B. supported the theoretical investigation. D.J.M. and R.M. supported the interpretation of the results. B.E.L. and S.T.C. designed and fabricated the integrated devices. All the authors contributed to the writing of the manuscript. A.P. supervised the research.
	
\subsection*{Acknowledgments}
We thank Dr Luke Peters and Dr Domenico Bongiovanni for insightful technical
discussions on the measurements and topological interpretation. The authors are
particularly indebted to Dr Bongiovanni for his careful reading of the
manuscript and his detailed comments.

The Emergent Photonics Research Centre acknowledges support from EPSRC-UKRI
(EP/Z533178/1; UKRI592; EP/Y004701/1; EP/W028344/1; EP/X012689/1);
from the ERC (851758 TELSCOMBE);  from the Leverhulme Trust, (RPG-2022-090; ECF-2023-315; ECF-2024-529)
 and from DEVCOM US Army Research Office, (W911NF231031). 
 D.J.M. acknowledges support from the ARC Centre of Excellence COMBS (CE230100006). 
 S.T.C. acknowledges support from City University of Hong Kong (MF-EXT 9678333). 
 G.-L.O. acknowledges support from the
 Computational Nonlinear and Quantum Optics group  at the University of Strathclyde. 
 R.M. acknowledges support from the Canada Research Chair program and from NSERC
(AQUA ALLRP 587602-23; QuEnSi ALLRP 578468-22; ALLRP 587352-23;
ALLRP 569583-21), and from FRQNT(328872).

\textbf{Correspondence and requests for materials} should be addressed to A.P.
	\bibliographystyle{naturemag}

\clearpage

	\section*{Supplementary Information}
	
	\section*{S1. Mean-Field Coupled Equations and Supermode Structures for a Nested Cavity System}

	\renewcommand{\theequation}{S1.\arabic{equation}}
	\setcounter{equation}{0}
	
	We briefly recall the normalised mean-field equations for the system in the case of quasi-perfectly matched cavities~\cite{Bao2019Laser}, where the free-spectral ranges of the two resonators are very close to an integer ratio
	
	\begin{align}
		\partial_t a &= \frac{i \zeta_a}{2}\,{\partial^{2}_{x}} a + i |a|^2 a - \kappa a + \sqrt{\kappa} \sum_{q=-N}^{N} b_q, \label{S1.1} \\
		\partial_t b_q &= \frac{i \zeta_b}{2}\,{\partial^{2}_{x}} b_q - \delta\,\partial_x b_q + 2\pi i (\Delta - q)\,b_q + \sigma\,{\partial^{6}_{x}} b_q + g\,b_q - \sum_{p=-N}^{N} b_p + \sqrt{\kappa}\,a. \label{S1.2}
	\end{align}
	
	Here, $a$ and $b_q$ represent the optical field envelopes for the microresonator and amplifying cavities, expressed as functions of the normalised time and space variables $(t,x)$ for $|x|<1/2$.  
	The indices $q$ range from $-N$ to $N$, defining $2N+1$ supermodes $b_q$, forming the truncated supermode basis of the main cavity.  
	The coefficients $\zeta_a,\zeta_b>0$ denote the normalised (anomalous) cavity dispersions; $\Delta$ is the detuning, while $\kappa$, $g$, and $\sigma$ represent the coupling, the saturated gain, and the spectral-filtering bandwidth, respectively.  
	
	To model spectral filtering accurately, we adopt a sixth-order spatial derivative term,
	${\partial^{6}_{x}} b_q$, in Eq.~(\ref{S1.2}), following our previous formulation~\cite{Rowley2022Self}, which yields a good fit to the experimentally observed soliton profiles.
	
	The model is defined with the following periodicity conditions for the microcavity and main-cavity fields:
	\begin{equation}
		\label{bc0}
		a(x,t)=a(x+1,t), \qquad b_q(x,t)=b_q(x+1,t).
	\end{equation}
	
	The equations are derived for quasi-perfectly matched cavities where the free-spectral range (FSR) of the microcavity $F_a$ is related to that of the amplifying cavity $F_b$ by
	\begin{equation}
		F_a = (M - \delta)\,F_b. \label{S1.3}
	\end{equation}
	
	Here $M$ is an integer and represents the maximum number of supermodes ($2N_{\mathrm{max}}+1=M$) that can be taken into account in the model, with $|\delta|<1/2$ representing the cavity-FSR mismatch.  
	For simplicity, we consider $M$ an odd number.  
	Note that the parameter $\delta$ in Eq.~(\ref{S1.2}) coincides with that in Eq.~(\ref{S1.3}), as the group-velocity mismatch in the model directly stems from the FSR mismatch between the two cavities.
	
	By considering the space $X$ and time $T$ variables in [m] and [s], respectively, in Eqs.~(\ref{S1.1}, \ref{S1.2}), time $t$ is normalised to the amplifying-cavity round trip so that $t = T F_b$ represents the slow propagation time of the system.  
	The space $x$ is normalised to the microcavity length and centred on the time frame of the pulse moving with velocity $v$, so that $x = (X/v - T)\,F_a$.  
	
	Since most of the supermodes are largely detuned from the centre of the microcavity resonance, the model does not require all $M$ supermodes for a proper description.  
	In this configuration, the supermode with $q=0$ has the strongest coupling,and is the closest to the centre of the microcavity resonance up to a small detuning.  Consequently, analytical considerations, including the calculation of stationary states, can be performed with a single leading supermode.  
	Higher-order supermodes primarily define the stability range of the background—practically, the lasing threshold—and influence the parameter regions corresponding to solution stability.  
	
	The number of supermodes required in practice depends on both the microcavity bandwidth $\Delta F_a$ and its relative value compared to $F_b$.  
	The coupling coefficient $\kappa = \pi \Delta F_a / F_b$ effectively accounts for this condition in the model.  
	In our numerical calculations, we use $\kappa \approx 1.5\pi$ in close agreement with the experimental conditions, and solutions converge for $N \ge 5$ \cite{Bao2019Laser}; hence, we fix $N=5$.
	
	\subsection*{S1.1 Matched-Cavity Derivation — Technical}
	
	For completeness, we briefly recall how the supermodes $b_q$ are defined in the cavity-matched case. We consider the perfectly matched configuration ${\delta} = 0$  for simplicity, and a full derivation is provided in Ref.~\cite{Bao2019Laser}.  
	
	The supermodes $b_q$ are obtained by grouping the oscillating modes with frequencies $f_n^{(b)}$ spaced by the microcavity FSR and amplitudes $\bar{b}_n$.  
	The $m$-th frequency mode of the microcavity is defined as
	\begin{equation}
		f_m^{(a)} = m F_a, \label{S1.4}
	\end{equation}
	and the $n$-th frequency mode of the main cavity as
	\begin{equation}
		f_n^{(b)} = n F_b - F_b \Delta, \label{S1.5}
	\end{equation}
	where $\Delta$ is a common detuning applied to all modes.  
	
	The two FSRs in Eq.~(\ref{S1.3}) are linked by an integer ratio through the parameter $M$.  
	Accordingly, we define the mode index $n$ as
	\begin{equation}
		n = M m + q, \label{S1.6}
	\end{equation}
	which allows the frequency modes $f_n^{(b)}$ to be grouped into sets (supermodes) spaced by $F_a$.  
	The coefficient $q \in \left[ -\tfrac{M-1}{2}, \tfrac{M-1}{2} \right]$ labels the specific supermode and acts as an additional detuning scaled by $F_b$.  
	The index $m$ spans $M$ values, effectively accounting for all modes within one microcavity FSR.  
	The corresponding frequency can be written as
	\begin{equation}
		f_n^{(b)} = f_{Mm+q}^{(b)} = m F_a + (q - \Delta) F_b. \label{S1.7}
	\end{equation}
	
	As detailed in Ref.~\cite{Bao2019Laser}, under these assumptions, Eqs.~(\ref{S1.1}) and~(\ref{S1.2}) can be derived directly from coupled-mode theory.  
	Without entering the full derivation—which includes both microcavity and laser-cavity modes with dispersion, gain, losses, nonlinearity, and coupling—it is useful to recall the linear detuning terms experienced by the modal amplitudes $\bar{b}_n(t)$ associated with the frequencies $f_n^{(b)}$ within the resonance $m F_a$.  
	Under such conditions, the oscillation equation for the modes $\bar{b}_n(t)$ reads
	\begin{equation}
		F_b \frac{d \bar{b}_n}{dt} = 2\pi (m F_a - f_n^{(b)}) \bar{b}_n, \label{S1.8}
	\end{equation}
	where time is normalised to the main-cavity FSR $F_b$.  
	
	Because of the periodicity conditions in Eq.~(\ref{bc0}), the supermodes can be expressed as
	\begin{equation}
		b_q(x,t) = \sum_{m=-\infty}^{\infty} \bar{b}_{mM+q}(t)\, \exp(2\pi i m x). \label{S1.9}
	\end{equation}
	Multiplying Eq.~(\ref{S1.8}) by $\exp(2\pi i m x)$, summing over $m$, and using Eq.~(\ref{S1.7}) gives
	\begin{equation}
		\partial_t b_q(x,t) = 2\pi i (\Delta - q)\, b_q(x,t), \label{S1.10}
	\end{equation}
	which corresponds to the detuning term in Eq.~(\ref{S1.2}).
	
	\subsection*{S1.2 Interleaved-Cavity Derivation — Technical}
	
	For the configuration discussed in this paper, the only element to modify in the model is the detuning relationship.  
	The simplified detuning condition clarifies how to adapt the system to the interleaved-cavity case.  
	Here, the relationship between the FSRs of the two cavities is expressed as
	\begin{equation}
		2 F_a = (2M - 1 - \bar{\delta}) F_b. \label{S1.11}
	\end{equation}
	We denote the FSR mismatch by $\bar{\delta}$ to distinguish it from the detuning introduced in Eq.~(\ref{S1.3}).  
	Formally, Eq.~(\ref{S1.3}) is equivalent to Eq.~(\ref{S1.11}) for $\delta = 1/2$ and $\bar{\delta} = 0$.  
	Under this equivalence, the mean-field equations (\ref{S1.1}, \ref{S1.2}) remain valid and can be used to model the system with $\delta = 1/2$.  
	However, this formulation provides an accurate numerical description only when the complete set of supermodes is retained (i.e., $2N + 1 = M$).  
	In such cases, the physical interpretation is lost, as it becomes impossible to isolate the key lasing frequencies within a single supermode.  
	This limitation motivates the development of an alternative mean-field model that restores interpretability by redefining the supermode structure.
	
	To obtain a mean-field model offering a clearer physical picture, we introduce an approach that captures the system behaviour with a single dominant supermode, assuming that the supermode grouping can be chosen arbitrarily.  
	Following the detuning relationship in Eq.~(\ref{S1.11}), the smallest FSR for a supermode with equally spaced lines is $2F_a$.  
	We therefore define
	\begin{equation}
		n = (2M - 1)m' + q', \label{S1.12}
	\end{equation}
	where $m'$ is an integer and $q'$ varies over a wide range, approximately covering $2F_a$ to include all possible modes, namely $q' \in \left[ -\tfrac{M-1}{2}, \tfrac{M-1}{2} + M \right]$.  
	Considering Eq.~(\ref{S1.11}), the resulting frequency for mode $n$ is
	\begin{equation}
		f_n^{(b)} = f_{(2M-1)m'+q'}^{(b)} = 2m' F_a + q' F_b - F_b \Delta. \label{S1.13}
	\end{equation}
	
	In the cavity-matched case (cf. Eqs.~\eqref{S1.3} and~\eqref{S1.6}), the mode with $q = 0$ corresponds to the best-coupled resonance of the microcavity.  
	This remains true here, as $q' = 0$ identifies the modes that couple best to the even microcavity resonances (up to a small detuning $\Delta$) and are spaced by $2m'F_a$.  
	However, we must also account for supermodes coupling efficiently to the odd resonances.  
	Such supermodes are identified by $q_e = q$ and $q_o = q + M$, with $q \in \left[ -\tfrac{M-1}{2}, \tfrac{M-1}{2} \right]$.  
	These indices directly discriminate the odd and even resonances while referring to the same index $q$, spanning a range $M$ that covers a single FSR $F_a$.
	
	The two families of frequencies can then be written explicitly.  
	The even frequencies result straightforwardly from Eq.~(\ref{S1.13}) and are defined by the common detuning $\Delta$ via
	\begin{equation}
		f_{(2M-1)m'+q_e}^{(b)} = 2m' F_a + q_e F_b - F_b \Delta = 2m' F_a + (q - \Delta)F_b,  \label{S1.14}
	\end{equation}
	which is analogous to the cavity-matched case in Eq.~(\ref{S1.7}) for $m = 2m'$.
	
	For the odd frequencies, the new configuration introduces an additional detuning term $F_b/2$ resulting in
	\begin{equation}
		f_{(2M-1)m'+q_o}^{(b)} = 2m' F_a + q_o F_b - F_b \Delta = (2m' + 1) F_a + q F_b - F_b \left( \Delta - \tfrac{1}{2} \right), \label{S1.15}
	\end{equation}
	where we have used $M F_b = F_a + F_b/2$ from Eq.~(\ref{S1.11}).
	
	This definition allows us to assign distinct normalised detunings to the best-coupled supermodes ($q=0$), namely
	\begin{equation}
		\Delta_e = \Delta, \qquad \Delta_o = \Delta - \tfrac{1}{2}, \label{newdelta0}
	\end{equation}
	for the even and odd modes, respectively, as illustrated in Fig.~S1.1.  
	To unify the notation, we introduce a single index $m$, defined as $m = 2m'$ for even values and $m = 2m' + 1$ for odd values.  
	Consequently, the index $n$ can be expressed as
	\begin{equation}
		n = \left( M - \tfrac{1}{2} \right)m + \tfrac{1}{4} - \tfrac{1}{4} \exp(i\pi m) + q, \label{S1.16}
	\end{equation}
	where $m$ spans the integers and $q \in \left[ -\tfrac{M-1}{2}, \tfrac{M-1}{2} \right]$ defines the supermode order.  
	Since $\exp(i\pi m) = \pm 1$ for even and odd $m$, we obtain a compact expression for the supermode frequencies
	\begin{equation}
		f_{m,q}^{(b)} = m F_a + q F_b - F_b \left( \Delta - \tfrac{1}{4} + \tfrac{1}{4}\exp(i\pi m) \right). \label{S1.17}
	\end{equation}
	
	Equation~(\ref{S1.17}) can be used to derive the corresponding mean-field system, analogously to the derivation of Eqs.~(\ref{S1.1}) and~(\ref{S1.2}) from coupled-mode theory, as outlined in Ref.~\cite{Bao2019Laser}.  
	As in the matched-cavity case, we focus on the detuning terms for simplicity.  
	Starting from Eq.~(\ref{S1.8}), we follow the same procedure used to derive Eq.~(\ref{S1.10}), but now incorporating the modification introduced by the interleaved configuration.  
	Using Eq.~(\ref{S1.17}) in Eq.~(\ref{S1.8}), the exponential term $\exp(i\pi m)$ effectively shifts the supermode amplitude $b_q(x,t)$ in space by half a period:
	\begin{equation}
		\sum_{m=-\infty}^{\infty} \bar{b}_{mM+q}(t) \exp(2\pi i m x + i \pi m) = b_q(x + \tfrac{1}{2}, t),
	\end{equation}
	leading to a modified form of Eq.~(\ref{S1.10}):
	\begin{equation}
		\partial_t b_q(x,t) = 2\pi i \left( \Delta - q - \tfrac{1}{4} \right)b_q(x,t) + \tfrac{2\pi i}{4}\, b_q(x + \tfrac{1}{2}, t). \label{S1.18}
	\end{equation}
	
	The right-hand side of Eq.~(\ref{S1.18}) represents the only required modification in the new model, 
	in full consistency with the procedure outlined in Ref.~\cite{Bao2019Laser}, 
	where the term $b_q(x+\tfrac{1}{2},t)$ denotes the supermode amplitude evaluated at the shifted position $x+\tfrac{1}{2}$. 
	This naturally leads to the introduction of a half-cavity shift operator, which underpins the \mobius geometry discussed in Section~S2.

	\begin{figure}[H]
		\centering
		\includegraphics[width=0.8\linewidth]{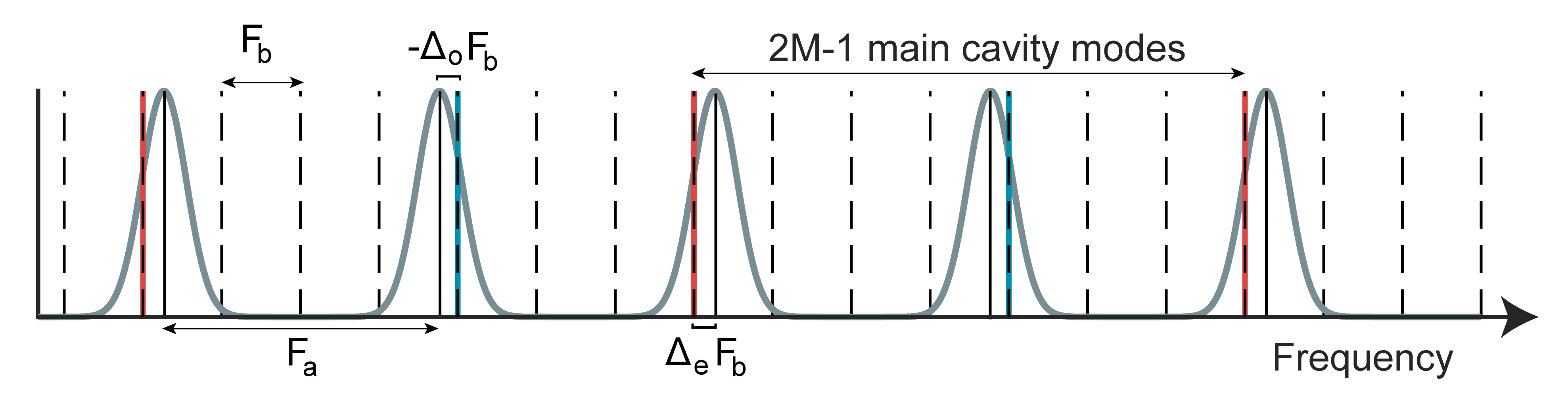}
		\caption{\small\textbf{Fig.~S1.1.} Grouping of supermodes. 
			Microcavity resonances (black solid) and main-cavity modes (black dashed). 
			Blue and red lines represent frequencies of the same supermode with detunings 
			$\Delta_e = \Delta$ and $\Delta_o = \Delta - \tfrac{1}{2}$, respectively.}
		\label{fig:S1.1}
	\end{figure}
	\newpage

	\section*{S2. Model Equations for the M\"obius System}
	\renewcommand{\theequation}{S2.\arabic{equation}}
	\setcounter{equation}{0}
	
	In our context, we define a one-dimensional \mobius geometry as a periodic system that combines a half-period translation with a $\pi$-phase inversion:
	\begin{equation*}
		f(x+\tfrac{1}{2}) = -f(x), \qquad f(x+1) = f(x),
	\end{equation*}
	for a spatial coordinate $x$ with unit period.
	
	In a microcavity system, this condition is realised by selecting resonator modes spaced by half the round-trip period of the microcavity. 
	The resulting interleaved sets of resonances satisfy periodic and antiperiodic conditions, respectively. As we show later in this section, these two mode families naturally define the two 
	independent subspaces of an operator closely connected to the M\"obius geometry: the 
	half-cavity shift operator. 
	
	To prevent the two mode families from merging
	into a single period-one state, however, the alignment between the even and odd microcavity resonances and the main-cavity modes must be offset. 
	The simplest configuration occurs when the ratio between the two free-spectral ranges (FSRs) is a half-integer. 
	This geometrical constraint leads directly to the modified relationship between the FSRs of the micro- and main cavities:
	\begin{equation*}
		2F_a = (2M - 1 - \delta)\,F_b,
	\end{equation*}
	which is equivalent to Eq.~\eqref{S1.11} and forms the basis of the extended mean-field model discussed below.  
	For clarity, we continue to denote by $\delta$ the detuning from this interleaved-cavity matching condition 
	(dropping the notation $\bar{\delta}$ used in Section~S1.2), {enabling a} direct comparison with the cavity-matched case described in Section~S1.
	
	Following the derivation reported in Section~S1.2—enforcing the half-integer FSR ratio in the mean-field equations of the matched-cavity system—Eqs.~\eqref{S1.1} and \eqref{S1.2} transform into the extended form:
	\begin{align}
		\partial_t a (x,t) &= \frac{i \zeta_a}{2} {\partial^{2}_{x}} a(x,t) 
		+ i |a(x,t)|^2 a(x,t) 
		- \kappa a(x,t) 
		+ \sqrt{\kappa} \sum_{q=-N}^{N} b_q(x,t), 
		\label{S2.1} \\
		\partial_t b_q (x,t) &= 
		\frac{i \zeta_b}{2}{\partial^{2}_{x}} b_q(x,t) 
		- \delta \, \partial_x b_q(x,t)
		+ 2\pi i \!\left(\Delta - q - \tfrac{1}{4}\right) b_q(x,t)
		+ \tfrac{2\pi i}{4}\, b_q\!\left(x+\tfrac{1}{2},t\right) \notag \\
		&\quad 
		+ g\, b_q(x,t) 
		+ \sigma\, {\partial^{6}_{x}} b_q(x,t)
		- \sum_{p=-N}^{N} b_p(x,t)
		+ \sqrt{\kappa}\, a(x,t). 
		\label{S2.2}
	\end{align}
	
	All quantities are defined as in Section~S1.1.  
	Eqs.~(\ref{S2.1}) and (\ref{S2.2}) are subject to the same periodicity conditions as the matched-cavity case 
	$a(x,t)=a(x+1,t)$ and $b_q(x,t)=b_q(x+1,t)$.  
	The modified spectral structure, {however}, introduces two additional terms in Eq.~\eqref{S2.2}: 
	a static frequency shift $-b_q(x,t)/4$ and a coupling term displaying a half-period spatial offset, 
	$b_q(x+\tfrac{1}{2},t)/4$. 
	Importantly, the latter enables the system to sustain solutions exhibiting a \mobius geometry—that is, 
	field configurations satisfying a half-period translation combined with a $\pi$-phase inversion.  
	Not all solutions display this property; however, since the equations admit such states, 
	we will generically refer to them as the \emph{\mobius system}. 
	For comparison, we refer to the matched-cavity case, Eqs.~\eqref{S1.1} and \eqref{S1.2}, as the \emph{trivial system}.
	
	\subsection*{S2.1 Fundamental symmetries of the \mobius system}
	
	To understand under which conditions a solution can effectively sustain a \mobius geometry, 
	it is convenient to analyse the fundamental symmetries of the governing equations. 
	Here we refer to \emph{symmetry} as \emph{an operator which, when applied to a solution of the equations, generates another valid solution of the same equations}. 
	The continuous and discrete symmetries relevant to our analysis, together with their associated actions, are summarised in Table 1. 
	
	\begin{table}[h]
		\centering
		\renewcommand{\arraystretch}{1.15}
		\begin{tabular}{|c|c|c| |c|c|c|}
			\hline
			\multicolumn{3}{|c| |}{\textbf{Continuous Symmetries}} &
			\multicolumn{3}{c|}{\textbf{Discrete $\mathbb{Z}_2$ Symmetries}} \\ 
			\hline
			\textbf{Operator} & \textbf{Symmetry} & \textbf{Action} &
			\textbf{Operator} & \textbf{Symmetry} & \textbf{Action} \\ 
			\hline
			$\mathcal U$ & Global phase & $t \mapsto t + t_0$ &
			$\mathcal P$ & Parity* &  $x \mapsto 2x_c - x$\\
			\hline
			$\mathcal T$ & Spatial translation & $x \mapsto x + x_0$ &
			$\mathcal S$ & Half-cavity shift & $x \mapsto x + \tfrac{1}{2}$ \\
			\hline
		\end{tabular}
		\caption{Symmetries of Eqs.~\eqref{S2.1}--\eqref{S2.2}. 
			Each continuous operator generates a neutral (Goldstone) mode. 
			The parity symmetry $\mathcal P$ holds only for $\delta=0$, while the {half-cavity shift} $\mathcal S$ defines the \mobius coupling between even and odd resonances. 
			For $\mathcal U$, the phase invariance is equivalent to a temporal translation, 
			as both correspond to a rotation in the complex plane.}
		\label{tab:sym_all}
	\end{table}
	
	In general, Eqs.~\eqref{S2.1} and \eqref{S2.2} inherit the continuous and discrete symmetries of the base model (Eqs.~\eqref{S1.1} and \eqref{S1.2}), while the new shifted term $b_q(x+\tfrac{1}{2},t)$ introduces the additional discrete symmetry $\mathcal S$.
	
	We briefly examine the continuous symmetries of the equations. 
	The first is the \emph{global phase} symmetry $\mathcal U$, 
	corresponding to a uniform rotation of the optical fields in the complex plane which, 
	since the system is autonomous, is equivalent to a temporal translation 
	$t \mapsto t + t_0$ that leaves Eqs.~\eqref{S2.1}--\eqref{S2.2} unchanged and which we highlight in the table.
	
	This invariance is characteristic of autonomous laser systems with homogeneous equations, in contrast to driven models such as the Lugiato--Lefever equation, where it is explicitly broken due to the presence of source terms. 
	
	When the dynamics are linearised around a stationary solution, the symmetry 
	$\mathcal{U}$ gives rise to an eigenmode with an eigenvalue at, or arbitrarily 
	close to, the origin—the \emph{Goldstone (or neutral) mode}—reflecting the 
	freedom to choose the overall optical phase, or equivalently the temporal 
	origin, of the solution.
	
	A second fundamental invariance is the \emph{spatial translation} operator $\mathcal T$, which expresses the invariance of the system under continuous spatial shifts. 
	Linearisation around a stationary state yields a \emph{Goldstone mode} also for this operator, 
	whose presence is directly linked to the conservation of the momentum operator $\hat{L}[f(x)]$ defined as
	\begin{equation*}
		\hat{L}[f(x)] = \int_{-1/2}^{1/2} \mathrm{Im}\,[f(x)^* \partial_x f(x)]\, dx,
	\end{equation*}
	which in our system corresponds physically to the mean frequency of the optical spectrum. 
	This translational mode governs the drift dynamics of the molecule and underpins the Ising--Bloch mechanism discussed later.
	
	The model also admits two important discrete $\mathbb{Z}_2$ symmetries. 
	The first is the \emph{parity operator} $\mathcal P$, corresponding to a reflection of the cavity axis around a generic point $x_c$, i.e. $x \mapsto 2x_c - x$. In our system, $\mathcal P(x_c)$ is conserved in the limit of vanishing group-velocity mismatch ($\delta=0$) and corresponds to a transformation of the solutions $(a(x), b_q(x))$ into $(a(2x_c - x), b_q(2x_c - x))$. Here we adopt the general definition of a parity operator. In the main text, for simplicity, we refer only to the specific choice $\mathcal P(0)$.
	
	The second discrete symmetry, which distinguishes Eqs.~\eqref{S2.1}--\eqref{S2.2} from Eqs.~\eqref{S1.1}--\eqref{S1.2}, 
	is the \emph{half-translation} symmetry $\mathcal S$, defined as $(a(x), b_q(x)) \mapsto (a(x+\tfrac{1}{2}), b_q(x+\tfrac{1}{2}))$. 
	This additional $\mathbb{Z}_2$ symmetry acts as a discrete half-period shift, giving rise to the characteristic \mobius coupling between even and odd resonances.
	
	Discrete symmetries are particularly useful because they can be used to define projection operators (projectors) that separate equations and solutions into their symmetric ($+$) and antisymmetric ($-$) components. 
	For the operator $\mathcal S$, in particular, we can define the symmetric and antisymmetric components as:
	\begin{equation}
		\Pi_{\pm}^{\mathcal S} = \tfrac{1}{2}(\mathbb{I} \pm \mathcal S), 
		\qquad
		\Pi_{\pm}^{\mathcal S} f(x) = \tfrac{1}{2}\bigl(f(x) \pm f(x+\tfrac{1}{2})\bigr),
		\label{prS}
	\end{equation}
	with $(\mathbb{I}f)(x) = f(x)$. 
	The projectors  satisfy 
	$(\Pi_{\pm}^{\mathcal S})^2 = \Pi_{\pm}^{\mathcal S}$ \;(idempotence), 
	$\Pi_+^{\mathcal S}\Pi_-^{\mathcal S} = 0$ \;(orthogonality), 
	and $\Pi_+^{\mathcal S} + \Pi_-^{\mathcal S} = \mathbb{I}$ \;(completeness), 
	providing an orthogonal decomposition of any field into two independent subspaces.  Specifically, we refer to $\Pi_{+}^{\mathcal S}$ as the \emph{even sector} and to $\Pi_{-}^{\mathcal S}$ as the \emph{odd sector}.
	Notably, this is precisely the action that separates the interleaved even and odd 
	families of modes discussed above. This property distinguishes the operator 
	$\mathcal S$ from a simple spatial translation $\mathcal T$ with $x_0 = 1/2$.
	
	Similarly to Eq.~\eqref{prS}, {a complete set of} projectors can also be defined for the parity symmetry operator $\Pi_{\pm}^{\mathcal P}$, 
	which generally {commute} with $\Pi_{\pm}^{\mathcal S}$, as $\mathcal S \mathcal P = \mathcal P \mathcal S$:
	\begin{equation*}
		\bigl[\Pi_{\pm}^{\mathcal P(x_c)},\,\Pi_{\pm}^{\mathcal S}\bigr] = 0.
	\end{equation*}
	Parity operators defined with different reflection centres $x_c$, conversely, \emph{do not} generally commute with each other.
	As discussed in the next subsection, the interplay between the half-cavity shift and suitable parity operators is particularly relevant to our system.
	
	As we will often refer to symmetry-breaking phenomena, we take this opportunity to specify our terminology. 
	We refer to \emph{explicit} or \emph{soft symmetry breaking} when the loss of symmetry occurs at the equation level, owing to the presence of a non-symmetric term. 
	For instance, the parity symmetry $\mathcal P$ is \emph{explicitly broken} in the presence of a finite group-velocity mismatch ($\delta \neq 0$), 
	which renders Eq.~\eqref{S2.2} asymmetric under the transformation $x \mapsto 2x_c - x$, as the term $-\delta \partial_x$ changes sign to $+\delta \partial_x$. We discuss this in Section S8.
	
	Conversely, we refer to \emph{spontaneous symmetry breaking} (SSB)~\cite{NonlinearDynamics2007} when the equations themselves remain invariant, but a specific state or solution no longer respects the {equation} symmetry. In this case, the symmetry is said to be broken by the solution rather than by the governing equations. 
	We will denote the occurrence of a SSB by adding an overbar across the operator (e.g. $\bcancel{\mathcal P}$), indicating that the symmetry is broken at the level of the solution but not of the equations.
	
	\subsection*{S2.2 Even–odd decomposition under the half-cavity shift operator $\mathcal S$}
	
	The projectors derived from the discrete symmetries provide a complete decomposition of any field into two orthogonal subspaces. Consequently, we can apply the half-cavity shift projectors $\Pi_{\pm}^{\mathcal S}$ defined in Eq.~\eqref{prS} to decompose the fields in their symmetric and antisymmetric contributions, which we denote as \emph{even} and \emph{odd} modes.
	With these definitions, the fields in Eqs.~(S2.1)–(S2.2) can be decomposed explicitly into even and odd components, as summarised in Table~\ref{S2.3}.
	For simplicity, we refer only to $a(x,t)$, implying an analogous definition for $b_q(x,t)$. 
	
	\begin{table}[H]
		\centering
		\renewcommand{\arraystretch}{2}
		\vspace{3pt}
		\begin{tabular}{c|c|c}
			\textbf{Full Field Solution} 
			& \textbf{$\Pi_\pm^{\mathcal S}$ Decomposition} 
			& \textbf{Periodicity Conditions} \\
			\hline
			$a(x) = a_e(x) + a_o(x)$ 
			& $a_e(x) = \tfrac{1}{2}\!\left[a(x) + a\!\left(x+\tfrac{1}{2}\right)\right]$ 
			& $a_e(x,t) = a_e\!\left(x+\tfrac{1}{2}, t\right)$ \\
			$a\!\left(x+\tfrac{1}{2}\right) = a_e(x) - a_o(x)$ 
			& $a_o(x) = \tfrac{1}{2}\!\left[a(x) - a\!\left(x+\tfrac{1}{2}\right)\right]$ 
			& $a_o(x,t) = -a_o\!\left(x+\tfrac{1}{2}, t\right)$ \\
		\end{tabular}
		\caption{Decomposition of the field under the half-cavity shift operator $\mathcal S$.}
		\renewcommand{\arraystretch}{1}
		\label{S2.3}
	\end{table}
	
	By computing $\partial_t a(x) + \partial_t a(x+\tfrac{1}{2}) = 2\,\partial_t a_e(x)$ and 
	$\partial_t a(x) - \partial_t a(x+\tfrac{1}{2}) = 2\,\partial_t a_o(x)$, 
	Eqs.~(S2.1)–(S2.2) can be rearranged to separate the {dynamics of the} even and odd modes.  
	Considering also the detuning definitions $\Delta_e = \Delta$ and $\Delta_o = \Delta - 1/2$ from Eq.~\eqref{newdelta0}, 
	we obtain the full set of dynamical equations for the even and odd components of the resonator and main-cavity fields:
	\begin{align}
		\partial_t a_e &= \frac{i \zeta_a}{2} {\partial^{2}_{x}} a_e + i|a_e|^2 a_e + 2i|a_o|^2 a_e + i\gamma\,a_e^*a_o^2 - \kappa a_e + \sqrt{\kappa} \sum_{q=-N}^{N} b_{q,e}, \label{S2.4} \\
		\partial_t b_{q,e} &= \frac{i \zeta_b}{2} {\partial^{2}_{x}} b_{q,e} - \delta \partial_x b_{q,e} + 2\pi i (\Delta_e - q) b_{q,e} + \sigma {\partial^{6}_{x}} b_{q,e} + g\, b_{q,e} - \sum_{p=-N}^{N} b_{p,e} + \sqrt{\kappa} a_e, \label{S2.5} \\
		\partial_t a_o &= \frac{i \zeta_a}{2} {\partial^{2}_{x}} a_o + i|a_o|^2 a_o + 2i|a_e|^2 a_o + i\gamma\,a_o^*a_e^2 - \kappa a_o + \sqrt{\kappa} \sum_{q=-N}^{N} b_{q,o}, \label{S2.6} \\
		\partial_t b_{q,o} &= \frac{i \zeta_b}{2} {\partial^{2}_{x}} b_{q,o} - \delta \partial_x b_{q,o} + 2\pi i (\Delta_o - q) b_{q,o} + \sigma {\partial^{6}_{x}} b_{q,o} + g\, b_{q,o} - \sum_{p=-N}^{N} b_{p,o} + \sqrt{\kappa} a_o. \label{S2.7}
	\end{align}
	
	It is evident that for $a_o=0$ ($a_e=0$), the equations for $a_e$ and $b_{q,e}$ ($a_o$ and $b_{q,o}$) reproduce exactly Eqs.~(S1.1)–(S1.2). 
	This means that the solutions of Eqs.~(S1.1)–(S1.2) and Eqs.~(S2.4)–(S2.5) 
	coincide, provided that the fields are restricted from the full, period-one 
	cavity ($|x|<1/2$) to a fundamental domain of period~$1/2$ ($|x|<1/4$), 
	with symmetric and antisymmetric periodicity for the even and odd components, 
	respectively, as defined in Table~\ref{S2.3}.

	Strictly speaking, Eqs.~(S2.4)–(S2.7) are equivalent to Eqs.~(S2.1)–(S2.2) only when the nonlinear parameter $\gamma$ is set to one. 
	In our derivation of the even–odd decomposed model, however, we {retain} $\gamma$ {as a mean} to systematically isolate the contribution arising solely from cross-phase-modulation (XPM) terms $2i|a_e|^2 a_o$ and $2i|a_o|^2 a_e$. 
	The case $\gamma = 0$, {in particular}, corresponds to the exclusion of four-wave-mixing (FWM) effects and, in the following, we will refer to Eqs.~(S2.4)–(S2.7) with $\gamma=0$ as the \emph{XPM-only} system.
	
	Since FWM is unavoidably present in our experiments, it is useful to clarify the relevance of the XPM-only system in our analysis. 
	By eliminating the parametric energy-exchange terms (FWM), Eqs.~(S2.4)–(S2.7) with $\gamma=0$ isolate the purely intensity-mediated interaction between the even and odd sectors. 
	From a practical perspective, the XPM-only model also separates the propagation frequencies of the even and odd solutions, 
	making it possible to project the two components onto pure stationary states that can then be identified numerically. 
	In the following sections, we use these stationary XPM-only states as reference points to compare with the full system with $\gamma=1$, 
	thereby clarifying whether they seed the soliton molecule.
	Quite importantly, we anticipate that the $\gamma=0$ and $\gamma=1$ systems yield essentially the same dynamics in the main regimes under study, 
	each triggered by distinct spontaneous symmetry-breaking transitions, which we briefly summarise in the next section. 
	
	Before closing this section, we highlight that an optical \mobius geometry was recently explored for CW-polarised modes in an externally driven fibre resonator~\cite{Lucas2024Polarization}. 
	In that configuration, the \mobius geometry was realised by introducing a phase defect in the cavity to sustain a spontaneous symmetry breaking of the coherent fields, 
	completing the \mobius loop over two round trips. 
	In contrast, in our system, the loop is completed within a single round trip of the cavity, 
	since the even–odd resonance division intrinsically doubles the repetition rate. 
	This key distinction enables the realisation of a \mobius geometry without requiring the inclusion of any physical phase defect.
	
	\subsection*{S2.3 Construction of the \mobius molecule: role of parity $\mathcal P$ and half-cavity shift $\mathcal S$}
	
	Having established the symmetry structure and the even/odd decomposition, we now discuss how their interplay governs the formation of the \mobius molecule. 
	The definition of a \mobius geometry requires a half-period translation combined with a $\pi$-phase inversion. 
	Hence, a solution that simultaneously populates both even and odd sectors naturally defines a relative \mobius geometry between them. 
	In this configuration, one component (for instance, the odd one) exhibits a \mobius geometry \emph{relative} to the other (the even component). 
	It is therefore useful to anticipate how symmetry breaking mediates the transitions that lead to a \emph{\mobius molecule state}, 
	which we identify as the formation of a non-trivial bound state populating the even and odd $\mathcal S$ sectors. The bound state is not simply a superposition of the even and odd solitons since they are intrinsically coupled by the nonlinear  terms in the microcavity.
	
	\begin{figure}[h!]
		\centering
		\includegraphics[width=0.7\linewidth]{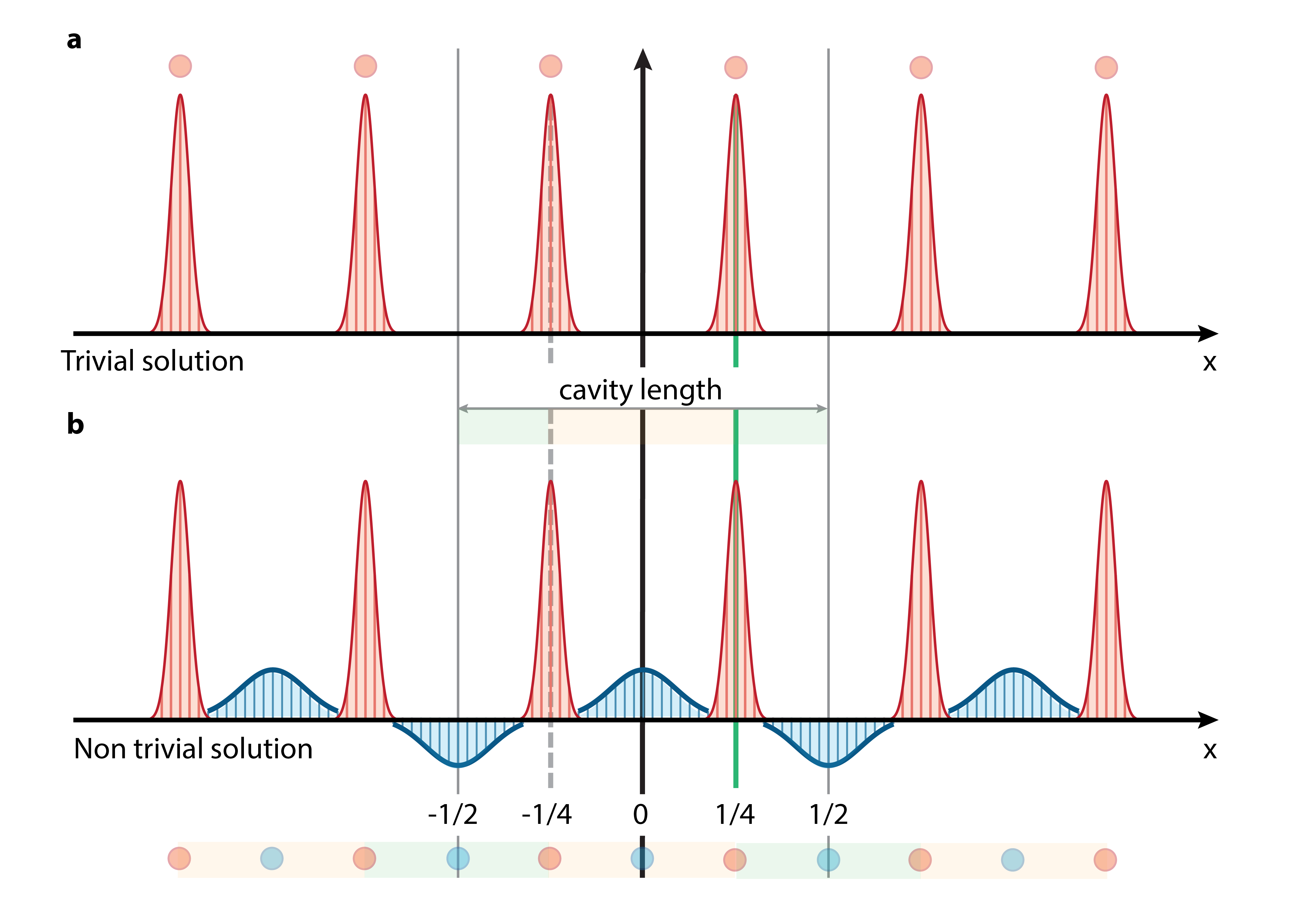}
		\caption{
			\textbf{{Figure S2.1.} Spatial symmetries and lattice centres of the even soliton train.}
			\emph{a}, The reference even two-soliton state defines a periodic lattice with parity centres 
			$\mathcal P(0)$ (midpoint between pulses) and $\mathcal P(\tfrac{1}{4})$ (at pulse peaks), 
			separated by half the lattice period ($\tfrac{1}{2}$). 
			\emph{b}, When the odd component appears, $\mathcal P(\tfrac{1}{4})$ is spontaneously broken 
			while $\mathcal P(0)$ is preserved, realising an effective \mobius geometry in which 
			the odd field populates the regions between neighbouring even solitons.}
		\label{fig:eventrain}
	\end{figure}
	
	{We summarise the key dynamical transitions observed in our system, and their key corresponding dynamical states, as follows:}
	
	\begin{itemize}
		\item \textbf{Initial state: even two-soliton state.} 
		Our starting point is a train of localised soliton pulses populating a single sector—here the even subspace of the half-cavity shift operator $\mathcal S$. We consider strongly localised solutions (solitons) whose profiles exhibit clear central symmetry, allowing us to associate the field with a discrete lattice of well-defined centres {(Fig. S2.1a)}. 
		This enables the definition of parity operators $\mathcal P$ that transform the solution into itself.
		We place the origin of the lattice midpoint between neighbouring pulses, defining the parity centre $\mathcal P(0)$. 
		A second relevant symmetry point is located at $\mathcal P(\tfrac{1}{4})$, 
		while additional centres are obtained by translating these by integer multiples of the lattice period $\tfrac{1}{2}$. 
		With these symmetry points defined, it is natural to construct the corresponding projectors 
		$\Pi_{\pm}^{\mathcal P(0)}$ and $\Pi_{\pm}^{\mathcal P(\tfrac{1}{4})}$.
		The two parity operators do not commute with each other, as expected for point symmetries, 
		but both commute with the half-cavity shift operator $\mathcal S$. 
		For either choice of $\mathcal P$, the reference even soliton train belongs to the positive-parity projector $\Pi_{+}$.

		\item \textbf{First transition: effective lasing threshold of the odd mode — SSB: $\bcancel{\mathcal S}, \bcancel{\mathcal P(\tfrac{1}{4})}, \bcancel{\mathcal U}$.} 
		The odd field $a_o(x)$ becomes nonzero, corresponding to the spontaneous
		breaking of the half-cavity–shift symmetry $\mathcal S$ via an instability of
		the pure–even modes, which acts as an effective lasing threshold.  In this regime, the molecule formation is no longer 
		symmetric or antisymmetric with respect to $\mathcal S$, but occupies both sectors. Because the molecule 
		originates from lasing on the background, the odd component develops in the regions between neighbouring even 
		solitons (Fig.~S2.1b).
		
		In parity terms, this corresponds to the spontaneous breaking of $\mathcal P(\tfrac{1}{4})$, while 
		$\mathcal P(0)$ remains preserved. This marks the realisation of an effective \mobius geometry: the odd sector 
		of $\mathcal S$ coincides with the odd sector of $\mathcal P(\tfrac{1}{4})$, ensuring that the mode does not 
		satisfy trivial (zero) boundary conditions at the even-soliton peaks.
		
		Note that this is fully consistent with the description in the main text, where we explained that the effective lasing 
		threshold originates from the breaking of $\mathcal S$ while the parity operator $\mathcal P(0)$ remains 
		preserved.
		
		In this scenario, the dominant even soliton acts as an effective periodic potential landscape for the odd field.
		As will be discussed in the next section, we have numerically verified that the dynamics in this regime are
		equivalent for $\gamma = 0$ and $\gamma = 1$ in Eqs.~\eqref{S2.4}--\eqref{S2.7}. This equivalence confirms that
		the energy required for the odd molecule originates from the gain rather than from parametric exchange, drawing a
		parallel with potential-induced mode locking.
		
		In such mode locking, the lasing modes evolve under an effective potential landscape that governs their interaction
		and phase locking~\cite{Bourgon2025Mode,Heckelmann2023Quantum}. In our system, a key difference from conventional
		potential mode locking is that the dominant even soliton introduces a local detuning of the cavity (i.e., in the
		vicinity of the pulse position) through cross-phase modulation. As a consequence, the local lasing threshold in its
		vicinity is increased, forcing the odd mode to grow only in the regions left empty by the even solitons, where the
		threshold is lower.
		
		In essence, the even solitons induce a preferential spatial distribution for the odd state that is
		gain-threshold-driven. This nonlinear mechanism underpins how the molecule is formed and sustained: the odd mode
		lases in the potential carved by the even soliton and not by parametric exchange. This physical picture is
		consistent with the symmetry-based interpretation.
		
		\item \textbf{Second transition (Ising--Bloch): destabilisation of the molecule drift — SSB: \bcancel{$\mathcal T$}, \bcancel{$\mathcal P(0)$}.} 
		This transition requires the coexistence of the two discrete point symmetries 
		($\mathcal S$ and $\mathcal P$) together with continuous translational invariance. 
		The Ising--Bloch bifurcation occurs when an antisymmetric eigenvector of $\mathcal P(0)$ 
		collides with the translational Goldstone mode of $\mathcal T$. 
		In the XPM-only model ($\gamma=0$), the even and odd stationary states remain exact solutions, 
		making it possible to analyse their eigenvalue spectra unambiguously. 
		This isolates the leading instability of the molecule: 
		the collision between the antisymmetric eigenmode and the translational neutral mode, 
		defining the nonequilibrium Ising--Bloch transition~\cite{Michaelis2001Universal} 
		that controls the onset of the molecule drift~\cite{Coullet1990Breaking,Michaelis2001Universal,DeValcarcel2002}.
		
		\item \textbf{Third transition: oscillatory instabilities — SSB: \bcancel{$\mathcal U$}.} 
		The onset of an Andronov--Hopf bifurcation signifies the breaking of the time-translation symmetry $\mathcal U$, 
		where the continuous invariance $t \mapsto t + t_0$ is replaced by a discrete time-translation symmetry 
		$t \mapsto t + nT$ $(n = \pm1, 2, \ldots)$ with period $T$, leading to $T$-periodic oscillations in the \mobius molecule. 
		Further instabilities in these oscillations arise from the disruption of this discrete time-translation symmetry, 
		modifying the temporal dynamics of the \mobius molecule. 
		We do not discuss in detail the subsequent interactions. 
		The XPM-only model ($\gamma=0$) still reproduces the main dynamical behaviour observed for $\gamma=1$ in this regime, 
		although discrepancies induced by the FWM terms become evident as the molecule approaches the route to chaos.
	\end{itemize}
	\subsection*{S2.4 Numerical parameters and physical considerations for the choice of $g$ and $\Delta$}
	
	In all numerical simulations, we use the following fixed parameters:
	$F_a \approx 50$~GHz and $F_b \approx 80$~MHz. 
	The normalised dispersions are 
	$\zeta_a = -\beta_a v_a T_b T_a^{-2} = 1.25 \times 10^{-4}$, 
	$\zeta_b = -\beta_b v_b T_b T_a^{-2} = 2.5 \times 10^{-4}$, 
	and the spectral filtering coefficient is 
	$\sigma = (2\pi \Delta F_F T_a )^{-6} = (1.5 \times 10^{-4})^3$. 
	Here, $\beta_a \approx -20$~ps$^2$km$^{-1}$ and 
	$\beta_b \approx -50$~ps$^2$km$^{-1}$ (see Ref.~\cite{Rowley2022Self}) 
	are the second-order anomalous dispersions of the waveguide and fibre, respectively; 
	$\Delta F_F = 650$~GHz is the band-pass filter bandwidth; and 
	$v_a \approx v_b \approx 2 \times 10^{8}$~m~s$^{-1}$ are the group velocities. 
	The microring resonance is $\Delta F_a \approx 120$~MHz, 
	so that the coupling constant is 
	$\kappa = \pi \Delta F_a F_b^{-1} = 1.5\pi$.
	
	\begin{table}[h]
		\centering
		\caption{Fixed parameters used in all numerical simulations.}
		\begin{tabular}{lcc}
			\hline\hline
			\textbf{Parameter} & \textbf{Symbol} & \textbf{Value} \\
			\hline
			Microring repetition rate & $F_a$ & $50~\text{GHz}$ \\
			Fibre repetition rate & $F_b$ & $80~\text{MHz}$ \\[3pt]
			
			Normalised dispersion (microring) & $\zeta_a$ & $1.25\times10^{-4}$ \\
			Normalised dispersion (fibre) & $\zeta_b$ & $2.5\times10^{-4}$ \\[3pt]
			
			Waveguide GVD & $\beta_a$ & $-20~\text{ps}^2\text{km}^{-1}$ \\
			Fibre GVD & $\beta_b$ & $-50~\text{ps}^2\text{km}^{-1}$ \\[3pt]
			
			Group velocities & $v_a \approx v_b$ & $2\times10^{8}~\text{m\,s}^{-1}$ \\[3pt]
			
			Filter bandwidth & $\Delta F_F$ & $650~\text{GHz}$ \\
			Spectral filtering coefficient & $\sigma$ & $(1.5\times10^{-4})^{3}$ \\[3pt]
			
			Microring resonance linewidth & $\Delta F_a$ & $120~\text{MHz}$ \\
			Coupling constant & $\kappa$ & $1.5\pi$ \\
			\hline\hline
		\end{tabular}
	\end{table}
	
	These parameters are determined by the physical setup and remain essentially unchanged across the experiments. 
	They reproduce the spectral shape of the experimental solitons in Ref.~\cite{Rowley2022Self}, which we retain unchanged here for consistency.
	
	The free parameters in the model are the detuning $\Delta$, the normalised gain $g$ (with $0 < g < 1$), and the group-velocity mismatch $\delta$. 
	Experimentally, $\Delta$ and $g$ are tuned by adjusting the EDFA pump power, while $\delta$ is controlled with the optical delay line. 
	In practice, $\delta$ is set to zero, since a nonzero value explicitly breaks the parity symmetry $\mathcal P$. 
	For this reason, we adopt $\delta = 0$ throughout the discussion of the topological states.
	When $\delta$ is reintroduced in Section S8, we consider $|\delta| \ll 1$, which softly breaks inversion symmetry but preserves the molecule state.
	
	The physical dependence of $g$ and $\Delta$ on the EDFA pump is central to our modelling. 
	Experimentally, soliton formation occurs spontaneously as the pump power increases, in analogy with the behaviour observed in matched cavities (Eqs.~S1.1–S1.2)~\cite{Rowley2022Self}. To describe this effect, Ref.~\cite{Rowley2022Self} introduced a slow dependence of $g$ and $\Delta$ on the field energy via standard Maxwell–Bloch arguments. 
	This approach captured the natural emergence of soliton states under slow energy-dependent variations of $g$ and $\Delta$. 
	After a soliton forms, $\Delta$ settles to a red-detuned value, as directly confirmed by the laser operating on the red slope of the resonance (see Fig.~2 in the main text). 
	As long as the soliton is maintained, an adiabatic change of the EDFA pump modifies the effective gain $g$ experienced by the soliton, 
	while also inducing small refractive-index shifts (thermal and carrier-induced) that affect $\Delta$. 
	These variations are weak and remain within the soliton stability region, as confirmed in~\cite{Rowley2022Self}.
	
	The same modelling strategy extends naturally to the \mobius system (Eqs.~S2.1–S2.2). 
	Once the even soliton state is established, both experiment and simulations show that its energy remains essentially constant 
	across the operating region of the \mobius molecule. 
	This indicates that the key features of the molecule—its formation, robustness, and bifurcations—are governed primarily by the intrinsic dynamics of the mean-field equations, and not by slow nonlinear drifts of $g$ and $\Delta$. 
	Consequently, we neglect slow energy-dependent corrections in the dynamical analysis that follows to isolate the intrinsic mechanism responsible for the formation and stability of the \mobius molecule.
	
	The primary goal of the next section is therefore to identify the relevant operating region in the $(\Delta,g)$ plane, 
	particularly the threshold for molecule formation, corresponding to the lasing of the odd modes. 
	Importantly, in the experiment, the EDFA pump acts as a single global control parameter that simultaneously tunes both $g$ and $\Delta$. 
	In the soliton regime, increasing the EDFA pump increases the effective gain $g$ 
	while also slightly shifting $\Delta$ towards larger red-detuned values. 
	We will show in Section~S5 that the observed sequence of transitions—molecule formation, group-velocity change, and oscillatory bifurcations—
	is consistent with a trajectory of this type in the $(\Delta,g)$ parameter space.
	
	Finally, the influence of slow parameter fluctuations remains essential to understanding the molecule’s response to noise. 
	Adiabatic variations of the parameter $g$, which is the most relevant, are analysed in Section~S7.
	
	\newpage
	\section*{S3. Stationary States and Instabilities}
	
	\renewcommand{\theequation}{S3.\arabic{equation}}
	\setcounter{equation}{0}
	
	In the previous section, we analysed the conditions for creating a \mobius molecule starting from a train of equally spaced solitons populating only the even modes—or, more generally, one of the subspaces of the half-cavity shift symmetry operator~$\mathcal S$. 
	The formation of the molecule requires a non-trivial population of the odd modes and thus a spontaneous breaking of the half-cavity shift symmetry~$\mathcal S$, in a configuration where the odd component grows in the gaps between neighbouring even solitons. 
	This process effectively breaks the parity operator~$\mathcal P$ with respect to a spatial coordinate centred on the even-soliton peaks.
	
	In this section, we identify the range of parameters $(\Delta,g)$ that {enable} this process to occur. 
	We begin by recalling where the system supports stable soliton solutions in the \emph{trivial system}, Eqs.~(\ref{S1.1},\ref{S1.2}). 
	The \emph{\mobius system}, Eqs.~(\ref{S2.1},\ref{S2.2}), sustains analogous solitons in the even and odd subspaces of~$\mathcal S$, as can be verified by setting $a_o=0$ or $a_e=0$ in Eqs.~(\ref{S2.4}–\ref{S2.7}). 
	These states define the elementary building blocks of the soliton train described in Section~S2.
	
	The next step is therefore to map the regions where the lasing background of one sector becomes unstable to perturbations of the opposite sector {(e.g., even states instabilities driven by odd, antisymmetric perturbations)}, since such an instability can seed field growth between adjacent solitons and may enable non-trivial symmetry breaking through the parity operator~$\mathcal P$ centred on the even-soliton peaks. 
	Whether this instability gives rise to a \mobius-type molecule must be verified numerically, {although} identifying its parameter domain provides the necessary foundation for the analysis that follows.
	
	\begin{figure}[H]
		\centering
		\includegraphics[width=0.98\linewidth]{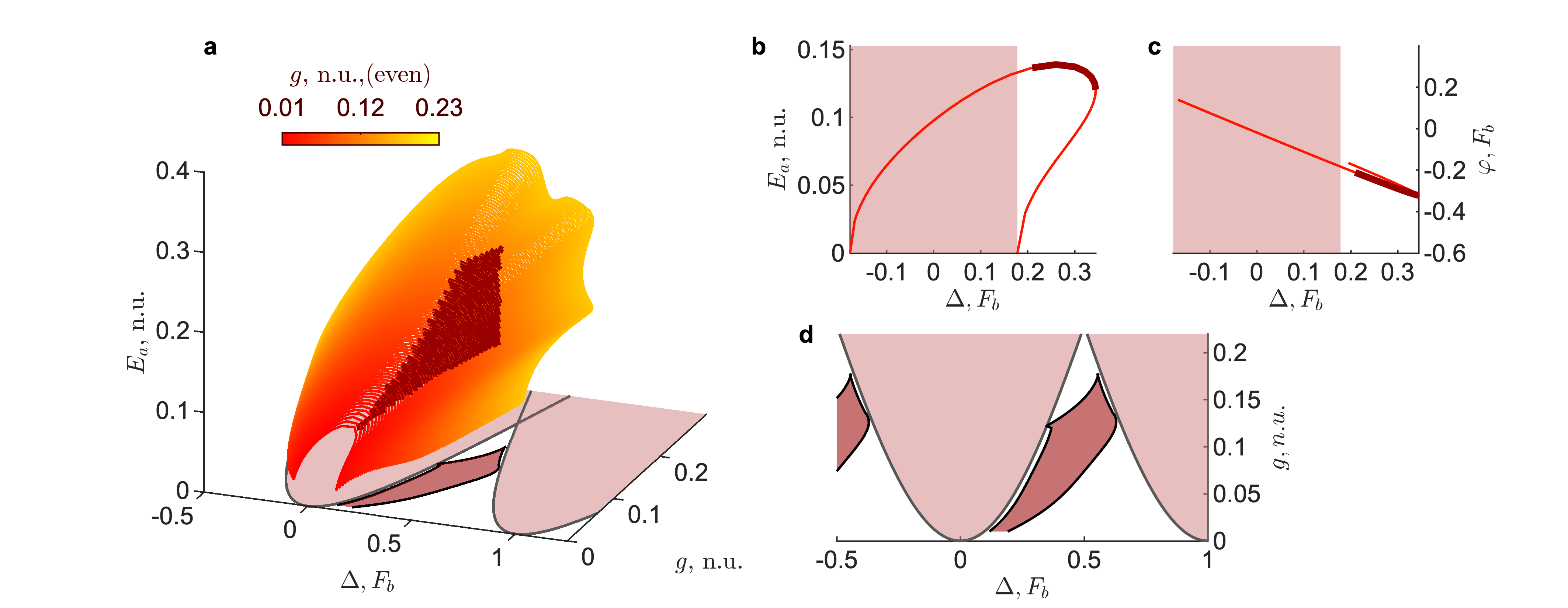}
		\caption{\small\textbf{Fig.~S3.1.} Soliton and background stability properties for Eqs.~(S1.1, S1.2): \textbf{a}, Energy of the soliton state versus detuning $\Delta$ and gain $g$, both expressed in units of the main-cavity FSR $F_b$ (continuous lines, red to yellow as $g$ increases). Thick lines correspond to stable solitons. On the $\Delta$–$g$ plane, the background-instability region is shown in pink and the soliton-stability region in dark red. \textbf{b}, Energy and \textbf{c}, phase of the soliton state versus detuning for a fixed gain value ($g=0.034$), with thick lines indicating stable states and background instability marked in pink. \textbf{d}, Instability regions of the background (pink) and soliton-stable region (dark red).}
		\label{fig:S3.1}
	\end{figure}
	
	We begin our analysis by summarising the properties of the stationary states of the trivial system described by Eqs.~(S1.1, S1.2). {As discussed in Section S2.2, these equations reproduce exactly the even and odd states \emph{in isolation}, i.e., when either $a_e=0$ or $a_o=0$}. 
	The stationary states are expressed as 
	$a = a_S(x)\exp(i\varphi t)$ and $b_q = b_{q,S}(x)\exp(i\varphi t)$, 
	where $a_S(x)$ and $b_{q,S}(x)$ are the spatial profiles and $\varphi$ is the frequency associated with the stationary state. 
	We first consider the background-instability regions. 
	In a laser system, these regions are essential because they determine where the system can spontaneously build up a {coherent} field from an empty cavity. 
	For perturbations of the form $\delta a = \delta a_0 e^{2\pi i f_{MI} x + \lambda t}$ and 
	$\delta b_q = \delta b_{q,0} e^{2\pi i f_{MI} x + \lambda t}$, 
	the zero state becomes modulationally unstable when the real part of the eigenvalue $\lambda$ is positive. 
	The corresponding instability region is given by
	\begin{equation}
		\left| \Delta - q - \frac{\zeta_b - \zeta_a}{4\pi} f_{MI}^2 \right| 
		< \frac{1}{2\pi}\sqrt{\frac{g}{1-g}}\,|1-g+\kappa|. 
		\tag{S3.1}
	\end{equation}
	for each modulational frequency $f_{MI} = 0, \pm1, \pm2$. The widest region occurs for $f_{MI}=0$.
	
	We now turn to the soliton states. Figure~{S3.1}a shows the energy $E_a = \int_{-1/2}^{1/2} |a|^2 dx$ of the stationary soliton as a function of $\Delta$ and $g$ (colour-coded from red to yellow as $g$ increases). 
	Thick lines mark stable solitons, which always lie outside the background-instability region, as is typical in these systems. 
	Because solitons are localised pulses, once the background becomes largely unstable, their tails destabilise and the state is destroyed. 
	This also implies that solitons cannot emerge directly from modulational instability of the background but require a separate initiation mechanism (as recalled in Section~S2.4 and discussed in~\cite{Rowley2022Self}). 
	Each soliton state is uniquely associated with a frequency $\varphi$. 
	Figure~{S3.1}b,c illustrate this for {a particular value of} $g=0.034$: 
	panel~b shows the energy versus detuning, while panel~c shows the soliton frequency $\varphi$, which follows the detuning at low energy but deviates at higher energy owing to the Kerr effect. 
	A summary of the different stability regions across the $(\Delta,g)$ plane is included in Fig. {S3.1}d, highlighting in dark red the soliton-stability region and in pink the background-instability region.
	
	\begin{figure}[H]
		\centering
		\includegraphics[width=0.95\linewidth]{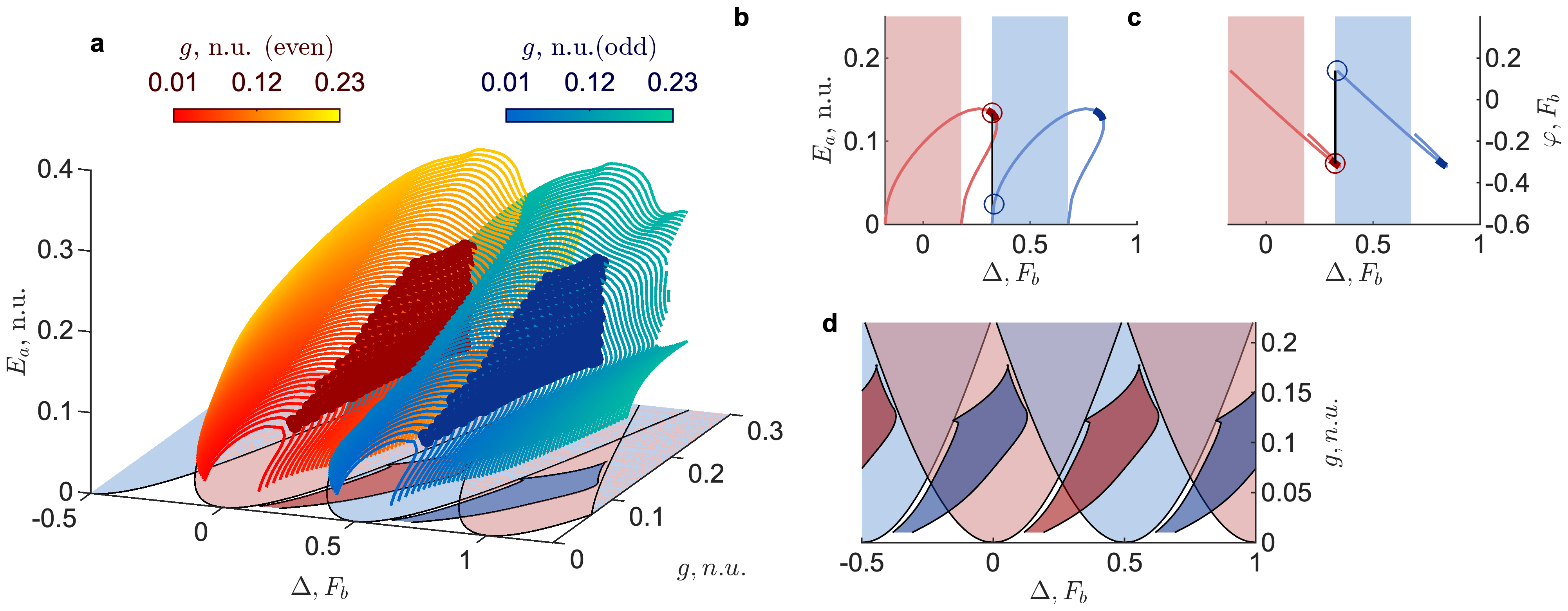}
		\caption{\small\textbf{Fig.~S3.2.} Summary of soliton-solution properties for Eqs.~(S2.1, S2.2). 
			\textbf{a}, Energy of the soliton state versus detuning and gain (continuous lines, red to yellow for even stationary states, blue to cyan for odd stationary states). 
			Thick lines denote solitons stable against perturbations belonging to the same sector. 
			On the detuning–gain plane, background-instability regions are shown in pink (even perturbations) and light blue (odd perturbations). 
			\textbf{b}, Energy and \textbf{c}, frequency $\varphi$ of the soliton state versus detuning for a fixed gain value ($g=0.034$), with red and blue for even and odd states, respectively, and with background-instability regions marked in pink and light blue. 
			\textbf{d}, Instability regions of the background (pink and light blue) and regions where solitons are stable against perturbations within the same sector (red and blue).}
		\label{fig:S3.2}
	\end{figure}
	
	We now focus on Eqs.~(S2.1, S2.2) to characterise how the stability of the stationary states changes for the \mobius system.
	As noted in Section~S2, Eqs.~(S2.4--S2.7) describe two nonlinearly coupled fields with effective detunings that differ by half of the main-cavity FSR: even states correspond to $\Delta_e = \Delta$, while odd states correspond to $\Delta_o = \Delta - 1/2$.
	When considering the stability of the homogeneous background state, the FSR mismatch implies that the modulational instability regions differ for even and odd sectors of the half-cavity shift operator~$\mathcal S$, leading to two distinct instability regions:
	\begin{align}
		\left| \Delta - q - \frac{\zeta_b - \zeta_a}{4\pi} f_{MI}^2 \right| 
		&< \frac{1}{2\pi} \sqrt{\frac{g}{1-g}}\,|1-g+\kappa|,
		\quad &&\text{for even ($\Pi_+^{\mathcal S}$) perturbation frequencies } f_{MI} = 0, \pm2, \ldots \tag{S3.2}\\[3pt]
		\left| \Delta - \tfrac{1}{2} - q - \frac{\zeta_b - \zeta_a}{4\pi} f_{MI}^2 \right| 
		&< \frac{1}{2\pi} \sqrt{\frac{g}{1-g}}\,|1-g+\kappa|,
		\quad &&\text{for odd ($\Pi_-^{\mathcal S}$) perturbation frequencies } f_{MI} = \pm1, \pm3, \ldots \tag{S3.3}
	\end{align}
	
	Turning to the soliton states, a purely even ($\Pi_+^{\mathcal S}$) stationary soliton solution $a_{S,e}(x)$ of Eqs.~(S1.1, S1.2) maps
	onto a superposition of two equidistant $a_S(x)$ solitons in Eqs.~(S2.4--S2.7):
	\begin{equation}
		a_{S,e}(x)e^{i\varphi_e t} = \bigl[a_S(x,\Delta) + a_S(x+1/2,\Delta)\bigr] e^{i\varphi_e t}, 
		\qquad \varphi_e = \varphi(\Delta).
		\tag{S3.4}
	\end{equation}
	Analogously, the odd-sector ($\Pi_-^{\mathcal S}$) solitons are described by
	\begin{equation}
		a_{S,o}(x)e^{i\varphi_o t}= \bigl[a_S(x,\Delta-1/2) - a_S(x+1/2,\Delta-1/2)\bigr] e^{i\varphi_o t}, 
		\qquad \varphi_o = \varphi(\Delta-1/2).
		\tag{S3.5}
	\end{equation}
	The {odd-sector solitons} effectively behave as combinations of two solitons of Eqs.~(S1.1, S1.2) detuned by $\Delta_o = \Delta - 1/2$.
	
	Both stationary states remain stable against perturbations belonging to the same sector of $\mathcal S$ as in Eqs.~(S1.1, S1.2), but this does not guarantee stability against perturbations from the opposite sector.
	
	Figure~S3.2 summarises the existence and stability properties of both families. 
	Similarly to Fig.~S3.1a, Fig.~S3.2a shows the soliton energy $E_a$ as a function of $\Delta$ and $g$, with lines colour-coded from red to yellow for even states and from blue to cyan for odd states as $g$ increases. 
	The background-instability regions are shaded pink for even perturbations and light blue for odd perturbations, corresponding to the thresholds of Eqs.~(S3.2) and (S3.3). 
	For clarity, these regions are also summarised in Fig.~S3.2d. 
	
	The analysis highlights that the stable region of the even soliton state under even perturbations (dark red in the $\Delta$–$g$ plane) overlaps with the background instability induced by odd perturbations (light blue). 
	This directly answers one of the key questions raised in Section~S2:  the molecule  originates from a background instability and 
	localises in the regions left empty by the even field.
	In the next section, we analyse how odd perturbations drive even states towards a stable molecule configuration. 
	By symmetry, the odd states exhibit equivalent behaviour: they become unstable against even perturbations once they enter the background instability associated with the even sector of $\mathcal S$. 
	In this paper, we focus on the destabilisation of even states {only} for simplicity.
	
	Before closing this section, it is useful to examine the stationary soliton states that can coexist in the system. 
	Figures~S3.2b and S3.2c show the energy and frequency of the even and odd soliton states for $g=0.034$. 
	The states at detuning $\Delta = 0.34$ are highlighted with circles and lie within the same-sector stability region of the even soliton. 
	At this {value of the} detuning, however, an odd soliton solution also exists, although it is unstable and characterised by a distinct frequency $\varphi$ (Fig.~S3.2c), corresponding experimentally to a different carrier-envelope offset (CEO). 
	When the even soliton is destabilised by odd perturbations, the dynamics of the odd sector naturally converge towards the odd-soliton branch, preserving consistent field and frequency {features}. 
	
	The frequency mismatch between the even and odd states generates a four-wave-mixing component at
	\[
	f_{CM} = 2F_b \Delta \varphi = 2F_b |\varphi_e - \varphi_o|,
	\]
	a feature that will also be investigated numerically. 
	Thanks to this correspondence, the stationary solutions Eqs.~(S3.4, S3.5) serve as useful initial guesses for the XPM-only system [Eqs.~(S2.4--S2.7)] with $\gamma = 0$, as anticipated in Section~S2 and employed in the following numerical analysis. Note that here we use a normalised definition of the CEO, differently from the Main.
	
	At this stage, we have identified the regions of background instability that can trigger coupling between the even and odd sectors.
	However, this mapping only provides the necessary conditions for the emergence of the \mobius molecule. 
	A full understanding of the instability mechanism requires a linear stability analysis of the molecule itself. 
	This provides a natural opportunity to introduce a more formal Bogoliubov–de Gennes framework, enabling a rigorous treatment of the geometrical and topological properties of the perturbations and the definition of a Zak \cite{Zak1989, Fan2022Topological, Li2023Zak} phase, supporting the picture developed in the symmetry discussion of Section~S2.
	We will carry out this analysis in Sections~S5 and S6. 
	Before doing so, we first verify that a molecule state can indeed be obtained from the destabilisation of the even two-soliton state, and summarise a few representative cases that are important for this work in Section~S4.
	\newpage

	\section*{S4. \mobius Molecule Formation: Summary of Relevant Transitions}
	
	It is useful to illustrate the nature, stability, and dynamics of the molecule emerging from the destabilisation of the even two-soliton state.
	In this Section, therefore, we summarise three representative dynamical scenarios observed in propagation simulations:
	\textbf{(i)} formation of a stationary \mobius molecule,
	\textbf{(ii)} a drifting molecule state, and
	\textbf{(iii)} an oscillating molecule state.
	These examples provide a phenomenological overview of the relevant states that dynamically occur in our system, and we introduce key figures of merit and measurable indicators of the topological character of the \mobius molecule.
	In the following sections, we will explicitly and quantitatively connect these dynamical cases to the system’s topological properties and provide a complete mapping of their existence in the $\Delta$–$g$ plane.
	
	\subsection*{S4.1 Lasing of the Odd State and Formation of a Stationary \mobius Molecule}
	\label{sec:S4.1}
	
	We start by examining an example of even-state propagation since, as discussed in Section~S3, the main formation mechanism of the \mobius molecule is the emergence of the odd soliton state from the instability of the homogeneous background. 
	From a fundamental point of view, this corresponds to the first spontaneous symmetry breaking of the half-cavity shift operator $\bcancel{\mathcal S}$, which populates the sector $\Pi^{(\mathcal S)}_{-}$ orthogonal to the even sector $\Pi^{(\mathcal S)}_{+}$ where the two-soliton even states naturally belong.
	
	We consider a system evolving with parameters $g = 0.0343$ and $\Delta = 0.34$, close to the instability threshold for odd-state generation, and propagate an even soliton state with zero background, as expressed by Eq.~(S3.4). 
	To clearly distinguish the numerical simulation results from the stationary states computed in Section~S3, we refer to the numerical propagation solution as the \emph{simulated field}, labelled $E_s$ and $O_s$ for the even and odd components, respectively. 
	These results are compared with the stationary states computed for the XPM-only system (Eqs.~(S2.4--S2.7) with $\gamma = 0$), referred to as the \emph{computed fields}, labelled $E_c$ and $O_c$. 
	As an initial guess for the stationary states, we used the solutions of Eqs.~(S1.1, S1.2), Eqs.~(S3.4)--(S3.5), with $\Delta_e = 0.34$ and $\Delta_o = 0.34 - 1/2$ for the even and odd components, respectively.
	
	We simulated the propagation of the variables $a(x)$ and $b_q(x)$, as described by Eqs.~(S2.1, S2.2), and extracted the even and odd modes by applying the projectors $\Pi^{\mathcal S}_{\pm}$. 
	The results are shown in Fig.~S4.1a--c, reporting the intensity of the total field $|a(x,t)|^2$ (panel~a) and its even (panel~b) and odd (panel~c) components. 
	Starting from a pure even soliton at $t=0$, the odd perturbations progressively grow, giving rise to a hybrid even--odd state that becomes visible around $t = 4000$. 
	The odd field exhibits a different CEO frequency (as also clearly visible in Fig.~S4.1g), making the global state effectively a soliton molecule in which the two sector components are superposed. 
	As anticipated in the previous section, the odd field occupies the spatial regions left empty by $a_e$ due to the local shaping of the odd-mode lasing threshold induced by the even soliton. 
	As illustrated by Fig.~S4.1d, the total spectrum is effectively composed of separate sets of lines corresponding to the even and odd components (labelled in red and blue for illustrative purposes, respectively), reflecting the interleaving of the spectral lines also observed experimentally (see Fig.~2 in the main text).
	
	The radio-frequency noise spectra of the two field components, computed from the average energies, are reported in Fig.~S4.1e,f, showing zooms around the baseband (DC, panel~e) and around the component at $f \sim F_b$ (AC, panel~f). 
	Specifically, we plot the power spectral density (PSD) of the average energy $E_{e,o}(t) = \int |a_{e,o}(x,t)|^2 dx$ for both even and odd components (red and blue lines), corresponding to the radio-frequency spectra measured in our experiments (see, e.g., Fig.~2 in the main text). 
	The results highlight the FWM CEO mismatch frequency $f_{CM} = 2F_b |\varphi_e - \varphi_o|$, confirming modulation at this frequency and the presence of FWM interactions consistent with experimental observations.
	
	To extract the soliton frequencies, we averaged the even and odd fields $\int a_{e,o}(x,t)\,dx$ and computed the corresponding Fourier transforms, whose PSDs are shown in Fig.~S4.1g (with Fig.~S4.1h zooming around the odd-soliton components). 
	Significantly, the resulting frequencies align well with the soliton frequencies of the stationary states, shown in dark blue ($\varphi_o=-0.129$) and red ($\varphi_e=0.322$). 
	A more detailed comparison of the output states is shown in Fig.~S4.1i--k, where we plot the spatial distribution of the final (asymptotic) fields in terms of intensity (panel~i), phase (panel~j), and Argand representation (panel~k), confirming that in this regime the system converges to a molecule composed of an even--odd soliton pair. 
	
	Further insight can be obtained from the temporal evolution of the Argand plot for the odd field, reported in Fig.~S4.1l for the last simulation period ($t\in [9600, 10000]$). 
	While completing a period, the field passes through the origin of the Argand plane at all propagation times, as shown in the zoom in Fig.~S4.1m, a singularity condition with zero field and undefined phase. 
	For the same time range, Fig.~S4.1n reports the Argand-plane evolution for the even mode. 
	The even trajectory does not complete a full period within the same time range, consistent with Fig.~S4.1g, where the soliton frequencies of the two stationary states differ. 
	This frequency mismatch confirms that the state is a molecule rather than a single-mode, fully stationary solution, as the two components evolve with different propagation frequencies. 
	The global state can therefore be considered a limit cycle characterised by the frequency $F_b |\varphi_e - \varphi_o|$, evidenced by the FWM CEO mismatch $f_{CM} = 2F_b |\varphi_e - \varphi_o|$.
	
	Notably, the soliton molecule described in Fig.~S4.1 is asymptotically stable and realises a topological \mobius molecule, a result that is demonstrated in Sections~S5 and~S6. 
	The physical signature of this topological behaviour lies in the phase of the odd mode, which displays two discontinuous jumps of $\pm\pi$ where the even solution has a peak (Fig.~S4.1j).
	
	Discontinuous $\pi$ phase jumps in cavity nonlinear optics are well known in connection with domain walls in optical parametric oscillators~\cite{Trillo1997Stable,Oppo1999From,Oppo2001,Englebert2024Topological} and in vectorial polarisation resonators~\cite{Gallego2000Self,Garbin2021Dissipative,Lucas2024Polarization}. 
	In our case, the intensity profile of the odd mode differs markedly from the flat shape of classical domain walls, yet the phase singularity provides comparable robustness and underpins the key dynamics. 
	Crucially, the $\pi$ phase jumps occurring at the peaks of the even soliton enforce the half-period inversion required by the \mobius condition $f(x+\tfrac{1}{2})=-f(x)$, enforcing a non-trivial configuration. 
	In other terms, the odd mode populating the sector $\Pi_-^{\mathcal S}$ originates from the breaking of the parity operator centred on the peak of the even soliton, $\mathcal P(\tfrac{1}{4})$, while the parity operator associated with the midpoint of the even-soliton doublet ($x_c=0$), $\mathcal P(0)$, remains conserved. 
	Physically, the two soliton sectors are shifted by half a cavity round trip and linked by this phase inversion, which closes the \mobius loop in a single round trip.
	
	\begin{figure}[H]
		\centering
		\includegraphics[width=.8\linewidth]{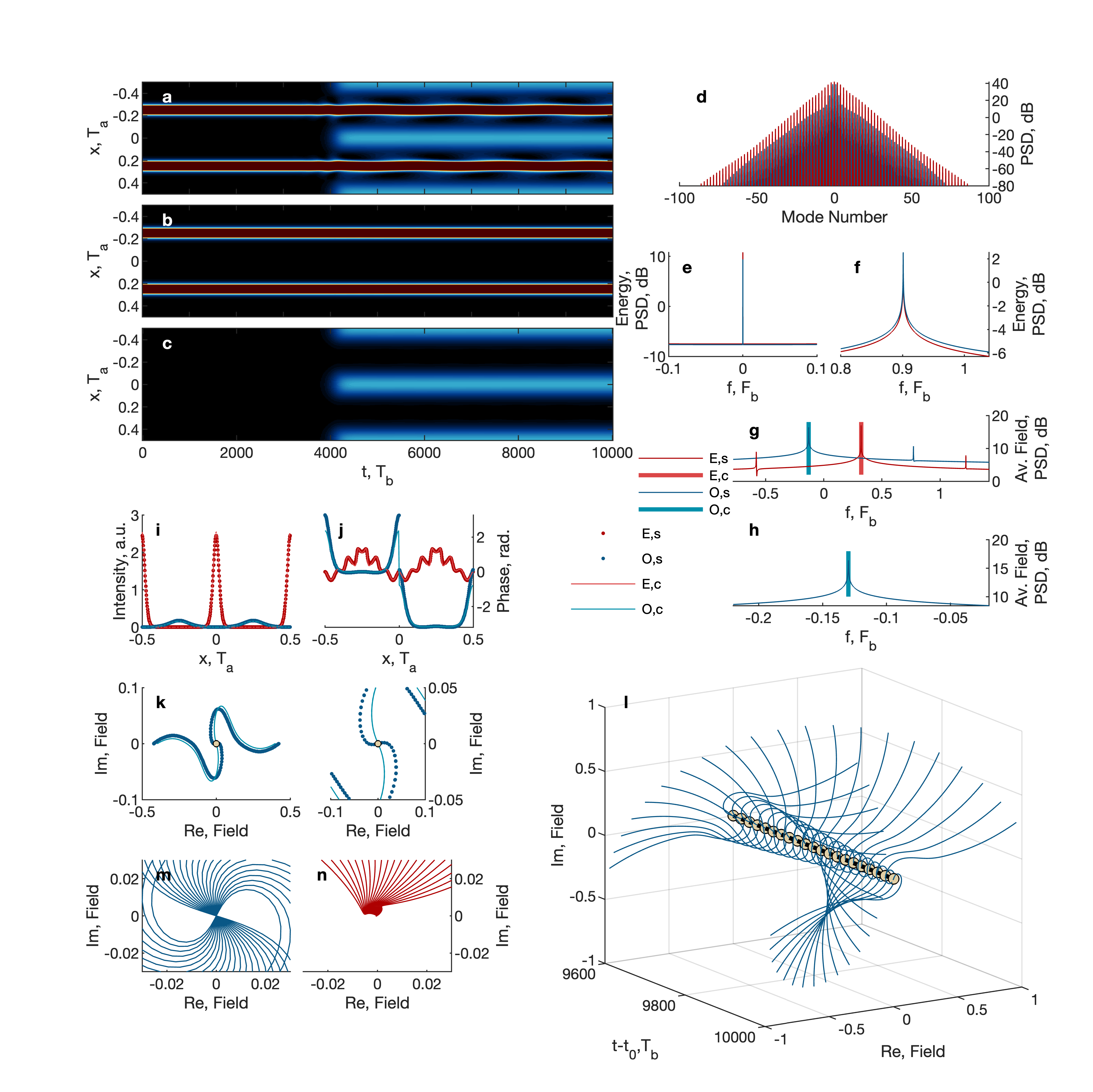}
		\caption{\small\textbf{Fig.~S4.1.} Formation of a \mobius molecule from an initially pure even soliton ($g=0.0343$, $\Delta=0.34$). 
			\textbf{a}, Time evolution of the total field intensity $|a(x,t)|^2$, showing the growth of the odd perturbation around $t\!\approx\!4000$ and convergence to a stable molecule. 
			\textbf{b}, \textbf{c}, Spatial distributions of the even (\textbf{b}) and odd (\textbf{c}) components, illustrating how the odd field populates the regions left empty by the even field. 
			\textbf{d}, Optical spectrum of the final state, with even (red) and odd (blue) components corresponding to different periodicities. 
			\textbf{e}, \textbf{f}, Radio-frequency noise spectra of the average energies $E_{e,o}(t)$ for even (red) and odd (blue) components, showing zooms around DC (\textbf{e}) and the AC component near $f\!\sim\!F_b$ (\textbf{f}). 
			\textbf{g}, Power spectral density of the average fields $\int a_{e,o}(x,t)\,dx$, with peaks aligned with the stationary-state soliton frequencies. 
			\textbf{h}, Zoom on the odd-component spectrum in (\textbf{g}). 
			\textbf{i}, Intensity profiles of the even (red) and odd (blue) output fields at $t=10000$: simulated fields ($E_s$, $O_s$, solid) and stationary-state solutions ($E_c$, $O_c$, dashed light red/light blue). 
			\textbf{j}, Corresponding field phase profiles. 
			\textbf{k}, Argand representation of the odd field $a_o(x,t)$ with zoom showing the zero crossing. 
			\textbf{l}, Temporal evolution of the odd field on the Argand plane during the last simulation period. 
			\textbf{m}, Zoomed Argand projection of the odd field around the node. 
			\textbf{n}, Same as (\textbf{m}), but for the even field.}
		\label{fig:S4.1}
	\end{figure}
	\subsection*{S4.2 Onset of Drift: Ising--Bloch-like Transition}
	\label{sec:S4.2}
	
	A second important example is summarised in Fig.~S4.2a, which shows simulations of Eqs.~(S2.1, S2.2) with parameters slightly higher than those in Fig.~S4.1 ($g=0.0366$, $\Delta=0.35$). 
	Starting from a pure double soliton in the even mode, the system first develops a molecule with odd-field components, followed by a transition to a drifting state in which the M\"obius molecule spontaneously acquires a finite group velocity.  
	
	From a fundamental perspective, this transition corresponds to the second spontaneous symmetry breaking: the loss of parity symmetry at the midpoint between the two soliton states ($x_c=0$). 
	The breaking of the second parity operator, $\bcancel{\mathcal P(0)}$, leads to the population of its antisymmetric sector, $\Pi_-^{\mathcal P(0)}$. 
	In this regime, the molecule no longer preserves any parity symmetry—that is, the solution cannot be transformed back into itself by reflection around any point~$x_c$. 
	As shown in Section~S5, this occurs through the interaction of the corresponding eigenvalue with the translational Goldstone neutral mode and the resulting breaking of $\bcancel{\mathcal T}$. 
	This defines the nonequilibrium Ising--Bloch transition, which here manifests as a drifting molecule state~\cite{Coullet1990Breaking,Michaelis2001Universal}.
	\begin{figure}[H]
		\centering
		\includegraphics[width=0.8\linewidth]{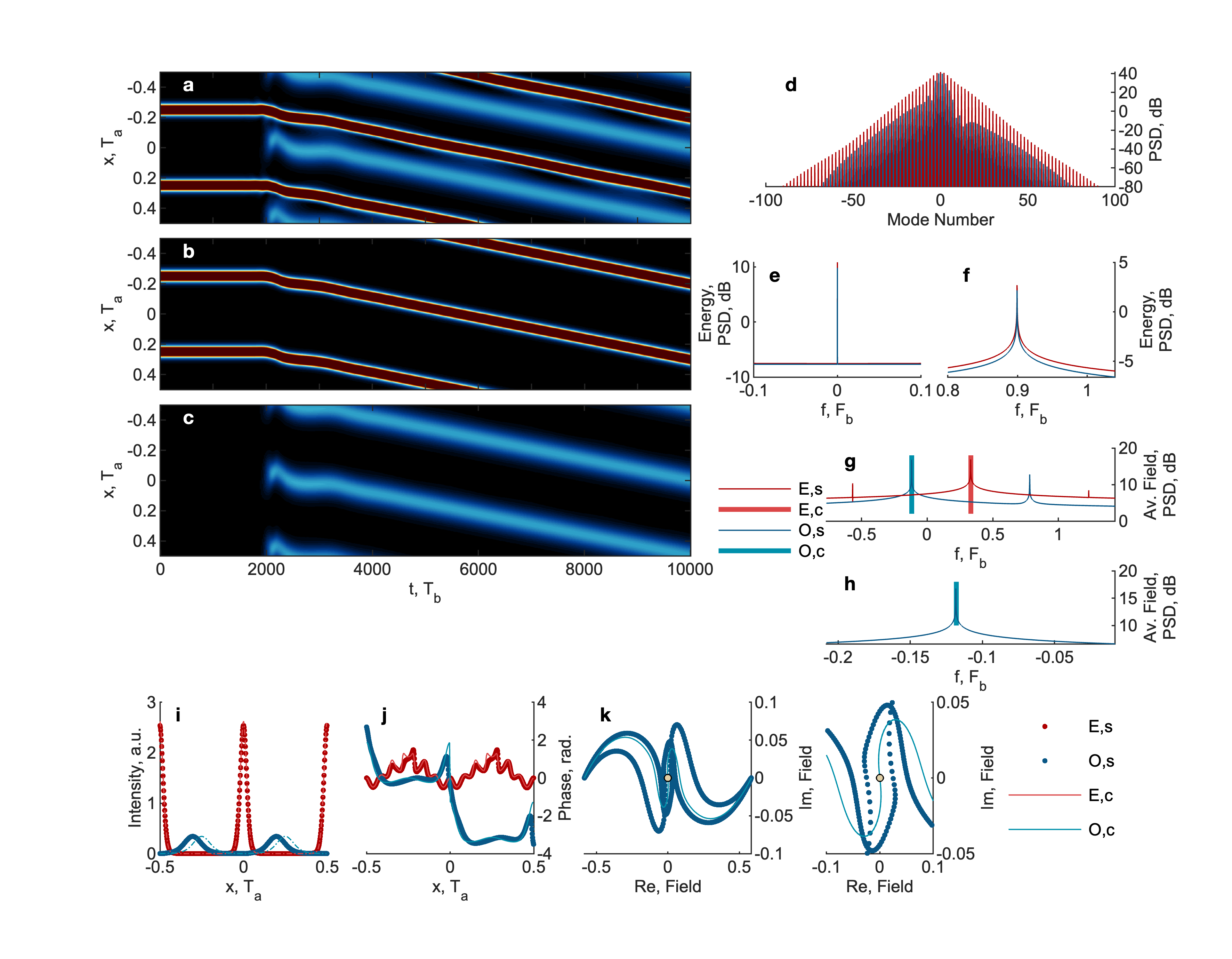}
		\caption{\small\textbf{Fig.~S4.2.} Formation of a drifting \mobius molecule from an initially pure even soliton ($g=0.0366$, $\Delta=0.35$), illustrating an Ising--Bloch-like transition. 
			\textbf{a}, Time evolution of the total field intensity $|a(x,t)|^2$, showing growth of the odd perturbation around $t\!\approx\!2000$ and subsequent formation of a stable molecule accompanied by a group-velocity transition. 
			\textbf{b}, \textbf{c}, Spatial distributions of the even (\textbf{b}) and odd (\textbf{c}) components, showing how the odd field populates the regions left empty by the even field. 
			\textbf{d}, Optical spectrum of the final state ($t=10000$), with even (red) and odd (blue) spectral lines corresponding to different periodicities. 
			\textbf{e}, \textbf{f}, Radio-frequency noise spectra computed as the power spectral density of the average energies $E_{e,o}(t)=\!\int |a_{e,o}(x,t)|^2dx$ for the even (red) and odd (blue) components. Panels show zooms around DC (\textbf{e}) and around the AC component near $f\!\sim\!F_b$ (\textbf{f}), confirming the FWM component at $f_{CM}=2F_b|\varphi_e-\varphi_o|$. 
			\textbf{g}, Power spectral density of the average fields $\int a_{e,o}(x,t)\,dx$, with peaks aligned with the stationary-state soliton frequencies. 
			\textbf{h}, Zoom on the odd-component spectrum in (\textbf{g}). 
			\textbf{i}, Intensity profiles of the even (red) and odd (blue) output fields at $t=10000$: simulated fields ($E_s$, $O_s$, solid lines) and stationary-state solutions ($E_c$, $O_c$, dashed light red/light blue). The odd intensity is no longer centred between the two even solitons, indicating the onset of drift. 
			\textbf{j}, Corresponding field-phase profiles. 
			\textbf{k}, Argand representation of the odd field $a_o(x,t)$, with zoom showing that the node has been removed.}
		\label{fig:S4.2}
	\end{figure}
	The field intensities for the total, even, and odd components are plotted in Fig.~S4.2a--c, as in Section~S4.1. 
	Figures~S4.2d--h confirm that the key soliton features of the molecule (spectral interleaving, RF spectra, soliton frequencies) remain consistent with Fig.~S4.1, with no additional spectral features emerging.
	However, for this parameter set, the molecule’s centre of mass is no longer stationary, and the odd state shifts closer to one of the even solitons (Fig.~S4.2i). 
	This is the first clear signature of the loss of $\mathcal{P}$ symmetry. 
	The change is also associated with a shift in the average position of the optical spectra in Fig.~S4.2d, quantified through the spectral-centroid momentum  \cite{Alberucci2007} for the field $a$:
	\[
	\hat{L}[a]=\int_{-1/2}^{1/2}\!\mathrm{Im}\,[a(x)^*\partial_x a(x)]\,dx,
	\qquad
	\hat{L}[a_e]=-0.0440\,F_a,\quad \hat{L}[a_o]=+0.1202\,F_a.
	\]
	an equivalent definition can be used for $b_q$, which we omit for simplicity.
	The two components thus acquire opposite momentum shifts, reflecting the imbalance of the molecule’s centre of mass. 
	This metric will be used in Section~S6 to identify the loss of parity symmetry.  
	
	Critically, the phase singularity disappears (Fig.~S4.2k, zoom), and the Argand plot no longer crosses the origin, indicating that the topological localisation of the molecule has been lost. 
	The sign of the resulting group velocity is random and depends on the initial noise realisation. 
	The onset of moving soliton molecules is reminiscent of nonequilibrium Ising–Bloch transitions of domain walls~\cite{Coullet1990Breaking,Michaelis2001Universal,DeValcarcel2002}, experimentally observed in nonlinear optics with photorefractive media in the transverse direction~\cite{EstebanMartin2005}.
	In those systems, the transition from static to dynamic domain walls marks the passage from an Ising-type to a Bloch-type configuration, and the resulting motion can occur in either direction depending on initial conditions or noise~\cite{Coullet1990Breaking,Michaelis2001Universal,DeValcarcel2002}.
	\begin{figure}[H]
		\centering
		\includegraphics[width=0.8\linewidth]{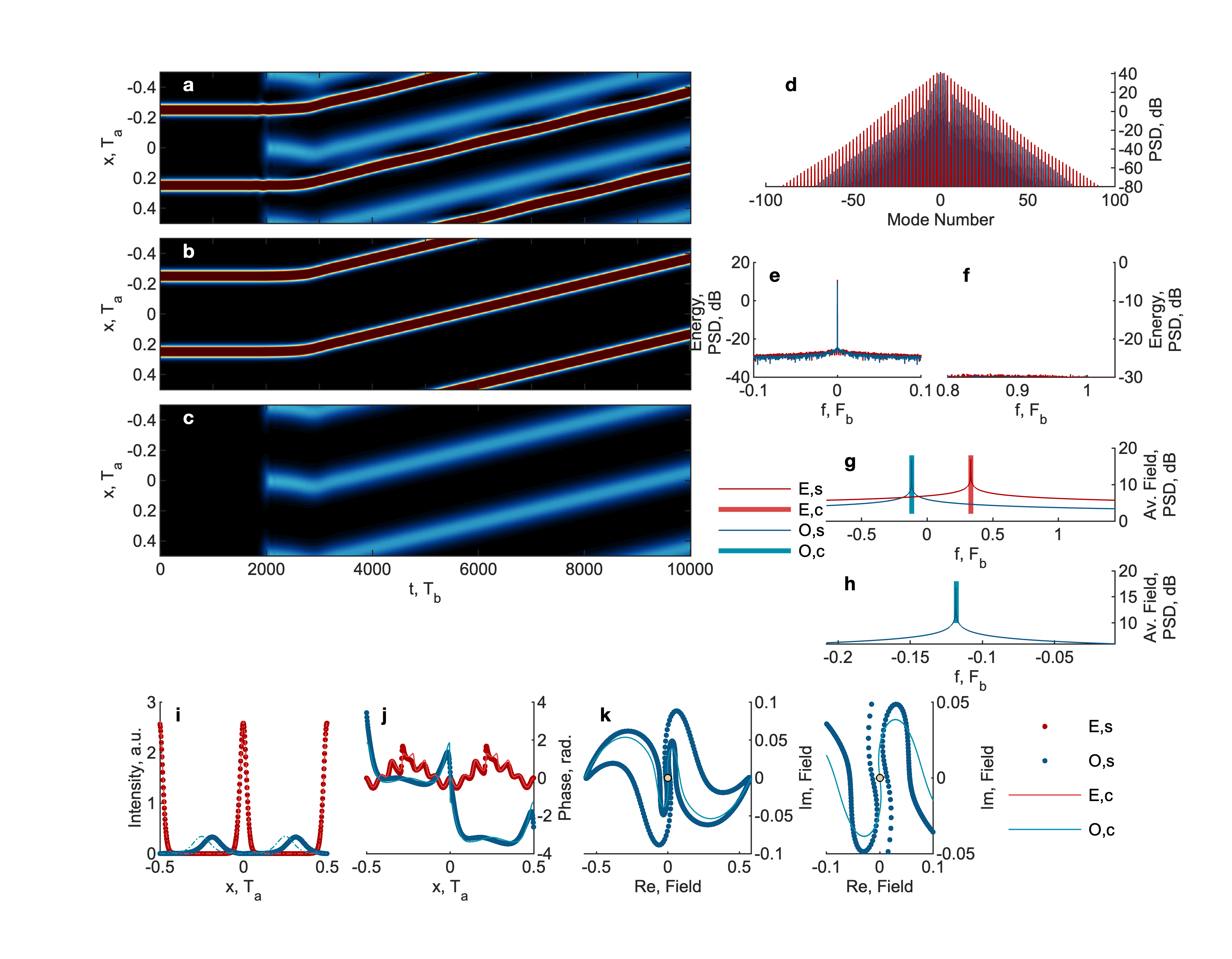}
		\caption{\small\textbf{Fig.~S4.3.} Formation of a drifting \mobius molecule in the purely XPM-only system (Eqs.~(S2.4--S2.7) with $\gamma=0$), for $g=0.0366$ and $\Delta=0.35$, illustrating an Ising--Bloch-like transition without FWM coupling. 
			\textbf{a}, Time evolution of the total field intensity $|a(x,t)|^2$, showing growth of the odd perturbation around $t\!\approx\!2000$ and the subsequent formation of a stable molecule with a transition in group velocity. 
			\textbf{b}, \textbf{c}, Spatial distributions of the even (\textbf{b}) and odd (\textbf{c}) components. 
			\textbf{d}, Optical spectrum of the final state ($t=10000$), with even (red) and odd (blue) spectral lines corresponding to distinct periodicities. 
			\textbf{e}, \textbf{f}, Radio-frequency noise spectra computed as the power spectral density of the average energies $E_{e,o}(t)=\!\int |a_{e,o}(x,t)|^2dx$ for the even (red) and odd (blue) components, zoomed around DC (\textbf{e}) and the AC component near $f\!\sim\!F_b$ (\textbf{f}). 
			As expected for the XPM-only system, no FWM component appears at $f_{CM}=2F_b|\varphi_e-\varphi_o|$. 
			\textbf{g}, Power spectral density of the average fields $\int a_{e,o}(x,t)\,dx$, with peaks aligned with the stationary-state soliton frequencies. 
			\textbf{h}, Zoom on the odd-component spectrum in (\textbf{g}). 
			\textbf{i}, Intensity profiles of the even (red) and odd (blue) output fields at $t=10000$: simulated fields ($E_s$, $O_s$, solid lines) and stationary-state solutions ($E_c$, $O_c$, dashed light red/light blue). The odd intensity is no longer centred between the two even solitons, indicating the onset of drift. 
			\textbf{j}, Corresponding field-phase profiles. 
			\textbf{k}, Argand representation of the odd field $a_o(x,t)$, with zoom showing that the phase singularity has been removed.}
		\label{fig:S4.3}
	\end{figure}
	
	This destabilisation mechanism will be analysed in detail in Sections~S5 and S6. 
	As discussed there, the strictly topological nature of the molecule exemplified in Fig.~S4.1 is lost. 
	Nonetheless, the centres of the even and odd solitons remain well separated, and a M\"obius geometry is still preserved in this configuration, with important implications for noise resilience. 
	
	To isolate the role of the FWM terms on the drifting-molecule dynamics, Fig.~S4.3 reports the same analysis as in Fig.~S4.2 for numerical simulations of the purely XPM-only system ($\gamma=0$). 
	Although the odd soliton component still forms, the characteristic spectral peak at $f\!\approx\!F_b$ in panel~f is absent, confirming the four-wave-mixing origin of this interaction. 
	Importantly, the solution again evolves into a molecule composed of the same soliton pair discussed in Section~S4 and undergoes the same Ising--Bloch-like transition as with the full nonlinear interaction. 
	Since the sign of the group-velocity change is purely random and depends on the initial noise conditions, Fig.~S4.3 shows an example where the direction of motion is opposite to that observed in Fig.~S4.2.
	
	\subsection*{S4.3 Onset of Oscillations: Andronov-Hopf Bifurcation}
	\label{sec:S4.3}
	
	\begin{figure}[H]
		\centering
		\includegraphics[width=0.8\linewidth]{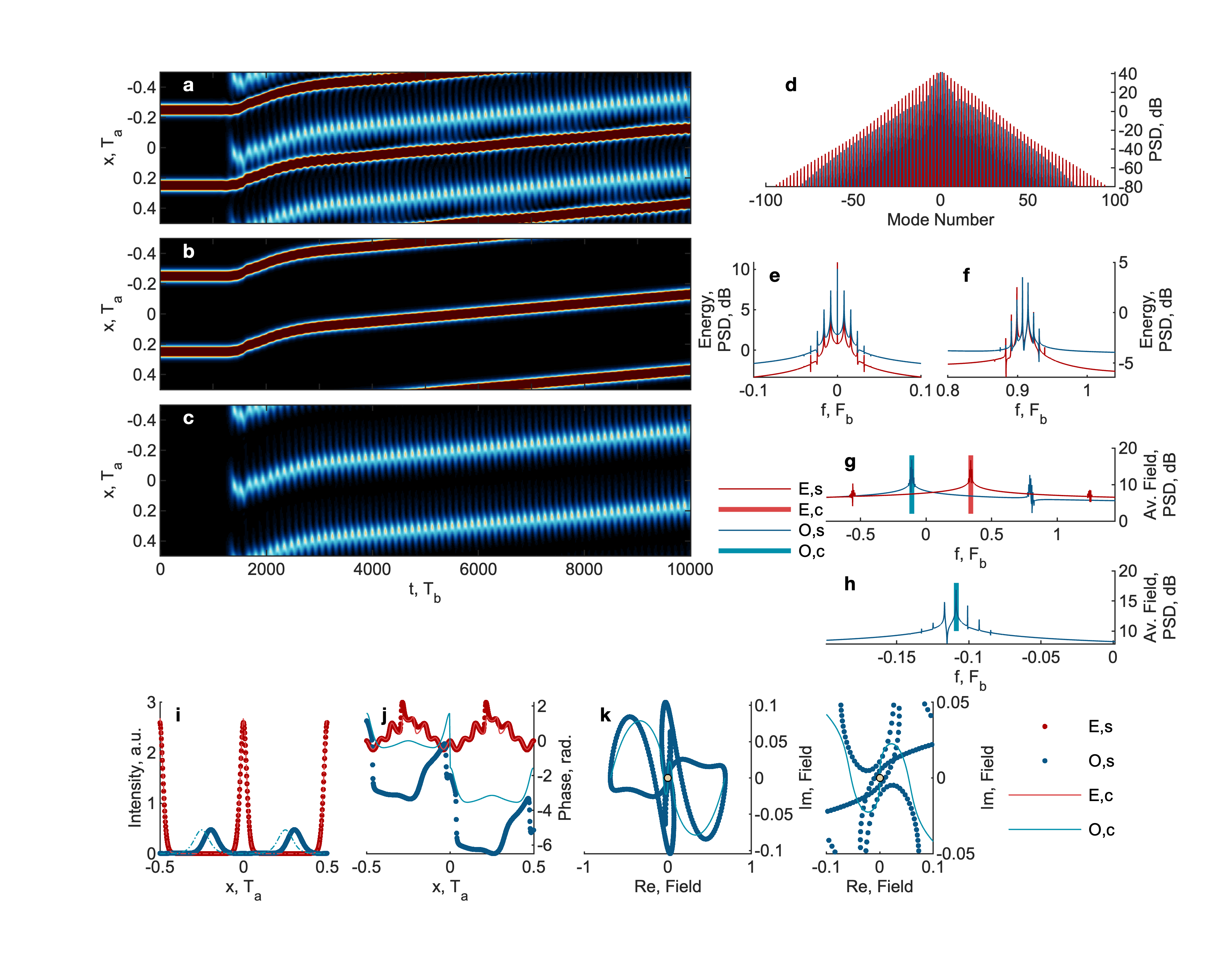}
		\caption{\small\textbf{Fig.~S4.4.} Formation of an oscillating \mobius molecule from an initially pure even soliton ($g=0.0386$, $\Delta=0.36$), illustrating the onset of a Andronov-Hopf  transition. 
			\textbf{a}, Time evolution of the total field intensity $|a(x,t)|^2$, showing growth of the odd perturbation around $t\!\approx\!2000$ and the subsequent onset of persistent oscillations. 
			\textbf{b}, \textbf{c}, Spatial distributions of the even (\textbf{b}) and odd (\textbf{c}) field components. 
			\textbf{d}, Optical spectrum of the final \mobius molecule ($t=10000$), with even (red) and odd (blue) spectral contributions reflecting distinct periodicities. 
			\textbf{e}, \textbf{f}, Radio-frequency noise spectra computed as the power spectral density of the average energies $E_{e,o}(t)=\int |a_{e,o}(x,t)|^2dx$, zoomed around the DC (\textbf{e}) and AC (\textbf{f}) components near $f\!\sim\!F_b$. 
			The modulation at this frequency confirms the presence of FWM at $f_{CM}=2F_b|\varphi_e-\varphi_o|$. 
			\textbf{g}, Power spectral density of the average fields $\int a_{e,o}(x,t)\,dx$, with peaks aligned to the stationary-state soliton frequencies; \textbf{h} shows a zoom on the odd-component spectrum. 
			\textbf{i}, Final intensity profiles of the even (red) and odd (blue) output fields, comparing simulated fields ($E_s$, $O_s$, solid lines) with stationary-state solutions ($E_c$, $O_c$, dashed light red/light blue lines). 
			\textbf{j}, Corresponding field-phase profiles. 
			\textbf{k}, Argand representation of the odd field $a_o(x,t)$, showing that the phase singularity has been removed as oscillations set in.}
		\label{fig:S4.4}
	\end{figure}
	At higher gain values, the system further develops its dynamics, undergoing additional bifurcations and dynamical transitions. 
	A typical evolution is the onset of oscillatory behaviour, illustrated in Fig.~S4.4. 
	This instability arises from an Andronov--Hopf bifurcation to a limit cycle~\cite{NonlinearDynamics2007}, 
	corresponding to a change in the stability of an eigenvector associated with a pair of complex-conjugate eigenvalues. 
	From a symmetry viewpoint, this marks the breaking of time-translation invariance, $\bcancel{\mathcal U}$. 
	As the gain increases further, the oscillatory regime experiences secondary instabilities that can eventually evolve into chaotic behaviour. 
	A full characterisation of the route to chaos lies beyond the scope of this work, 
	as our focus remains on the primary instability and the associated \mobius molecule formation. 
	Nonetheless, preliminary simulations confirm that the molecule persists deep into the chaotic regime, 
	underscoring the central role of the \mobius geometry in maintaining soliton-molecule integrity 
	against both primary and higher-order dynamical transitions.

	\subsection*{S4.4. On the Topological Nature of the Phase Jump within the Odd-Molecule Component}
	
	The Ising--Bloch (IB) transition in dissipative optical systems has long been associated with domain walls and dark solitons~\cite{Coullet1990Breaking,Michaelis2001Universal,DeValcarcel2002}. 
	In such systems, the topological character is encoded in the $\pi$ phase discontinuity of the field, which acts as a one-dimensional analogue of a {two-dimensional} vortex core. 
	This correspondence allows the phase jump to be treated as a localised topological defect. 
	The same feature is directly visible in our M\"obius molecule: the odd component, which identifies the non-trivial topology of the state, carries a characteristic $\pi$ phase inversion superimposed on the even-soliton background. 
	
	The approach that follows, which assigns a topological character to the molecule, focuses on the geometric properties of the full even and odd components in momentum space. 
	It links the dynamical regimes identified above to symmetry-breaking topological transitions, and it assigns a quantised geometrical (Zak) phase to the resulting nonlinear topological state.
	It is, however, important to assess whether the $\pi$ phase inversion itself constitutes a \emph{topological signature}, as commonly accepted for domain walls and for two-dimensional optical vortices~\cite{Coullet89a,Coullet89b}.
	
	To make this correspondence explicit, we adopt the classical approach used in the literature on dissipative domain walls~\cite{Coullet89a,Coullet89b,Zhao21}, in which a one-dimensional phase discontinuity is analytically continued into a second, synthetic spatial dimension. 
	This construction provides a compact and visual method for classifying nonlinear solutions directly within their dissipative regime and is complementary in spirit to the Zak-phase classification introduced later.
	
	Starting from the one-dimensional odd component of the M\"obius molecule in Fig.~S4.5, we generate a two-dimensional field by analytic continuation in the synthetic dimension~$y$: 
	\begin{equation*}
		A(x,y) = |a(x)|\,\exp[i(\phi_a(x) + \arctan(y/x))], 
		\qquad 
		B(x,y) = |b(x)|\,\exp[i(\phi_b(x) + \arctan(y/x))],
	\end{equation*}
	where $\arctan(y/x)$ denotes the four-quadrant inverse tangent between $-\pi$ and $\pi$.  
	This operation converts the one-dimensional phase discontinuity into a continuous $2\pi$ winding in the 2D plane, thereby defining a topological vortex.  
	The analytic continuation omits the soliton periodicity and its coupling to the even field, retaining only the local phase defect to highlight the vortex-like topology. 
	The next step is to determine whether the extended equations support this configuration as a stable solution.
	\begin{figure}[H]
		\centering
		\begin{overpic}[width=0.43\linewidth]{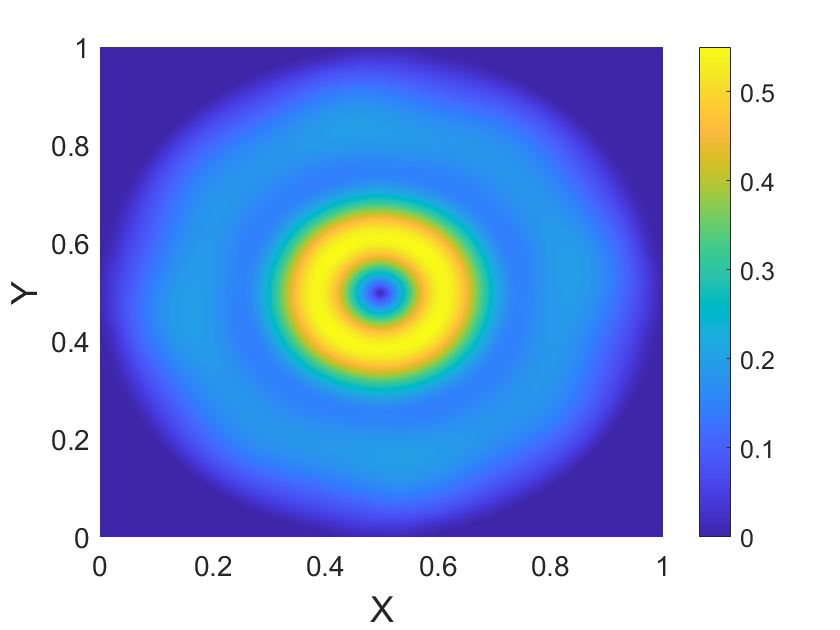}
			\put(2,72){\small\textbf{a}}
		\end{overpic}
		\hspace{0.04\linewidth}
		\begin{overpic}[width=0.43\linewidth]{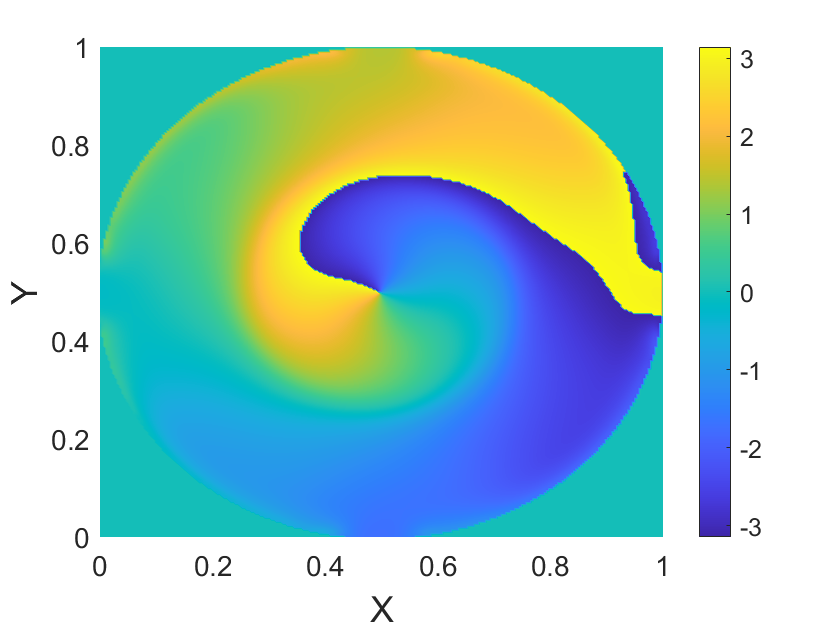}
			\put(2,72){\small\textbf{b}}
		\end{overpic}
		\caption{\small\textbf{Fig.~S4.5.} Two-dimensional analytic extension of the 1D odd-mode solution of Fig.~\ref{fig:S4.1}, obtained from the 2D generalisation of Eqs.~(S2.1)--(S2.2) after a slow-time propagation of $t=1240$. 
			\textbf{a} Amplitude of the field $a(x,y)$. 
			\textbf{b} Phase distribution showing a $2\pi$ winding around the core, confirming the vortex topology associated with the 1D phase jump.}
		\label{fig:Vortice}
	\end{figure}
	Accordingly, we propagate the fields $A(x,y)$ and $B(x,y)$ using the two-dimensional generalisation of Eqs.~(S2.1)--(S2.2), where the Laplacian ${\partial^{2}_{x}}$ is replaced by ${\partial^{2}_{x}+\partial^{2}_{y}}$ for a single longitudinal mode ($q=0$) and $\delta=0$. 
	The cross-linking term $b(x+\tfrac{1}{2},t)$ is evaluated as 
	$B\Big((\sqrt{x^2+y^2}+\tfrac{1}{2})\,e^{i(\arctan(y/x)+\pi)},\,t\Big)$
	ensuring the correct \mobius mapping across the synthetic plane. 
	Specifically, this transforms the equations into
	
	\begin{align*}
		\partial_t A 
		&= \frac{i\zeta_a}{2}\nabla_\perp^2 A 
		+ i|A|^2 A 
		- \kappa A 
		+ \sqrt{\kappa} B, \\[4pt]
		\partial_t B 
		&= \left( \frac{i\zeta_b}{2} + \sigma \right)\nabla_\perp^2 B 
		- \delta\,\partial_x B
		+ 2\pi i(\Delta - \tfrac{1}{4})\, B 
		+ \frac{2\pi i}{4}\, \mathcal{R}_\pi[B]
		+ g\, B 
		- B
		+ \sqrt{\kappa}\, A,
	\end{align*}
	where the \mobius half-turn operator $\mathcal{R}_\pi$ acts as
	\begin{equation*}
		\mathcal{R}_\pi[F](x,y) =
		\begin{cases}
			F(-x,-y), & x^2 + y^2 < \tfrac{1}{2}, \\[4pt]
			F(x,y), & x^2 + y^2 \ge \tfrac{1}{2}.
		\end{cases}
	\end{equation*}
	Figure~S4.5 shows the amplitude and phase of the propagated field after a slow-time evolution ($t=1240$). 
	The persistence of the phase singularity at the vortex core and the total phase winding of $2\pi$ confirm the topological charge of the defect. 
	This demonstrates that the $\pi$ phase jump of the 1D \mobius molecule can indeed be interpreted as the one-dimensional projection of a stable topological vortex. 
	Note that we do not aim here to reproduce the periodicity or the interaction of the odd mode with the even soliton. Our analysis is purely directed towards a classification of the local phase defect of the odd mode.
	
	This equivalence with the two-dimensional vortex construction demonstrates that the M\"obius molecule carries a topological signature that is intrinsically embedded in its local phase jump. 
	It provides an important conceptual bridge: although the molecule forms within a periodic cavity system, the phase defect itself corresponds to a nonlinear topological structure. 
	This behaviour is well established in the literature on optical vortices and, in one dimension, on domain walls and dark solitons. 
	In all these cases, topology originates from a quantised phase discontinuity and remains robust under continuous deformations of the system parameters. 
	As we discuss in the next sections, this robustness manifests over a broad parameter region in which the M\"obius molecule persists as long as the parity symmetry $\mathcal{P}(0)$ remains unbroken, confirming that the observed phase jump represents a genuine topological signature in its own right.
	\subsection*{S4.5 Full Parameter-Space View of the Bifurcation Structure — Technical}
	
	\begin{figure}[H]
		\centering
		\includegraphics[width=0.8\linewidth]{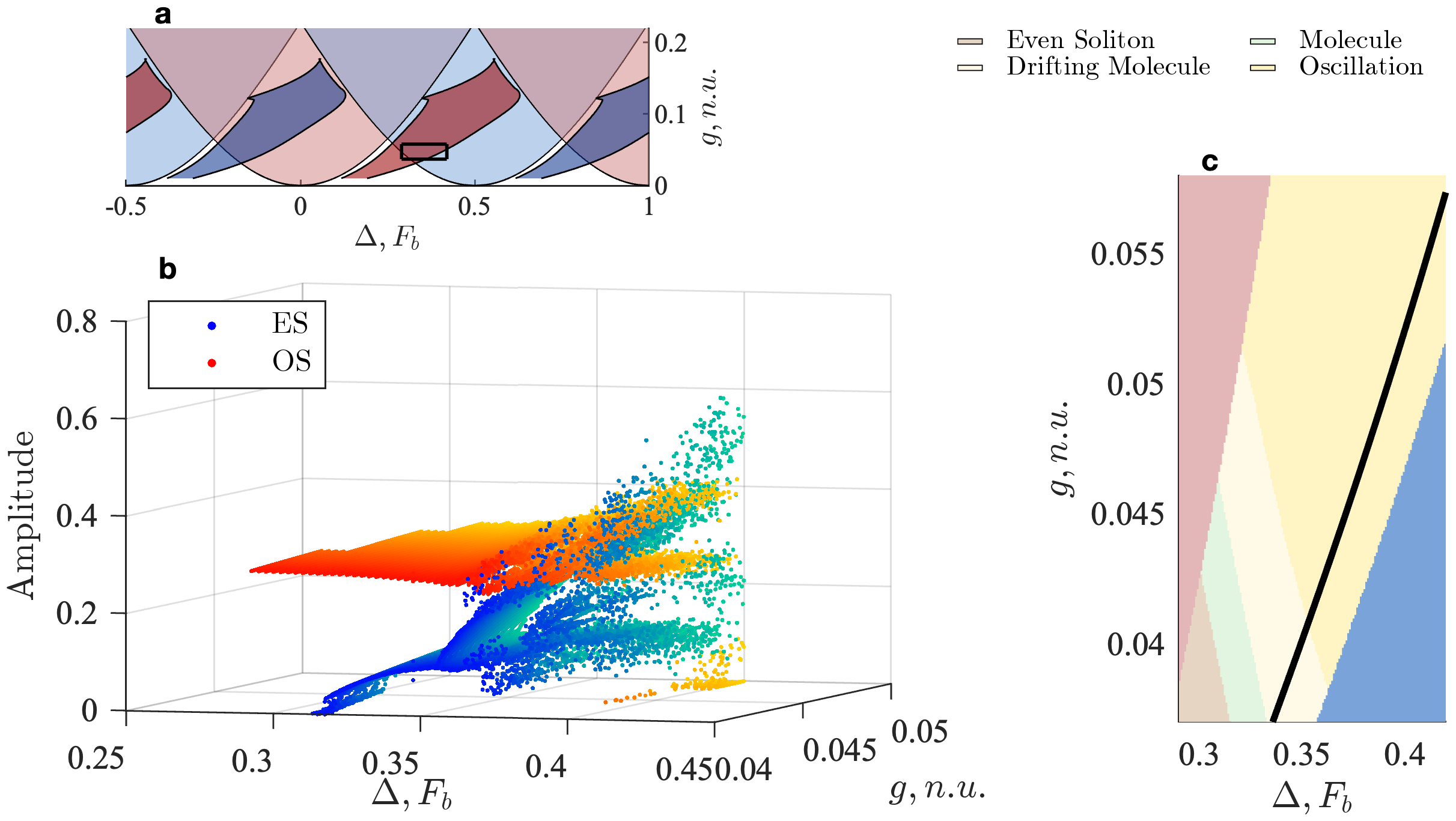}
		\caption{\small\textbf{Fig.~S4.6. Global bifurcation structure across the parameter plane.} 
			\textbf{a}, Instability map of the even and odd backgrounds, reproduced from Section~S3, indicating the regions where coupling between the two sectors can occur. 
			The black square marks the region numerically investigated in this work.  
			\textbf{b}, Phenomenological bifurcation diagram of the soliton amplitude versus detuning $\Delta$ and gain $g$, showing the even soliton (E$_s$, blue to green for increasing gain) and the emerging odd soliton (O$_s$, red to orange for increasing gain). 
			\textbf{c}, Stability map in the $(\Delta,g)$ plane summarising the dynamical regimes observed in simulations: 
			even-soliton lasing, stationary topological molecule, drifting molecule, and Hopf-induced oscillatory or chaotic states.  
			Both Ising--Bloch and Hopf bifurcations occur throughout the full stability domain of the even soliton, 
			confirming that the transitions described in Section~S4 are structural features of the system.  
			The trajectories analysed along the \emph{reference path} (black) traverse these regions sequentially, 
			forming the numerical basis for the detailed analyses presented in Sections~S5 and~S6.}
		\label{fig:Fig_grid}
	\end{figure}
	
	Having identified the three main dynamical regimes of interest, we now verify that they are broadly present across the parameter space. 
	To ensure representativeness, we extended the analysis to the entire $(\Delta,g)$ plane, following the guidelines of Section~S3, 
	where the regions of background instability triggering coupling between even and odd sectors were identified.  
	
	These regions are reproduced for convenience in Fig.~S4.6a.  
	Figure~S4.6b summarises the results from the propagation simulations.  
	For each simulated field, the even and odd components were separated, their soliton peaks located, 
	and the corresponding maxima and minima collected, constructing an effective phenomenological bifurcation diagram.  
	This representation provides immediate visual confirmation that molecule formation occurs throughout the $(\Delta,g)$ domain predicted in Fig.~S4.6a.  
	
	At higher energies, the drifting molecule undergoes an Andronov--Hopf bifurcation and proceeds through a route to chaos, eventually collapsing.  
	The Ising--Bloch bifurcation was identified through the emergence of a finite group velocity; 
	although not explicitly shown here, it will be discussed in detail in Sections~S5 and~S6, as exemplified by the cases in Sections~S4.2 and~S4.3.  
	Figure~S4.6c summarises the dynamical regimes observed in simulations: 
	even-soliton lasing, stationary topological molecule, drifting molecule, and the Andronov--Hopf regime leading to oscillations and chaos. 
	
	These results confirm that the stationary, drifting, and oscillating molecule states occur consistently 
	throughout the entire region identified in Section~S3 where the even solitons are stable against even 
	perturbations, up to gains of about $g \approx 0.055$. For larger gains, the even soliton tends to collapse and a 
	pure odd state remains. This numerical set establishes that the observed dynamical transitions are intrinsic 
	structural features of the system.
	
	To gain deeper insight into their mechanisms, we next focus on a specific trajectory in the $(\Delta,g)$ plane—%
	the \emph{reference path}.  
	As discussed in Section~S2, in the experiment, the EDFA pump acts as a single global control parameter 
	that simultaneously tunes both $g$ and $\Delta$.  
	In the soliton regime, increasing the EDFA pump increases the effective gain $g$ 
	while slightly shifting $\Delta$ toward larger red-detuned values.  
	The reference path follows this same correspondence, ensuring a consistent match with the experimental tuning sequence.  
	Trajectories intersecting the same bifurcation lines exhibit equivalent dynamical transitions, as shown in Sections~S5 and~S6. 
	All examples presented in Section~S4 were obtained along this reference path.
	
	\newpage

	
	\section*{S5. Topological Framework of the M\"obius Molecule and Mathematical Formalism }
	
	Having established in Section~S4 that the \mobius molecule arises from the
	destabilisation of the even-soliton branch by odd perturbations, we now seek to 
	rigorously characterise its topological nature.  
	Our aim is to connect our system with canonical concepts of periodic
	topological structures—originally developed in condensed-matter physics and now 
	widely adopted in photonics~\cite{Su1979,Zak1989,Berry1984,Resta1994,Ozawa2019,Smirnova2020}.  
	Among recent contributions, we highlight the work by Fan \emph{et al.}~\cite{Fan2022Topological} 
	on topological soliton metacrystals in microcombs, which provides a directly
	relevant methodology.
	
	It is useful to recall the seminal works that established the physical framework 
	for topological invariants such as Berry phases, and in particular the 
	one-dimensional formulation introduced by Zak~\cite{Zak1989}.  
	In condensed-matter and photonic systems, topology is defined in periodic 
	configurations where states acquire a quantised geometric phase determined by 
	the symmetries of the underlying lattice or field structure.  
	These geometric phases remain invariant under continuous deformations as long as 
	the protecting symmetries are preserved, which in the reciprocal lattice 
	(Bloch momentum space) manifests itself through the persistence of an open 
	energy bandgap.
	
	It is essential to clarify that our periodic system satisfies closed boundary 
	conditions; hence, we do not discuss boundary-localised edge states here~\cite{Hasan2010}.  
	In closed or periodic systems, topological invariants are intrinsic properties 
	of the bulk symmetry and remain unchanged under any continuous deformation, 
	unless the spectral gap closes~\cite{Resta1994,Ozawa2019,Smirnova2020,Fan2022Topological}.
	
	Our demonstration of a topological regime for the molecule proceeds in two 
	complementary steps, which will be discussed in Section~S6:
	
	\begin{itemize}
		
		\item We calculate a geometric topological invariant for the molecule (denoted here 
		\emph{Zak-NL}).  
		In his seminal work, Zak introduced a method to evaluate the Berry phase
		in one-dimensional periodic systems by extending the parameter space
		through an auxiliary Bloch momentum~\cite{Zak1989,Resta1994}.
		For systems possessing the relevant symmetries, this geometric phase
		takes quantised values ($0,\pi$) and thus serves as a topological
		invariant.  
		This construction is standard in condensed-matter physics and has since been
		generalised to nonlinear waves, including nonlinear optical systems~\cite{Li2023Zak,Smirnova2020Nonlinear}.  
		The resulting invariant provides a direct measure of the geometric phase
		of the molecule field—computed separately for its even and odd soliton
		components—and enables direct comparison with the regimes identified
		through symmetry-breaking analysis.
		
		\item We then identify the generating and protecting symmetries that
		govern the topological behaviour of the molecule.  
		As discussed in Section~S2, the formation of the topological molecule requires
		spontaneous symmetry breakings that populate the odd sector of the
		\mobius operator~$\mathcal S$.  
		This occurs through the breaking of one of the two point-parity symmetries 
		of the soliton train:  
		$\mathcal P(\tfrac{1}{4})$, centred on the soliton peaks, or 
		$\mathcal P(0)$, centred at the midpoint between them.  
		Once the molecule forms, exactly one of these parity symmetries remains
		unbroken; this surviving symmetry protects the topological character of the molecule.
		
	\end{itemize}
	
	To rigorously connect these transitions to bandgap closure and to a 
	topological phase inversion, we use the natural extension of this framework 
	to nonlinear systems through the linearisation of specific stationary states, 
	yielding the Bogoliubov--de~Gennes (BdG) problem~\cite{Resta1994,Ozawa2019,Smirnova2020}.  
	This approach characterises the topological nature of a state by analysing the 
	Berry phase of its \emph{perturbations} rather than that of the state itself, 
	and by detecting transitions between topological sectors.  
	The BdG eigenmodes inherit the periodicity of the base state and can therefore 
	be assigned Zak phases (here labelled as \emph{Zak-BdG}, to distinguish them from the 
	geometric phase of the nonlinear solution, \emph{Zak-NL}), enabling the 
	identification of transitions with a topological origin.  
	As the states are periodic, the linearised eigenvalue spectrum acquires a 
	corresponding band structure.  
	When the system undergoes a topological transition, the Zak phase of its 
	perturbations changes accordingly.  
	Such transitions are typically associated with symmetry breaking and with the 
	closure of a bandgap when the protecting symmetry is lost; in bifurcation 
	diagrams, they appear as \emph{gap-closure} points accompanied by a 
	\emph{Zak-phase inversion}.  
	This formulation is particularly suitable for nonlinear and dissipative systems.
	
	To identify these transitions, we construct a soliton molecule starting from the 
	stationary states of the XPM-only system [Eqs.~(\ref{S2.4})--(\ref{S2.7}) with 
	$\gamma=0$], compute its BdG spectrum and corresponding band structure, and 
	detect the relevant transitions through bandgap closure and Zak-phase inversion.  
	This analysis specifically identifies the lasing threshold and the 
	Ising--Bloch bifurcation.  
	
	It is important to select the appropriate optical field to build the soliton 
	molecule—particularly the correct continuation for the odd mode.  
	As introduced in Section~S3, we use the soliton solutions of the trivial 
	problem defined by their respective even and odd detunings.  
	We have already seen in Section~S4 that this choice reproduces the molecule 
	emerging dynamically from the destabilisation of the even-soliton train.  
	Here, we connect this construction to the instability spectrum of the 
	even-soliton branch.  
	The real parts of the eigenvalues identify the unstable modes with 
	$\mathrm{Re}(\lambda)>0$, which act as seeds for the odd perturbations giving 
	rise to the molecule.  
	These eigenmodes are then compared with those arising from the BdG analysis 
	of the full molecule, establishing the correspondence between the unstable 
	even-soliton mode and the emerging topological branch.
	
	Before introducing these results, Section~S5.1 summarises the mathematical 
	framework used to compute Zak phases and Berry connections.
	
	\subsection*{S5.1 Zak Phase Extraction and Berry Connection — Technical}
	\renewcommand{\theequation}{S5.\arabic{equation}}
	\setcounter{equation}{0}
	
	Since we deal with a one-dimensional case, we follow Zak's original construction~\cite{Zak1989}, adapted to the
	Bogoliubov--de Gennes (BdG) framework~\cite{Resta1994,Ozawa2019,Smirnova2020}. A closely related implementation for
	soliton microcombs was presented by Fan \emph{et al.}~\cite{Fan2022Topological}, whose
	approach we parallel here. In Zak's method, the problem is \emph{extended} by introducing a slowly
	varying Bloch momentum $l\in[-\tfrac12,\tfrac12]$. In the microcomb context,
	this corresponds to multiplying the field by the phase factor $\exp(2\pi i\,l\,x)$,
	representing an angular-momentum shift along the ring~\cite{Fan2022Topological}.
	Adiabatically sweeping $l$ across one Brillouin zone reconstructs the
	geometric phase of an eigenmode, which is quantised for topological bands.
	
	For our equations, the extended linearisation around the even state
	$a_{S,e}(x)$ (Eq.~S3.4) is obtained by replacing spatial derivatives with
	covariant derivatives $D_{\pm l}=\partial_x\pm2\pi i\,l$, while retaining the
	original boundary conditions, including the term $b_q(x+1/2)$ enforcing the
	half-cavity translation operator $\mathcal{S}$ and its even/odd structure.
	For each fixed $l$, we linearise around a stationary soliton and define the
	perturbation vector
	\[
	\mathbf{u}_l(x,t)=
	\begin{pmatrix}
		\delta a(x,t;l)\\[2pt]
		\delta a^{*}(x,t;-l)\\[2pt]
		\delta b(x,t;l)\\[2pt]
		\delta b^{*}(x,t;-l)
	\end{pmatrix},
	\qquad
	D_{\pm l}:=\partial_x \pm 2\pi i\,l.
	\]
	
	For simplicity, we display the $N=0$ case (single supermode). The linearised
	evolution then takes the BdG form
	\begin{equation}
		i\,\partial_t \mathbf{u}_l
		\;=\; \eta\,\mathcal{H}(l)\,\mathbf{u}_l
		\;-\; i\,\mathcal{Q}(l)\,\mathbf{u}_l,
		\qquad
		\eta=\mathrm{diag}(1,-1,1,-1),
	\end{equation}
	where $\eta$ is the Bogoliubov metric (pairing each perturbation with its
	conjugate), $\mathcal{H}$ collects the conservative blocks, and
	$\mathcal{Q}$ the dissipative/driving terms. In the numerics, we solved the
	$N=5$ system, consistent with the rest of the paper. The conservative block reads
	\[
	\mathcal{H}(l)=
	\begin{pmatrix}
		\mathcal{H}_a(l) & 0\\
		0 & \mathcal{H}_b(l)
	\end{pmatrix},
	\qquad
	\mathcal{H}_a(l)=
	\begin{pmatrix}
		h_a(l) & \Delta_a(x)\\
		\Delta_a^*(x) & h_a(-l)
	\end{pmatrix},
	\]
	with
	\[
	h_a(\pm l)=-\frac{\zeta_a}{2}\,D_{\pm l}^2 + 2|a_{S,e}(x)|^2,
	\qquad
	\Delta_a(x)=[a_{S,e}(x)]^2.
	\]
	
	For a single main–cavity mode ($q=0$) in the auxiliary sector, we have
	\[
	\mathcal{H}_b(l)=
	\begin{pmatrix}
		h_b^{(+)}(l) & 0\\
		0 & h_b^{(-)}(l)
	\end{pmatrix},
	\qquad
	\begin{aligned}
		h_b^{(+)}(l)&=-\frac{\zeta_b}{2}\,D_{+l}^2 - 2\pi\!\left(\Delta-\tfrac14\right) - \frac{2\pi}{4}\,\mathcal{S},\\
		h_b^{(-)}(l)&=-\frac{\zeta_b}{2}\,D_{-l}^2 - 2\pi\!\left(\Delta-\tfrac14\right) - \frac{2\pi}{4}\,\mathcal{S}^\dagger.
	\end{aligned}
	\]
	
	\medskip
	The dissipative block is
	\[
	\mathcal{Q}(l)=
	\begin{pmatrix}
		\kappa & 0 & -\sqrt{\kappa} & 0\\
		0 & \kappa & 0 & -\sqrt{\kappa}\\
		-\sqrt{\kappa} & 0 & q_b^{(+)}(l) & 0\\
		0 & -\sqrt{\kappa} & 0 & q_b^{(-)}(l)
	\end{pmatrix},
	\qquad
	q_b^{(\pm)}(l)= (1-g)-\sigma\,D_{\pm l}^6+\delta\,D_{\pm l}.
	\]
	The problem for Eqs.~\ref{S2.4}–\ref{S2.7} is constructed analogously by
	building a perturbation vector
	\[
	\mathbf{u}_l(x,t)=
	\begin{pmatrix}
		\delta a_e(x,t;l)\\[2pt]
		\delta a_e^{*}(x,t;-l)\\[2pt]
		\delta b_e(x,t;l)\\[2pt]
		\delta b_e^{*}(x,t;-l)\\[2pt]
		\delta a_o(x,t;l)\\[2pt]
		\delta a_o^{*}(x,t;-l)\\[2pt]
		\delta b_o(x,t;l)\\[2pt]
		\delta b_o^{*}(x,t;-l)
	\end{pmatrix}.
	\]
	
	\medskip
	\noindent\textbf{Berry connection and Wilson loop.}
	BdG eigenmodes $\mathbf{u}_l=U_{n,l}\,e^{\lambda_{n,l} t}$ inherit the system
	periodicity and admit a geometric (Zak-BdG) phase. We use the $\eta$–inner product
	\begin{equation}
		\langle u, v\rangle_{\eta}
		= \int_{-1/2}^{1/2}\! u(x)^\dagger\,\eta\,v(x)\,dx,
		\label{eq:eta_ip}
	\end{equation}
	defining the (non-Hermitian) Berry/Zak connection~\cite{Fan2022Topological} as
	\begin{equation}
		\mathcal{A}_n(l)=i\langle U_{n,l},\,\partial_l U_{n,l}\rangle_{\eta},
		\label{eq:zak_conn}
	\end{equation}
	so that the Zak-BdG phase over the Brillouin zone reads
	\begin{equation}
		\mathcal{Z}_n = i \int_{-\tfrac12}^{\tfrac12}
		\langle U_{n,l},\,\partial_l U_{n,l}\rangle_{\eta}\, dl.
		\label{eq:zak_int}
	\end{equation}
	
	Numerically, we evaluate $\mathcal{Z}_n$ via a discrete Wilson loop over a mesh
	$\{l_j\}_{j=1}^{K}$ spanning $[-\tfrac12,\tfrac12]$, corresponding to one
	Brillouin zone of the cavity periodicity. After solving the eigenproblem at
	each $l_j$ and selecting band $n$ by continuation, we normalise the modes as
	\begin{equation}
		\langle u_{n}(l_j),\, \eta\, u_{n}(l_j) \rangle = 1,
		\label{eq:Norm}
	\end{equation}
	and compute
	\begin{equation}
		e^{\,i\mathcal{Z}_n}
		\approx
		\prod_{j=1}^{K}
		\frac{\langle u_{n}(l_j),\, \eta\, u_{n}(l_{j+1}) \rangle}
		{|\langle u_{n}(l_j),\, \eta\, u_{n}(l_{j+1})\rangle|},
		\qquad l_{K+1}\equiv l_1,
		\label{eq:wloop_prod}
	\end{equation}
	or equivalently
	\begin{equation}
		\mathcal{Z}_n
		\approx
		\arg\!\Bigg(\prod_{j=1}^{K}
		\frac{\langle u_{n}(l_j),\, \eta\, u_{n}(l_{j+1})\rangle}
		{|\langle u_{n}(l_j),\, \eta\, u_{n}(l_{j+1})\rangle|}\Bigg),
		\qquad l_{K+1}\equiv l_1.
		\label{eq:wloop_sum}
	\end{equation}
	This yields the quantised Zak-BdG phases discussed in Section~S6.3.
	
	\medskip
	\noindent\textbf{Zak phase of the nonlinear soliton state.}
	The previous derivation links the BdG problem to the topological character of its perturbations. To study instead the geometric phase of the stationary soliton field itself, we follow the standard approach in nonlinear and photonic systems~\cite{Zak1989,Smirnova2020}, where Zak’s construction is extended directly to the nonlinear equations of motion (see also the experimental implementation by Li \emph{et al.}~\cite{Li2023Zak}). Specifically, we introduce a covariant Bloch momentum $l\in[-\tfrac12,\tfrac12]$ by replacing all spatial derivatives with $D_{\pm l}=\partial_x \pm 2\pi i\,l$ in the original equations and, for each fixed $l$, continue the stationary solution $u_l(x)=(a_l(x),\,b_{q,l}(x))$ of Eqs.~(S2.1)--(S2.2). In this way, one obtains a smooth loop across the Brillouin zone with the physical periodicity of the ring, in the same spirit as the rigorous analysis above.  
	
	Since this is a stationary solution (not a BdG mode), the geometric phase is computed with the standard $L^2$ inner product,
	\begin{equation}
		\langle \Phi, u\rangle = \int_{-1/2}^{1/2} \Phi(x)^\ast\, u(x)\,dx,
	\end{equation}
	and the Berry/Zak connection reads
	\begin{equation}
		\mathcal{A}(l)=i\langle u_l,\,\partial_l u_l\rangle,
	\end{equation}
	so that the Zak-NL phase of the soliton branch is
	\begin{equation}
		\mathcal{Z} = i\int_{-\tfrac12}^{\tfrac12}
		\langle u_l,\,\partial_l u_l\rangle\,dl \;\; \mod(2\pi).
		\label{Berry}
	\end{equation}
	
	Numerically, we normalise the modes as
	\begin{equation}
		\int_{-1/2}^{1/2} u_{l_j}^*(x)\, u_{l_j}(x)\,dx = 1,
	\end{equation}
	and evaluate $\mathcal{Z}$ via the discrete Wilson loop
	\begin{equation}
		\mathcal{Z}
		\approx
		\arg\!\Bigg(\prod_{j=1}^{K}\frac{\int_{-1/2}^{1/2} u_{l_j}^*(x)\,u_{l_{j+1}}(x)\,dx }{\big| \int_{-1/2}^{1/2} u_{l_j}^*(x)\,u_{l_{j+1}}(x)\,dx\big|}\Bigg),
		\qquad l_{K+1}\equiv l_1,
		\label{Zak}
	\end{equation}
	with the overlap
	\begin{equation}
		\int_{-1/2}^{1/2} u_{l_j}^*(x)\,u_{l_{j+1}}(x)\,dx
		=
		\int_{-1/2}^{1/2} a_{l_j}^*(x)\,a_{l_{j+1}}(x)\,dx
		+
		\sum_{q=-N}^{N} \int_{-1/2}^{1/2} b_{q,l_j}^*(x)\,b_{q,l_{j+1}}(x)\,dx.
	\end{equation}
	
	This procedure is widely used in nonlinear topological photonics~\cite{Smirnova2020} to extract the geometric phase of nonlinear states. When analysing symmetry-resolved components (e.g., $a_e, a_o$), we apply the projector $\Pi^{\mathcal{S}}_{\pm}$ to $u_l$ at each $l$ and evaluate the Wilson loop on the projected fields, thereby assigning geometric phases to the parity-resolved contributions of the soliton independently (yielding $\mathcal{Z}_e$ and $\mathcal{Z}_o$), which we discuss in Section~S6.2.
	
	\newpage
	\section*{S6. Topological Nature  of the M\"obius Molecule}
	The aim of this section is to reconstruct, step by step, how the \mobius molecule emerges from the destabilisation of the even–soliton state and how its topological properties evolve across the key bifurcations of the system.
	Our analysis proceeds as follows.
	In S6.1, we first analyse the eigenvalue spectrum of the even–soliton train, identifying which parity symmetry is broken at threshold and which odd mode grows to seed the molecule.
	In S6.2, we then study the dynamical formation and evolution of the molecule, mapping its bifurcation structure—stationary, drifting, and oscillatory regimes—and assigning the Zak-NL phase of the nonlinear state.
	In S6.3, finally, we analyse the BdG spectrum of the stationary molecule, obtaining the band structure of its perturbations and detecting the symmetry-breaking topological transitions through bandgap closures and Zak-BdG inversions.
	\subsection*{S6.1. Perturbations of the Even Soliton State}
	
	We begin our analysis with the eigenvalue spectrum of the system obtained by
	linearising Eqs.~(S2.1)--(S2.2) around the even-soliton state $a_{S,e}(x)$
	defined in Section S3 (Eq.~(S3.4)) and extracting the eigenvalues with the largest real part.
	Since symmetry plays a central role in this problem, we apply the projectors
	$\Pi^{\mathcal S}_{\pm}$ to separate the perturbations into even and odd
	components. When perturbing a purely even state, the perturbations remain purely
	even or purely odd; the subspaces ${\mathcal S}_{\pm}$ are therefore fully
	separated.
	
	This separation is evident in Fig.~S6.1a, which shows only the eigenvalues with
	the largest real part in the $(g,\Delta)$ plane, coloured red and blue for the
	even and odd sectors, respectively.
	Following the eigenvalues with the largest real part, the
	even perturbations reproduce the behaviour of Eqs.~(S1.1)--(S1.2) (red surface),
	where the Möbius symmetry is absent. As anticipated in Section~S3, a state
	belonging to a given sector maintains the same instability regions as the
	trivial problem under perturbations of the same symmetry. The stability region
	of the trivial problem is outlined with black lines in panel~(a) for reference.
	It is clear that the new odd bands acquire positive real eigenvalues,
	$\mathrm{Re}(\lambda)>0$, indicating that the corresponding odd states can grow.
	
	It is now important to examine the spatial form of the perturbations that
	determine the formation of new odd states. The number of unstable modes (denoted
	by $\#\mathrm{Re}(\lambda)$), all associated with odd perturbations, is
	summarised in Fig.~S6.1b. The grey line marks the modulational-instability
	threshold of the zero state defined by Eq.~(S3.2), which lies close to the lower
	boundary of the instability region. For clarity, we focus on a specific
	trajectory in the $(\Delta,g)$ plane, plotted in yellow. This is the
	\emph{reference path} introduced earlier. Because the eigenvalue spectrum
	displays broad, nearly parallel bands in the $(\Delta,g)$ plane, any trajectory
	crossing the same bifurcation lines exhibits equivalent dynamical transitions.
	
	\begin{figure}[H]
		\centering
		\includegraphics[width=.8\linewidth]{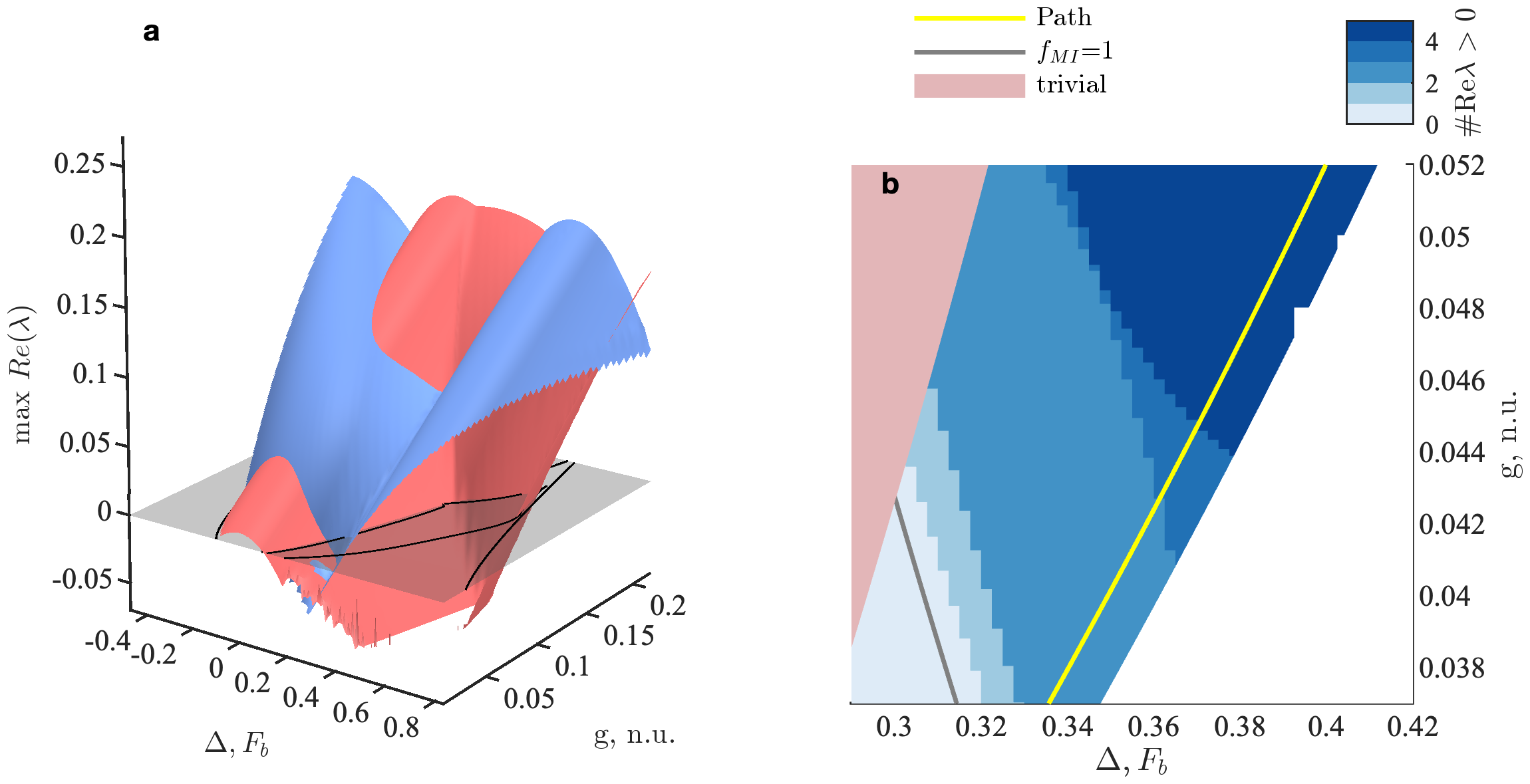}
		\caption{\textbf{Fig.~S6.1. Instability map of the even--soliton state.}  
			\textbf{a}, Maximum real part of the eigenvalues $\mathrm{Re}(\lambda)$ as a function of $(\Delta,g)$.  
			Even--sector perturbations form the red surface, odd--sector perturbations the blue surface.  
			The grey plane marks the $\mathrm{Re}(\lambda)=0$ stability threshold.  
			\textbf{b}, Two--dimensional instability map in the $(\Delta,g)$ plane.  
			The blue shading encodes the number of unstable odd modes ($\mathrm{Re}(\lambda)>0$), while the red region corresponds to the instability of the trivial even--soliton problem.  
			The grey line indicates the modulational--instability boundary from Eq.~(S3.2).  
			The yellow curve shows the reference path used throughout Section~S6.}
		\label{fig:BdG}
	\end{figure}
	
	The perturbations along the reference path as $g$ varies are summarised in
	Fig.~S6.2. Panels~(a) and~(b) show, respectively, the real and
	imaginary parts of the eigenvalues with the largest real part. We focus on the
	low-gain region of the spectrum, where only two families of eigenvalues cross
	the instability thresholds, highlighted in two shades of blue.
	
	To illustrate the characteristic features of the resulting perturbations, we fix
	$g = 0.0343$ (marked by solid circles in Fig.~S6.2a,b) and
	plot the corresponding modes in Fig.~S6.2(I,II), ordered by decreasing real part
	of their eigenvalues.
	
	The perturbed even-soliton state used as reference is shown in panels~(c)
	and~(d). For each family, panels~(e) and~(h) show the
	intensity profiles of the eigenvectors (blue), with the intensity of the
	even-soliton state represented as a red shaded background. Panels~(f)
	and~(i) give the corresponding phase profiles, and
	panels~(g) and~(j) show the Argand representations of the fields.
	
	All modes along this path belong to the odd sector of $\mathcal S$ and are
	characterised by a $\pi$ phase jump. The modes are generally complex conjugates
	but possess one dominant eigenmode that seeds the dynamics, shown in
	Fig.~S6.2~I,II.
	
	It is useful to recall that the periodic cell of the system contains two symmetry
	points for the parity operator~$\mathcal P$: one located at the soliton centre
	($x=\tfrac14$, $\mathcal P(\tfrac14)$) and one at the midpoint between the
	solitons ($x=0$, $\mathcal P(0)$). These two points define the \emph{on-site}
	and \emph{off-site} parity axes of the field, respectively. Following
	Section~S2, populating the odd sector of~$\mathcal S$ requires the breaking of
	one of these two parity symmetries.
	
	\begin{figure}[H]
		\centering
		\includegraphics[width=1\linewidth]{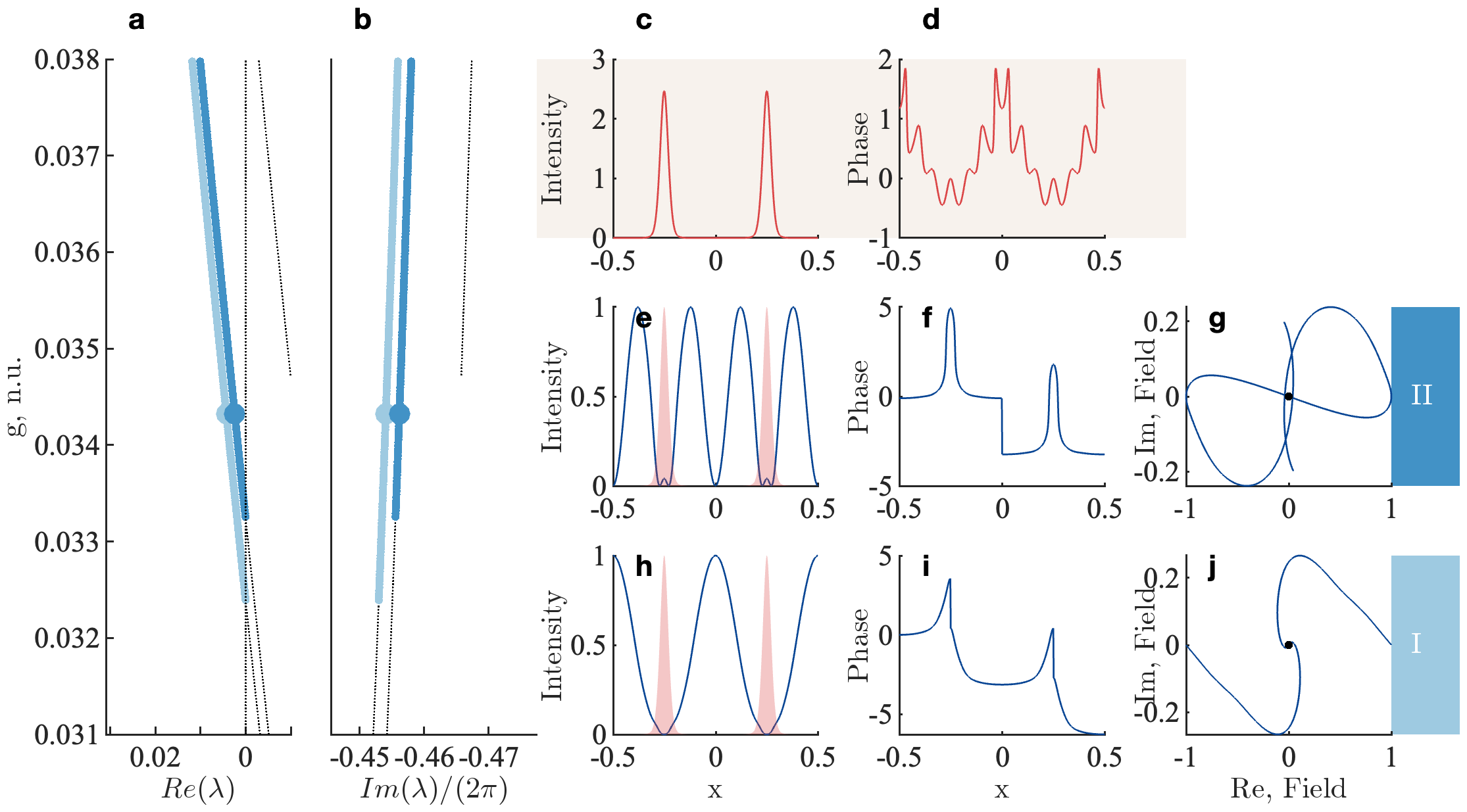}
		\caption{\textbf{Fig.~S6.2. Stability of the even-soliton eigenmodes along the $(\Delta,g)$ path.}
			Eigenvalues and eigenmodes for the even-soliton state are shown as a function of the gain~$g$ along
			the reference path (yellow line in Fig.~S6.1b). 
			\textbf{a}, Real parts of the eigenvalues, with unstable branches ($\mathrm{Re}(\lambda)>0$) highlighted
			in two shades of blue for the two families with the largest values. 
			\textbf{b}, Corresponding imaginary parts of the same eigenvalues. 
			\textbf{c,d}, Intensity (c) and phase (d) of the perturbed even-soliton state (shaded background). 
			\textbf{I--II}, Representative eigenvectors belonging to the two families of modes crossing the
			instability thresholds, ordered by decreasing $\mathrm{Re}(\lambda)$ at $g=0.0343$, indicated by the
			dots in panels~a,b. 
			\textbf{e,h}, Intensity profile of the eigenvector (blue), with the shaded red background showing the
			intensity of the even-soliton state being perturbed.  
			\textbf{f,i}, Phase profile of the eigenvector.  
			\textbf{g,j}, Argand diagram of the eigenvector in the complex plane.}
		\label{fig:FS5eig}
	\end{figure}
	
	The first odd perturbation (Panel~I) breaks the symmetry at the soliton peak,
	$\bcancel{\mathcal P(\tfrac14)}$, where the intensity of the even--soliton
	state is maximal. In this case, the field amplitude remains continuous, but
	the phase undergoes a $\pi$ inversion across the soliton centre.
	Importantly, this mode preserves the operator parity $\mathcal P(0)$, which
	is the symmetry protecting the topology~\cite{Su1979,Zak1989,Berry1984,
		Resta1994,Ozawa2019,Smirnova2020}.
	
	The second odd perturbation (Panel~II) instead breaks the symmetry at the
	midpoint between solitons, $\bcancel{\mathcal P(0)}$. The $\pi$ jump
	characteristic of the odd $\mathcal S$ sector therefore occurs in a region
	where the even--soliton intensity is close to zero. The mode still exhibits
	two localised phase jumps within the even--soliton regions.
	
	Examining these mode shapes reveals a strong correspondence between the mode
	in Panel~I and the odd mode discussed in Fig.~S4.1\,i--k. Recalling
	the results of Section~S4.1, the destabilisation of the even soliton
	($g=0.0343$, above the instability threshold) can lead to a translationally
	invariant molecule state. Such a molecule displays dominant features that
	also belong to the leading eigenvector, shown here in Panels~I (e--g).
	
	The odd component of the molecule exhibits the same zero crossing in the
	Argand representation. Its imaginary part ($\sim|0.45|$ in units of $2\pi$)
	corresponds to the frequency shift of the odd component $\varphi_o$ relative
	to the even--soliton frequency $\varphi_e$, as shown in Fig.~S4.1g.
	
	Most importantly, this provides the first indication that the molecule
	populates the odd sector of $\mathcal S$ by breaking $\bcancel{\mathcal
		P(\tfrac14)}$. We will see in the next section that the influence of the
	second mode (Panel~II) becomes apparent only at larger gain values
	($g \approx 0.036$): once the odd mode associated with Panel~I is generated,
	its presence suppresses the instability of the second mode. As a result, the
	molecule preserves the $\mathcal P(0)$ symmetry in this regime.
	
	Conversely, the results of Section~S4.2 show an odd component that deviates
	from this picture. In that regime the molecule breaks $\mathcal P(0)$, and
	the displacement of the odd soliton from its symmetric position reflects the
	growing influence of the mode in Panel~II.
	\newpage
	
	\subsection*{S6.2. Dynamical Evolution of the Molecule and Zak Phase for the Nonlinear State }
	\renewcommand{\theequation}{S6.\arabic{equation}}
	\setcounter{equation}{0}
	
	In the previous section S6.1, we clarified the link between the unstable
	eigenmodes of the even–soliton state and the formation of the molecule.
	We now use this connection to track how the molecule evolves dynamically as gain
	and detuning vary along the reference path in Fig.~S6.1b, and to identify when
	and how its topological character emerges.
	
	\begin{figure}[H]
		\centering
		\includegraphics[width=\linewidth]{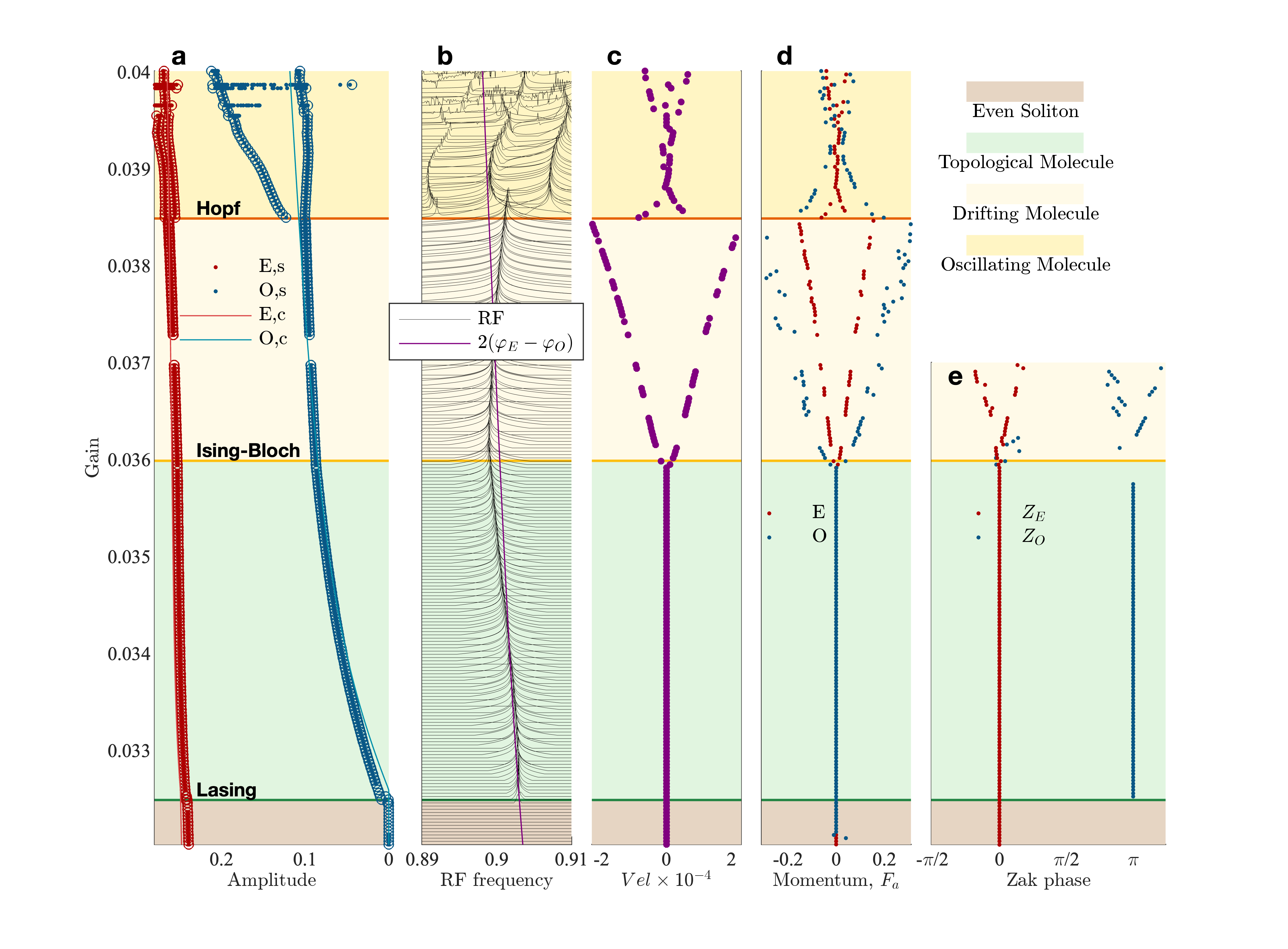}
		\caption{\small\textbf{Fig.~S6.3. Bifurcation structure and symmetry/topology indicators of the M\"obius molecule.} 
			\textbf{a}~Bifurcation diagram of the average energy for the high-energy even soliton (red markers, E$_s$) and the emerging low-energy odd soliton (blue markers, O$_s$). 
			Solid red (E$_c$) and blue (O$_c$) curves correspond to stationary-state energies calculated from Eqs.~(S2.4--S2.7) for $\gamma=0$. 
			Background shading marks the main dynamical regimes: even soliton, topological molecule, drifting molecule, and oscillating molecule. 
			Horizontal lines indicate the lasing threshold, the Ising--Bloch (IB) bifurcation, and the Hopf bifurcation. 
			\textbf{b}~Radio-frequency spectrum versus gain, showing the emergence of sidebands. 
			The purple line tracks the frequency difference $f_{CM}=2(\varphi_E-\varphi_O)$ between even and odd components. 
			\textbf{c}~Soliton velocity versus gain; the IB corresponds to the spontaneous splitting of the velocity from zero. 
			\textbf{d}~Average momentum of the even (red) and odd (blue) components versus gain. 
			\textbf{e}~Zak-NL phases of the even ($\mathcal{Z}_E$) and odd ($\mathcal{Z}_O$) branches versus gain. 
			Their separation remains close to $-\pi$ across the IB bifurcation, confirming the persistent topological distinction between the two symmetry sectors.}
		\label{fig:BifZakRand}
	\end{figure}
	
	For clarity, we focus on the reference path indicated in Fig.~S6.1b and perform
	a set of simulations under the same conditions as in Section~S4, while tracking
	the stability eigenvalues associated with the molecule to identify relevant
	transitions.
	
	Section~S4 already provided strong evidence for three main dynamical regimes: a
	topological stationary state above the odd-mode lasing threshold, a drifting
	state, and an oscillating state. These regimes are summarised in Fig.~S6.3, with
	the corresponding parameter regions highlighted in green, light yellow, and dark
	yellow, respectively. For completeness, we also include in brown the region
	corresponding to the purely even-soliton state preceding the onset of
	instability at the line labelled ``lasing'', as it marks the effective
	lasing generation of the odd soliton.
	
	Recalling the results of Section~S4, we track specific observables extracted
	from the simulations for each parameter point along this path.
	
	Figure~S6.3a reports the energies
	$\int_{-1/2}^{1/2}|a_{e,o}(x)|^2\,dx$ of the even and odd components for the
	stationary soliton states computed in the XPM-only model ($\gamma=0$), shown as
	continuous red and blue lines ($E_c$ and $O_c$), together with the maximum and
	minimum energies from the time-domain simulations ($E_s$ and $O_s$), shown as
	markers.
	
	As anticipated in Section~S4, the energies computed within the XPM-only
	approximation follow closely those obtained from the full model across the
	entire topological region. This agreement persists in the drifting regime and
	continues to match the average energy of the oscillating fields. This confirms a
	key result of Section~S4: even in the dynamical regime, the molecule is not
	destroyed but remains well preserved up to the onset of chaos.
	
	To track the carrier--envelope offset between the two components of the molecule,
	Fig.~S6.3b displays the radio-frequency spectrum (black) around
	$f_{CM}=2|\varphi_{E}-\varphi_{O}|$ (purple), calculated from the soliton
	frequencies $\varphi_{E,O}$ as in Figs.~S4.1f and S4.2f. Panel~(c) reports the
	group velocity, while to analyse the breaking of the parity symmetry
	$\mathcal{P}(0)$ we monitor the average angular momentum of the even and odd
	projections,
	\begin{equation}
		\hat{L}[a_{(e,o)}] = \int_{-1/2}^{1/2} \mathrm{Im}\!\left(a_{(e,o)}^*(x)\,\partial_x a_{(e,o)}(x)\right)\,dx.
	\end{equation}
	In general, a change in Berry phase is associated with a change in
	momentum~\cite{Su1979,Zak1989,Berry1984,Resta1994,Ozawa2019,Smirnova2020}, which
	makes $\hat{L}$ a particularly relevant observable for comparison with the Zak phase.
	
	The first key transition occurs at the odd-mode lasing threshold, just above
	$g = 0.0326$. This signifies the breaking of the odd-parity symmetry
	$\Pi^{(\mathcal S)}_{-}$, which aligns well with the threshold where the first
	even-soliton eigenmode loses stability as shown in Fig.~S6.2a.
	
	Beyond this point, the system converges to a topological molecule. The Zak-NL
	phases (see Section~S5.1) of the parity-resolved components reflect this behaviour:  
	the even branch sets to $0$, while the odd branch sets to $\pi$.  
	Notably, the Zak-NL phases remain perfectly quantised for each sector independently.  
	It is the odd mode that carries the non-trivial $\pi$ topological signature, directly
	linked to the perturbation that preserves $\mathcal{P}(0)$ while breaking $\mathcal{S}$.
	
	In the parameter interval highlighted in green in Fig.~S6.3a--e, and before the
	next transition at $g \approx 0.036$, the molecule closely follows the
	stationary states computed within the XPM-only model. To investigate its subsequent
	destabilisation, we highlight a few points to compare with the BdG analysis of
	Section~S6.3. We recall that the molecule displays features of the eigenmode
	that breaks the operator $\mathcal P(\tfrac14)$, while the symmetry
	$\mathcal P(0)$ remains preserved throughout this region up to the instability
	threshold at $g = 0.036$.
	
	The breaking of $\mathcal{P}(0)$ symmetry at the Ising--Bloch (IB) transition
	lifts the constraint that pins the soliton. Since the drift direction is not
	enforced by any system symmetry, it is selected by the specific initial
	conditions. This naturally produces the two branches observed in Fig.~S6.3c.
	
	After the IB, the centre of mass of the molecule shifts: the odd soliton moves
	closer to one of the even solitons (see Fig.~S4.2). This displacement is also
	reflected in the optical spectrum, where the average angular momenta of the even
	and odd components change in opposite directions to maintain overall momentum
	balance.
	
	The Zak-NL phase in Fig.~S6.3e remains quantised up to $g = 0.0357$.  
	Interestingly, just beyond this value we observe a temporary collapse of the  
	odd-sector Zak-NL phase from $\pi$ to $0$, after which the quantisation is  
	restored at the instability threshold $g = 0.036$. We return to this behaviour  
	in the next section.
	
	Our computation of the Zak-NL phase relies on continuation of the stationary  
	solution under variation of the extended momentum $l$ (Section~S5.1).  
	This procedure is only reliable near stationary regimes, so we do not track it  
	far beyond the IB transition. Nevertheless, we observe an adiabatic variation of  
	the Zak phase that mirrors the momentum evolution near the transition, consistent  
	with the known correspondence between Berry phase and  
	momentum~\cite{Resta1994}. Crucially, the difference  
	$\mathcal{Z}_E - \mathcal{Z}_O$ remains close to $\pi$ across the transition.
	
	This analysis clarifies the role of the symmetries in our system. The
	non-trivial topological state is primarily induced by the breaking of the
	half-cavity shift operator $\mathcal S$ via the breaking of $\mathcal
	P(\tfrac14)$, which seeds the odd perturbations of the even soliton and leads to
	the formation of the topological molecule at threshold through the removal of
	the $\Pi^{(\mathcal S)}_{-}=0$ condition. A clear signature of this topology is
	the discrete $\pi$ phase jump in the odd component, which our results of
	Section~S4.4 associate with a phase defect. Remarkably, it is this sector that
	carries the M\"obius geometry and is associated with a Zak-NL phase of $\pi$, typical
	of non-trivial topologies~\cite{Zak1989,Resta1994}.
	
	In contrast, the reflection symmetry $\mathcal P(0)$ is the symmetry lost at the
	IB transition. As a consequence, the odd soliton shifts from its symmetrical
	position between the even soliton peaks, its phase singularity disappears (see
	Section~S4), and the Zak-NL phase undergoes a slow evolution, losing its
	quantisation.
	
	\begin{figure}[H]
		\centering
		\includegraphics[width=1\linewidth]{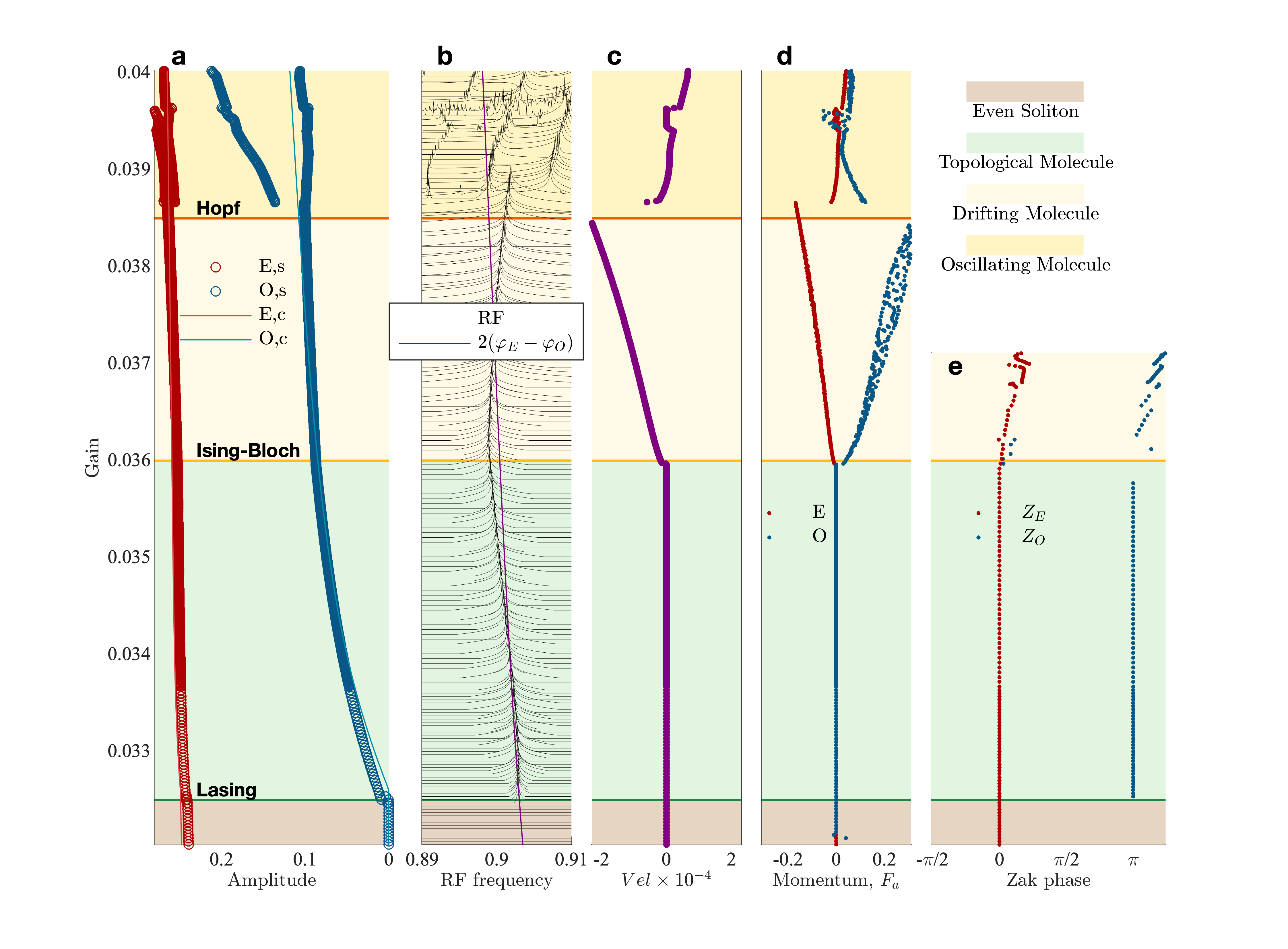}
		\caption{\small\textbf{Fig.~S6.4. Bifurcation structure and symmetry/topology indicators of the M\"obius molecule, obtained by continuation from below the lasing threshold.} 
			The plotting conventions and labels are the same as in Fig.~S6.3. 
			In this case, the diagram is obtained by starting from a solution at a gain value below the lasing threshold and using it as the initial condition for each consecutive gain step. 
			After the Ising--Bloch (IB) transition, the system follows a single drift branch, reproducing the experimental situation where only one drift direction is selected.}
		\label{fig:BifZakcont3}
	\end{figure}

	Nonetheless, although the IB transition breaks the reflection symmetry
	$\mathcal{P}(0)$ and therefore lifts the strict quantisation of the $\pi$ Zak
	phase, the M\"obius geometry continues to operate within the two sectors of the
	operator $\mathcal S$. Interestingly, the difference between the Zak phases of
	the even and odd soliton components remains close to $\pi$, consistent with the
	persistence of the M\"obius structure.  
	Indeed, applying $\mathcal P(0)$ after the IB maps a solution with velocity $v$
	to one with velocity $-v$, while the half-cavity translation operator
	$\mathcal S$ maps it back to a solution with the same velocity. This remapping
	mechanism is fully consistent with the scenario described in
	Ref.~\cite{Michaelis2001Universal}.
	
	In parallel, Fig.~S6.3b shows the evolution of the radio-frequency spectrum. The
	IB manifests as a knee-like feature in the beating frequency $f_{CM}$ and by the
	formation of prominent sidebands. Above $g \approx 0.0385$, a secondary
	instability develops, marked by the proliferation of harmonics in the
	radio-frequency spectrum, typical of quasi-periodic oscillations resulting from
	an Andronov--Hopf bifurcation for a periodic orbit (torus bifurcation). This
	gives rise to periodic oscillations of M\"obius molecules. Further increasing $g$
	eventually produces multi-branch solutions and initiates a route to chaos.
	
	A detailed analysis of the molecule states along this route to chaos lies beyond
	the scope of this work. Here we simply note that, even at energies beyond the
	chaotic region, the M\"obius molecule persists over a wide parameter range, with
	representative examples shown in Extended Fig.~E2.
	
	As a final note of this section, Figs.~S6.4 and~S6.5 report bifurcation diagrams
	analogous to Fig.~S6.3, obtained by starting from a solution at a gain value
	below the lasing threshold and using it as the initial condition for each
	consecutive gain step. Figures~3 and~E2 in the main text use the same dataset.
	In this way, the parameter space is explored in a manner closer to experimental
	practice, with the system following a single solution branch after the
	Ising--Bloch transition. The resulting diagrams not only reproduce the
	bifurcation structure of Fig.~S6.3 but also clearly distinguish the two
	Ising--Bloch families corresponding to opposite drift directions,
	reinforcing the identification of the IB transition as a parity--symmetry--breaking
	bifurcation.
	
	The results in Figs.~S6.4 and~S6.5 also highlight a key aspect of the
	molecule’s behaviour: \emph{its robustness under adiabatic parameter
		variations}. By slowly varying gain and detuning, the topological molecule
	consistently preserves its quantised phase up to $g=0.036$, where the protecting
	symmetry is lost. This adiabatic robustness is a fundamental criterion for the
	definition of a topological state and is reflected in the persistence of the
	quantised Zak phase as long as the protecting symmetry remains intact.
	
	In the next section, we formalise this behaviour by analysing the perturbations
	of the molecule itself, identifying precisely how symmetry breaking induces the
	loss of topological quantisation and the onset of drift.
	\begin{figure}[H]
		\centering
		\includegraphics[width=1\linewidth]{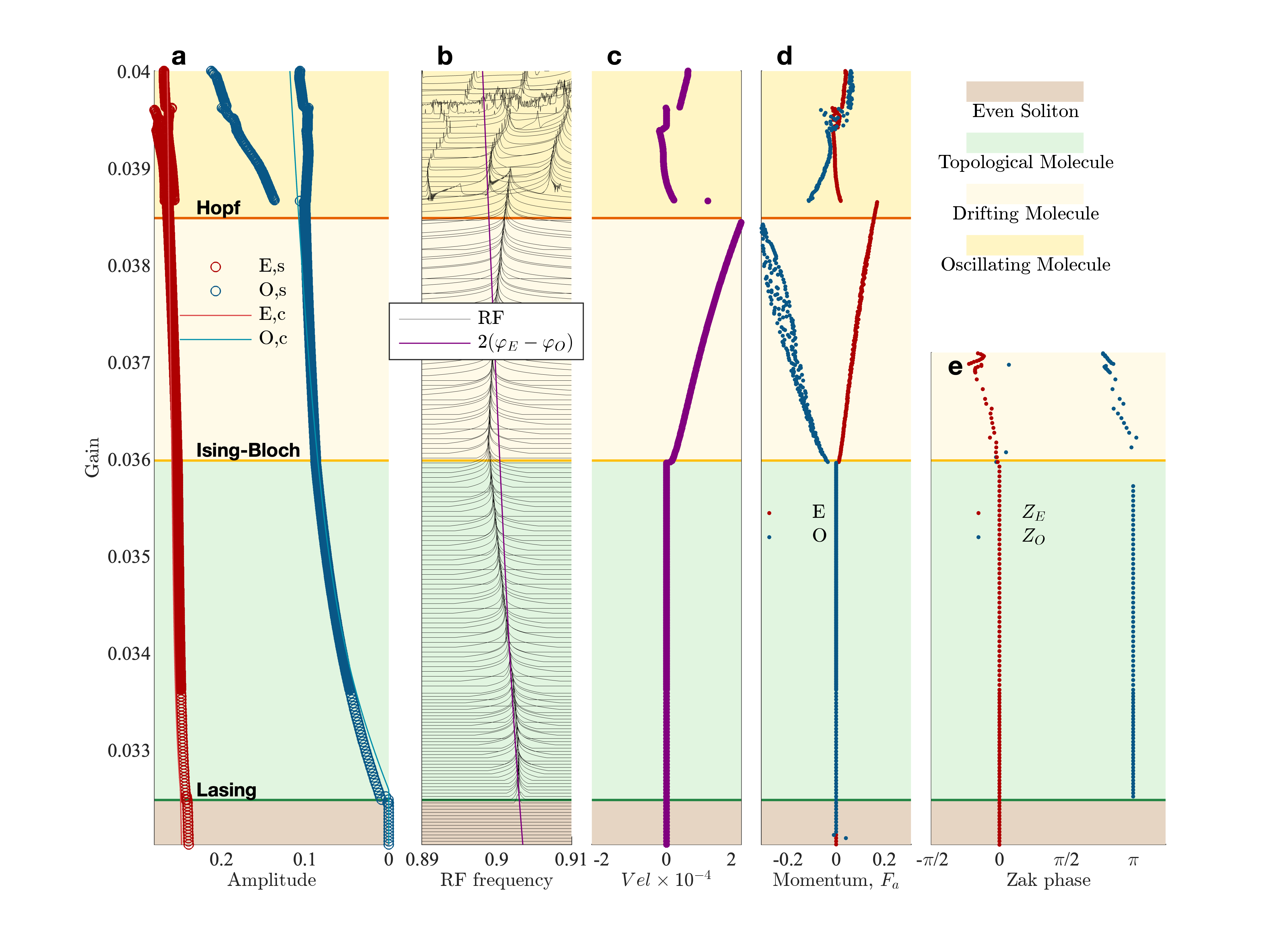}
		\caption{\small\textbf{Fig.~S6.5. Bifurcation structure corresponding to the opposite drift branch.} 
			The same conventions as in Fig.~S6.4 apply. 
			This symmetric branch confirms the coexistence of the two Ising--Bloch families, as already indicated in Fig.~S6.3.}
		\label{fig:BifZakcont2}
	\end{figure}
	
	\newpage
	\subsection*{S6.3 Linear Stability Analysis (BdG) of the Molecule and Identification of the Topological Transition}
	
	To investigate the topological transitions, we analyse the spectrum of the molecule’s linearised perturbations within a Bogoliubov--de Gennes (BdG) framework. 
	
	As introduced earlier, we compute the stationary molecule using the XPM-only model, which accurately reproduces Eqs.~(\ref{S2.1})--(\ref{S2.2}) in the formation/topological and drifting regimes.
	
	Practically, the molecule is obtained by analytically continuing the stationary even and odd solitons of the trivial model [Eqs.~(\ref{S1.1})--(\ref{S1.2})] at detunings $\Delta_e$ and $\Delta_o$, respectively, into Eqs.~(\ref{S2.4})--(\ref{S2.7}) with $\gamma=0$, where it remains a strictly stationary solution of the equations.
	
	\begin{figure}[H]
		\centering
		\includegraphics[width=\linewidth]{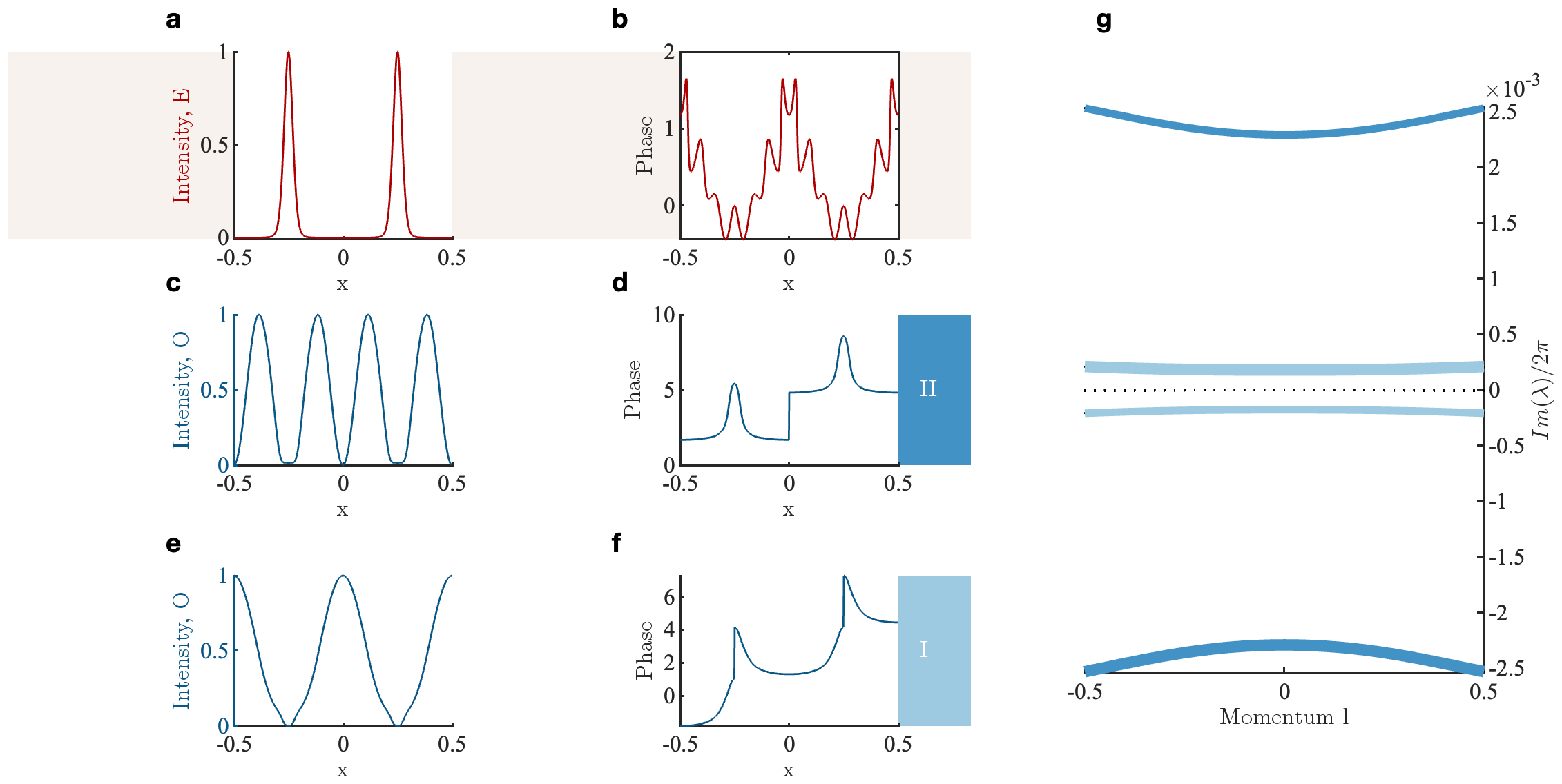}
		\caption{\textbf{Fig.~S6.6. M\"obius molecule, eigenmodes and bandgap structure.}
			\textbf{a,b}, Stationary molecule computed in the XPM-only model
			[Eqs.~(\ref{S2.4})--(\ref{S2.7}) with $\gamma=0$] at $g=0.0323$, below the
			even--soliton instability threshold; only the even component (red) is visible.
			\textbf{c--f}, Representative eigenmodes of the linearised molecule (BdG analysis)
			corresponding to the two families identified in Fig.~S6.2. Family~I is reported in (e,f), and family~II in (c,d).
			\textbf{g}, Imaginary parts of the corresponding eigenvalues across the Brillouin
			zone. Thick lines indicate the principal branch; thin lines show the complex
			conjugates. Both families exhibit an open bandgap.}
		\label{fig:FigEigMolBd}
	\end{figure}
	We begin by examining the molecule just below the instability threshold of the even--soliton train, as discussed in the previous section, with a representative example in Fig.~S6.6a,b for $g=0.0323$.  
	The two dominant eigenmodes of this molecule correspond directly to the same families identified in Fig.~S6.2 (Panels~I and~II) and are shown in Fig.~S6.6e,f and Fig.~S6.6c,d, respectively.
	
	A key element of the BdG analysis is the identification of a bandgap. As discussed in Section~S5.1, in a dissipative periodic system the linearisation yields a complex BdG spectrum $\lambda(l)$, whose real and imaginary parts depend on the Bloch momentum $l$, introduced through Bloch boundary conditions.  
	We plot  the imaginary part of the eigenvalues, $\operatorname{Im}\lambda(l)$, as a function of $l$ to illustrate the \emph{band structure}.

	Figure~S6.6g shows $\operatorname{Im}\lambda(l)$ 
	across the Brillouin zone for the two eigenmodes of Fig.~S6.6c,e (thick curves), together with their complex conjugates (thin curves).  
	A technical point is relevant here: as in any BdG linearisation, the reference frequency is the nonlinear frequency of the stationary state.  
	For the even--soliton train of Section~S6.1, this reference is $\varphi_e$, while in the XPM-only model the even and odd fields oscillate at different frequencies, $\varphi_e$ and $\varphi_o$.  
	Thus, the imaginary parts of the odd eigenmodes are measured relative to $\varphi_o$ and, although numerically smaller than in Section~S6.1, they represent the same physical quantity.  
	The momentum range $l \in [-\tfrac12,\tfrac12]$ spans the Brillouin zone associated with the cavity periodicity.
	
	Both branches remain \emph{gapped} across the full zone in this example.  
	These gaps play a central role: transitions are detected through the closure of the imaginary--part bandgap and the associated inversion of the Zak-BdG phase, as developed in Section~S5.1.  
	We recall that, since here we are investigating the Zak phase within the BdG framework, we refer to this quantity as the Zak-BdG phase to distinguish it from the Zak-NL phase calculated for the nonlinear states discussed in Section~S6.2.  
	For this reason, in the following analysis along the reference path we report not only the imaginary parts of the eigenvalues at $l=0$, but also at $l=\tfrac12$, which bound the band edges.  
	This allows us to track the complete behaviour of the gap, which in our case varies monotonically between the zone centre and the zone boundary.

	Figure~S6.7a summarises the behaviour of the stationary state, showing the energy branches of the even and odd components, as also reported in the previous section.  
	The results of the BdG spectrum of the molecule are shown in Figs.~S6.7b,c for the real and imaginary parts, respectively.  
	In line with Fig.~S6.6 and Section~S6.2, we emphasise the dominant eigenmodes crucial for forming the soliton molecule using light and dark blue. For completeness, the eigenvalues of the remaining eigenmodes are shown with dotted grey lines.  
	The imaginary parts are plotted for both $l=0$ and $l=\tfrac12$ to analyse the gap closure.  
	Panel~(b) shows that, throughout the region below $g=0.036$, the real parts of the eigenvalues across the full spectrum remain negative.  
	This allows us to establish the nature of the topological transition by directly analysing the spectral properties of the molecule, which in this range remains stationary.

	The Zak–BdG phases for these modes, shown in Fig.~S6.7d, confirm that the spectrum is quantised, yielding $0$ and $\pi$ for each complex–conjugate pair, as expected in a BdG problem.  
	Here we are interested in detecting \emph{transitions} in the Zak–BdG phases, which signal topological changes in the structure~\cite{Fan2022Topological}.
	\begin{figure}[H]
		\centering
		\includegraphics[width=0.98\linewidth]{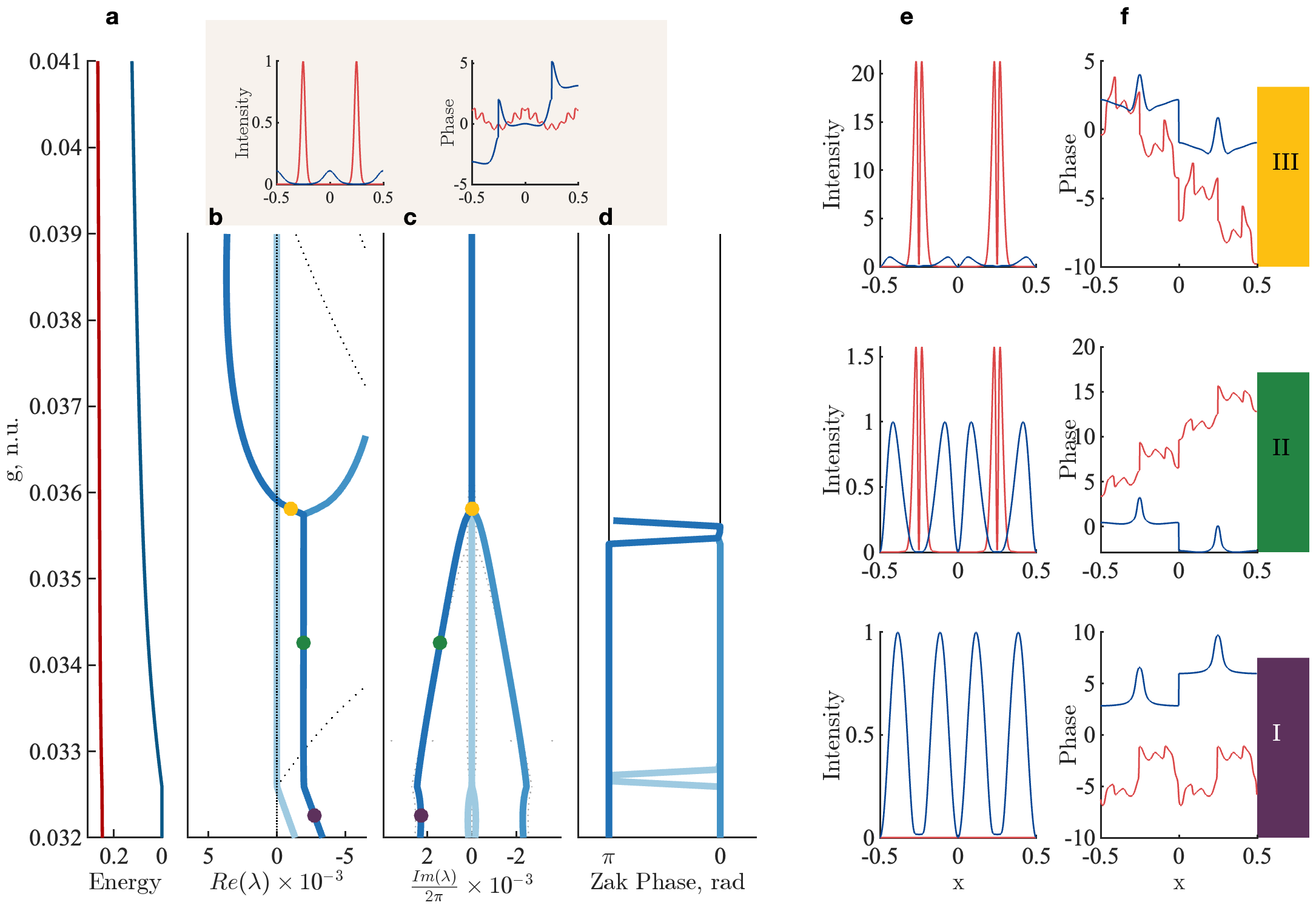}
		\caption{%
			\textbf{Fig.~S6.7. Summary of the molecule BdG spectrum.}
			Results are evaluated along the reference path of Fig.~S6.1b (only $g$ is shown).
			\textbf{a} Energies of the even (red) and odd (blue) components computed in the XPM-only model. 
			The inset shows the stationary molecule at $g=0.0362$.
			\textbf{b} Real and \textbf{c} imaginary parts of the eigenvalues.
			Dotted grey lines indicate the full spectrum. 
			Dark and light blue curves track the continuation of the two dominant eigenmodes.
			In panel~(c) both $\mathrm{Im}(\lambda)$ at $l=0$ (solid) and $l=\tfrac12$ (dotted) are shown.
			\textbf{d} Zak-BdG phases extracted from the corresponding eigenvectors.%
			The light--blue jump at $g\approx0.0326$ identifies breaking of $\mathcal{P}(\tfrac14)$;%
			the dark--blue jump at $g\approx0.0357$ marks breaking of $\mathcal{P}(0)$.
			\textbf{e--f} Intensity and phase of the second--family eigenmode (dark blue) at three representative gains:
			(I) $g=0.0321$, (II) $g=0.0343$, (III) $g=0.0358$.
			This mode breaks $\mathcal{P}(0)$ and becomes the translational Goldstone mode at the Ising--Bloch transition.
			The bandgap closure in the light--blue branch at $g\approx0.0326$ and in the dark--blue branch at $g\approx0.0357$ marks the two topological transitions.
		}
		\label{fig:S6.7}
	\end{figure}
	
	The first key transition occurs at the destabilisation threshold of the even--soliton train in the M\"obius model ($g \approx 0.0326$), where the bandgap of the light--blue branch closes permanently (Fig.~S6.7c), and the Zak–BdG phase of this eigenmode undergoes a $\pi$ jump (Fig.~S6.7d).  
	The corresponding eigenmode evolves smoothly into the neutral mode of the global–phase operator~$\mathcal{U}$.  
	This marks a genuine symmetry-breaking transition (light–blue branch), originating from the mode that in Section~S6.1 broke $\mathcal{P}(\tfrac14)$ (light–blue line).

	This transition strongly affects the second family (dark--blue branch), corresponding to the eigenmode that preserves $\mathcal{P}(\tfrac14)$ and breaks $\mathcal{P}(0)$.  
	It remains latent throughout the stability region ($0.0326 < g < 0.0357$), with the imaginary part of the eigenvalue exhibiting an open gap for all~$l$.  
	The dotted lines calculated for $l = \tfrac12$ (in panel~c) evidence that the gap remains open across the full Brillouin zone.  
	The real part, importantly, remains flat and negative in this region, indicating that the mode is stabilised.  
	The stabilisation of this eigenmode after molecule formation implies that the molecule retains the $\mathcal{P}(0)$ symmetry that protects its topological state.  
	At $g \simeq 0.0357$, this branch undergoes a dynamical transition anticipating the next symmetry breaking of the molecule: the complex--conjugate pair of eigenvalues coalesces into two purely real, negative eigenvalues.
	
	The effective Ising--Bloch transition, which further breaks symmetry and destabilises the molecule with a drift, occurs when the eigenvalue with the largest real part crosses zero at $g \simeq 0.036$.  
	Following the universal criterion of Michaelis \textit{et al.}~\cite{Michaelis2001Universal}, the nonequilibrium Ising--Bloch transition occurs when the antisymmetric eigenmode becomes degenerate with the translational Goldstone mode.  
	In our case this is directly visible from the shape of the corresponding mode --- the antisymmetric eigenvector breaking $\mathcal{P}(0)$ --- shown in Figs.~S6.7e,f, which converges at $g = 0.036$ to the \emph{derivative} of the molecule, the shape of the translational Goldstone mode.
	
	As a further signature of the topological nature of this transition, the Zak-BdG phase of this mode exhibits an abrupt $\pi$ jump at the bifurcation ($g \simeq 0.0357$), indicating the loss of the non-trivial topological character of the molecule.  
	Beyond $g = 0.036$, the system enters the dynamical regime introduced in Section~S6.2.  
	The loss of $\mathcal{P}(0)$ marks the onset of the molecule's drift; as anticipated in Section~S6.2, the gap closure at $g \simeq 0.0357$ is also visible in the Zak-phase evolution in Figs.~S6.3--S6.5.

	In summary, the BdG analysis reveals two distinct topological transitions that define the life cycle of the M\"obius-molecule topology.  
	The first occurs at the lasing threshold of the odd component, where breaking $\mathcal{P}(\tfrac14)$---or equivalently populating the odd sector of $\mathcal{S}$---creates the molecule.  
	This transition is marked by a quantised $\pi$ jump in the Zak phase of the corresponding eigenmode and its complex conjugate, together with a permanent closure of the associated bandgap, providing direct evidence that the molecule originates from a symmetry-breaking topological bifurcation.  
	Between this threshold and the Ising--Bloch point, the molecule remains a stationary, noise-suppressed state protected by the unbroken $\mathcal{P}(0)$ symmetry, which guarantees the quantisation of its geometric phase (as further shown in Section~S6.2).  
	The second transition, at the Ising--Bloch bifurcation, corresponds to the loss of this protection: the antisymmetric mode breaking $\mathcal{P}(0)$ coalesces with the translational Goldstone mode, and the Zak phase undergoes a $\pi$ inversion associated with this symmetry loss.  
	This sequence demonstrates unambiguously that the molecule forms through a symmetry-breaking topological transition and remains  protected until the onset of the drift instability.  
	Beyond this point, as shown in Section~S6.2, remnants of the M\"obius symmetry persist in the dynamical regime, maintaining the characteristic even--odd structure of the molecule even after the loss of parity protection.
	\newpage

	\section*{S7. Noise Response and Robustness of the M\"obius Molecule}
	\renewcommand{\theequation}{S7.\arabic{equation}}
	\setcounter{equation}{0}
	The analysis in Section~S6 identified the primary instabilities and bifurcations that lead to the formation of the M\"obius molecule and its dynamical regimes as the gain~$g$ varies, and clarified their topological character through the associated Zak phase. 
	In this section, we examine the robustness of these regimes against noise.
	
	Topology is often associated with robustness against disturbances and defects.  
	In integrated photonics, topological systems preserve their essential features under continuous deformations, showing enhanced resilience to disorder and device imperfections compared with conventional photonic architectures~\cite{khanikaev_alù_2024}.  
	In our context, the key question is how the topological structure of the M\"obius molecule enhances resilience against physical noise sources in a microcomb system, and in particular how it influences the stability of the \emph{repetition rate}, the primary observable to be stabilised.
	
	Previous sections identified two fundamental symmetries underlying the M\"obius molecule.  
	The half-period translation symmetry $\mathcal{S}$ induces the M\"obius identification and partitions the system into two orthogonal invariant subspaces (even and odd).  
	Molecule formation requires breaking this symmetry, which couples these subspaces and ensures nonzero population in both, thereby making the M\"obius geometry dynamically manifest.
	
	Once formed, the molecule is further stabilised by the parity operation $\mathcal P(0)$ enforcing reflection symmetry about the midpoint between the two even solitons.  
	Breaking this symmetry removes the quantised Zak phase and causes the molecule to drift with a velocity that depends on system parameters.  
	Despite this, the even and odd sectors remain cleanly separated, preserving the structural integrity of the molecule.  
	These features imply that perturbations influence the molecule differently across its dynamical regimes.  
	To analyse this behaviour, we performed numerical simulations incorporating the dominant noise sources present in microcomb laser systems.
	
	In our theoretical analysis, we assessed the robustness of the system against both \emph{external field noise} and \emph{slow structural drifts}.  
	In our context, optical field fluctuations are associated with spontaneous emission or Schawlow--Townes quantum noise.  
	To simulate this, we added a randomly distributed complex noise source $S_q(x,t)$ into the fibre--cavity equations~(S2.2).
	
	For adiabatic, slow drifts of the cavity parameters, we focus in particular on the gain~$g$, imposing that it varies according to a slowly time-varying stochastic process.
	
	We monitored the system response by computing the \emph{Allan deviation} of the soliton velocity, extracted from the time evolution of the even–soliton position $x_e(t)$,
	\[
	v(t) = \frac{dx_e(t)}{dt}.
	\]
	
	Given a discrete time series of averaged velocities $\{ \bar{v}_i \}$ obtained over sampling intervals of duration~$\tau$, the Allan deviation is defined as
	\begin{equation}
		\sigma_y(\tau)
		= \sqrt{
			\frac{1}{2(N-1)}
			\sum_{i=1}^{N-1}
			\left(
			\bar{v}_{i+1} - \bar{v}_{i}
			\right)^2
		},
		\label{eq:Allan}
	\end{equation}
	where $N$ is the number of samples and $\bar{v}_i$ denotes the mean velocity over each interval.  
	
	The Allan deviation is the standard metrological measure of frequency stability across different timescales: unlike a simple variance, it quantifies how the average value of a signal changes between consecutive intervals of duration~$\tau$.  
	By varying~$\tau$, one can distinguish fast optical fluctuations (short~$\tau$) from slow parameter drifts (long~$\tau$), yielding a time–resolved characterisation of the stability directly linked to the comb repetition rate.
	
	\subsection*{S7.1. Response to Optical Noise}
	
	\begin{figure}[ht]
		\centering
		\includegraphics[width=0.8\linewidth]{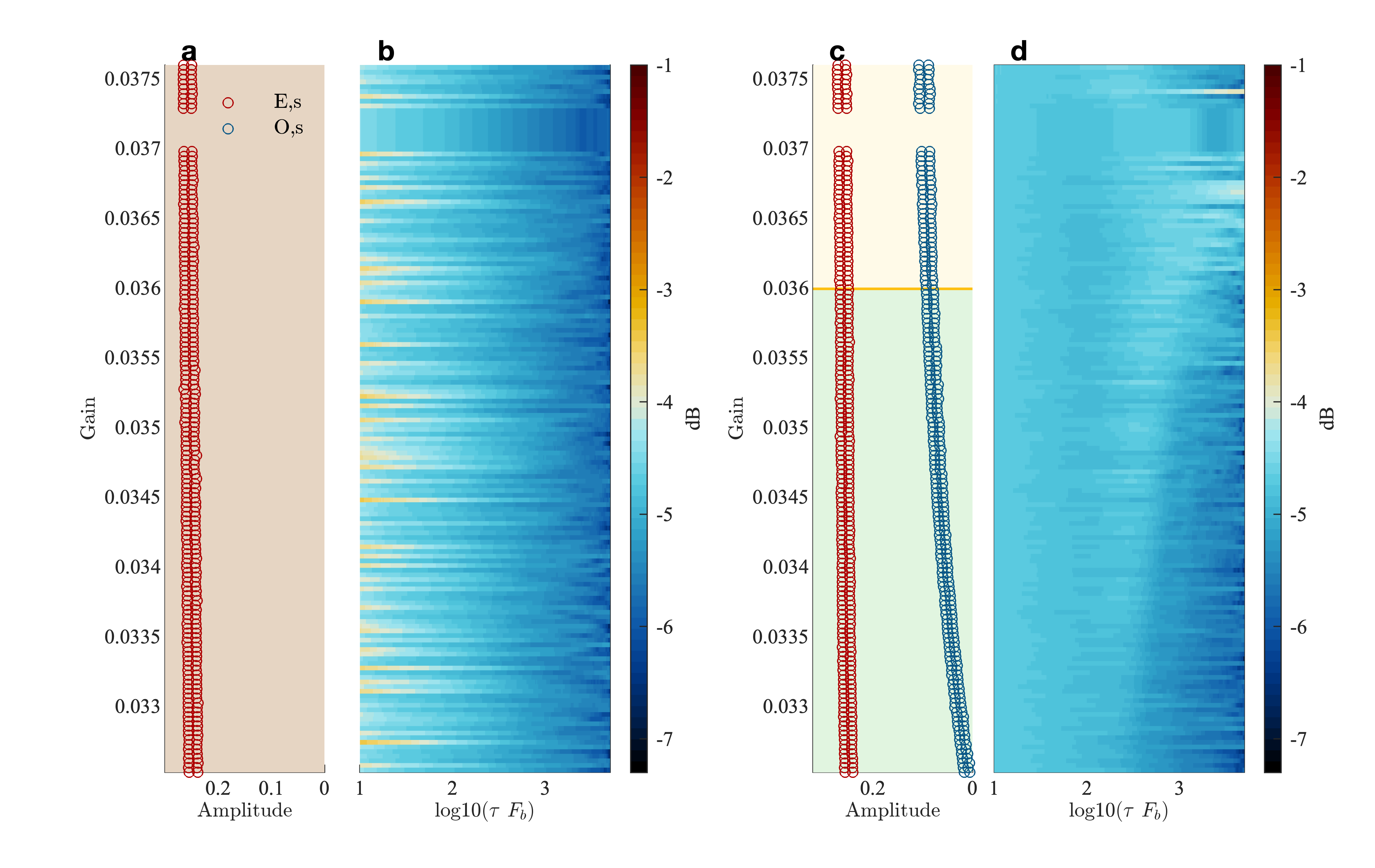}
		\caption{\small \textbf{Fig.~S7.1.} Theoretical analysis of the M\"obius molecule under optical noise. 
			Results are shown as a function of gain along the \emph{reference path} of Fig.~S6.1b. 
			The system is initialised from a pure high-energy soliton state, propagated to its stationary condition, and then subjected to optical noise. 
			Integration parameters: $h_t = 0.025$, $h_x = 0.002$, with $\epsilon = 0.6325$.  
			\textbf{a},\textbf{b}: Even-soliton case for the trivial equations (Eqs.~S1.1--S1.2).  
			\textbf{a}, Bifurcation map of the average power;  
			\textbf{b}, Allan deviation of the group velocity.  
			\textbf{c},\textbf{d}: Corresponding analysis for the M\"obius molecule, which remains stable and shows no collapse within the same noise regime.The background colour in c denotes the same dynamical regimes as in Figure S6.5.}
		\label{fig:S7.1}
	\end{figure}
	
	In our simulations, the noise term $S_q(x,t)$ was modelled as a zero-mean,
	$\delta$-correlated complex Gaussian process with correlation strength $\epsilon$
	(in normalised units), regenerated independently at each propagation step
	(i.e.\ white and temporally uncorrelated).  
	All stochastic simulations were performed using the Euler--Maruyama scheme
	(It\={o} interpretation), with temporal and spatial steps $h_t = 0.025$ and
	$h_x = 0.002$.  
	To implement the It\={o} calculus, the stochastic increment was scaled
	by $\sqrt{h_t}$~\cite{GarciaOjalvo1999}.
	
	This type of noise acts simultaneously in time and space, providing a meaningful
	assessment of robustness to localised perturbations.  
	As a first test, we propagated an even-soliton pair in the trivial system
	(Eqs.~S1.1--S1.2), adding the noise term $S_q(x,t)$ to Eq.~(S1.2).  
	We used $\epsilon = 0.6325$ and evaluated the system response under the same
	noise conditions in both the trivial and M\"obius configurations.
	
	To probe the dynamical stability of the soliton and molecular states, we tracked
	the peak field amplitude (Fig.~S7.1a,c), with the even and odd molecular
	components shown in red and blue, respectively.  
	We then computed the Allan deviation~(\ref{eq:Allan}) of the extracted group
	velocity (Fig.~S7.1b,d) as a function of the gate time~$\tau$ on a logarithmic
	scale, enabling a direct comparison between the trivial and M\"obius
	configurations.  
	The system was evolved up to a dimensionless time $t = 10^4$ ($\sim
	0.1\,\mathrm{ms}$ experimentally), covering the timescales over which optical
	amplitude and phase noise dominate the microcomb dynamics.
	
	At the chosen noise level, $\epsilon = 0.6325$, the even soliton in the trivial
	configuration (Eqs.~\eqref{S1.1}--\eqref{S1.2}) is on the brink of collapse.
	This is evidenced by sharp spikes in the Allan deviation (Fig.~S7.1b), revealing
	its weak resilience to optical perturbations.  
	In contrast, the M\"obius molecule remains robust (Figs.~S7.1c,d), exhibiting a
	substantially reduced Allan deviation across a wide range of~$\tau$.  
	Residual long-term modulations in the Allan deviation appear only above the
	Ising--Bloch transition, indicating the onset of slow drift in the field
	profile but without disrupting the molecular state.  
	These results demonstrate that the M\"obius geometry confers intrinsic
	robustness against optical noise.
	
	\begin{figure}[ht]
		\centering
		\includegraphics[width=1\linewidth]{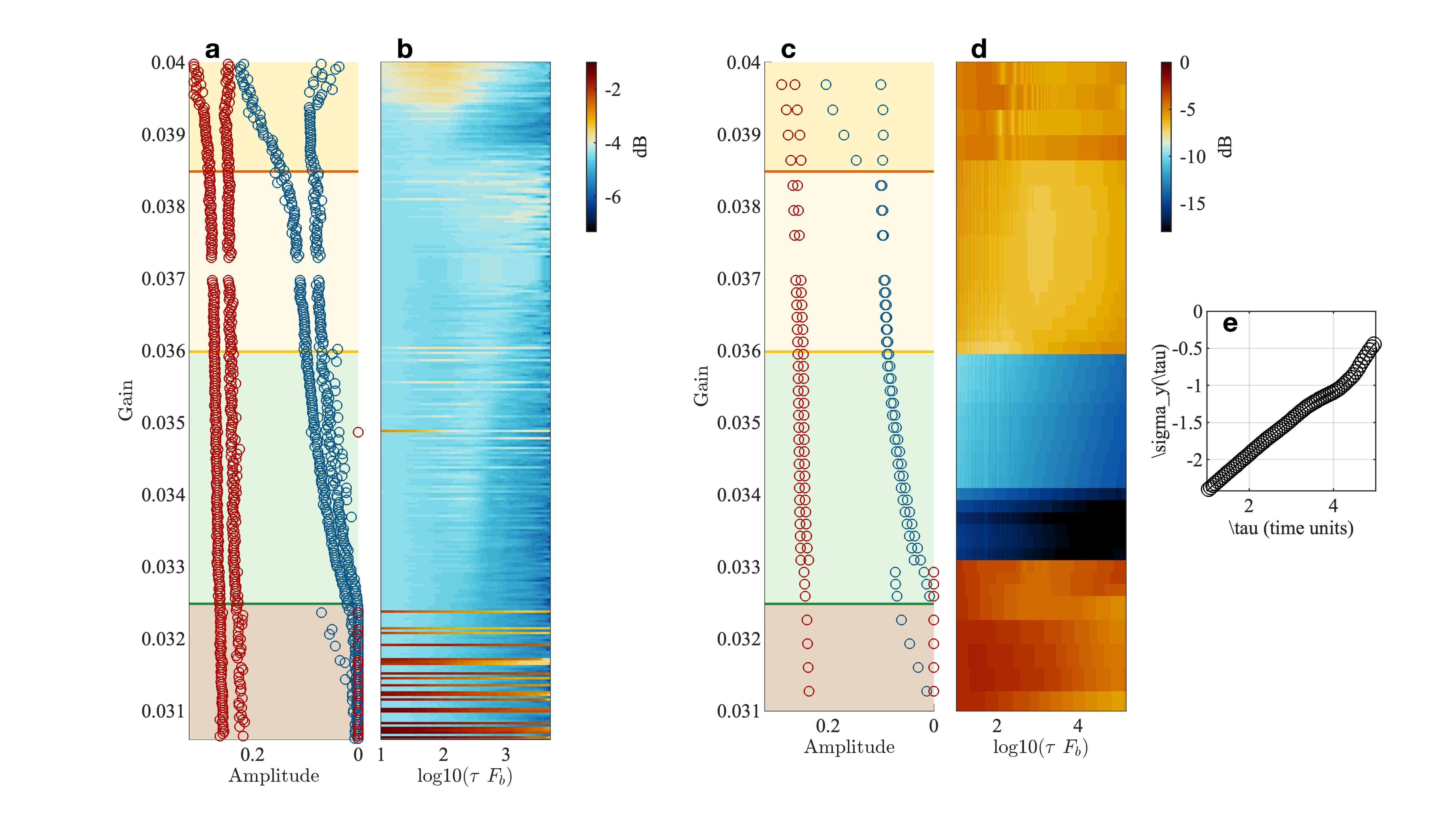}
		\caption{\small \textbf{Fig.~S7.2.} Response of the M\"obius molecule to optical and adiabatic noise. 
			Results are shown as a function of gain along the \emph{reference path} of Fig.~S6.1b. 
			The system is initialised from a pure high-energy soliton state, propagated to its stationary condition, and then subjected to noise. 
			Integration parameters: $h_t = 0.025$, $h_x = 0.002$.The background colour in a and  c denotes the same dynamical regimes as in Figure S6.5.
			\textbf{a},\textbf{b}: Response of the M\"obius molecule under optical noise with amplitude $\epsilon = 1.2649$.  
			\textbf{a}, Bifurcation map of the average power.  
			\textbf{b}, Allan deviation of the group velocity.  
			High noise levels drive the system toward collapse before the molecule-formation threshold.
			\textbf{c},\textbf{d}: Response of the M\"obius molecule to adiabatic parameter noise with pure random-walk components.  
			\textbf{c}, Bifurcation map under slow gain fluctuations.  
			\textbf{d}, Corresponding Allan deviation of the group velocity.
			\textbf{e}, Stochastic noise sequences used as seeds for the adiabatic-gain perturbations.  
			The trace shows the normalised temporal profile for a pure random walk, highlighting the slow long-term drift behaviour.}
		\label{fig:S7.2}
	\end{figure}
	
	To further clarify the solitons' response, we drive the molecule at high noise levels ($\epsilon = 1.2649$, Figs.~S7.2a,b).  
	At this noise intensity the molecule experiences stronger perturbations dominated by spatial fluctuations.  
	Under the same noise level, the soliton in the trivial equations collapses completely (we do not report this case, as the simulation output no longer contains solitons), whereas the M\"obius molecule remains dynamically sustained.  
	
	Below the molecule-formation threshold ($g \approx 0.032$), the even mode undergoes frequent collapses, evidenced by the strongly reduced average power (red low-amplitude points in Fig.~S7.2a) and by intermittent spikes in the Allan deviation (Fig.~S7.2b).  
	Once the M\"obius molecule is formed, however, the system becomes markedly more resilient: the Allan deviation remains low over a broad range of gain values, and collapse events become exceedingly rare.
	
	\subsection*{S7.2. Response to Adiabatic Parameter Noise}
	
	We now analyse how the system responds to \emph{adiabatic parameter noise}—that is, slow fluctuations of the gain~$g$, which appear to be an important source of noise in the experiments.  
	These drifts represent long-term  instabilities that gradually perturb the operating point of the laser system.
	
	A realistic model of adiabatic noise should include both a broadband (white-noise) component and a slow random-walk drift, reflecting the short-term fluctuations and long-term drifts typically present in optical cavities. However, in numerical simulations, incorporating broadband noise over long evolution times would require prohibitively small integration steps, severely restricting the realistically accessible temporal range. For this reason, in the present analysis we focus exclusively on the slow random-walk component.  
	
	The system’s sensitivity to fast, white optical noise has already been characterised in Section~S7.1, therefore here we concentrate on long-term parameter variations, where adiabaticity is the dominant effect. This approximation allows us to probe significantly longer temporal windows while retaining the physically relevant slow-drift dynamics.
	
	For the simulations with parametric noise, we used integration steps of $h_t = 0.05$ and $h_x = 0.002$, which ensure good convergence of the numerical method.  
	We pre-generated a stochastic sequence and applied it to the gain~$g$ with amplitude $\epsilon_g = 5\times10^{-4}$, with its Allan deviation reported in Fig.~S7.2e.  
	To preserve a consistent relative scaling among the parameter fluctuations, we employed normalised amplitudes chosen to lie within realistic experimental ranges.
	
	Figure~S7.2c,d summarises the results of these simulations, showing the field amplitudes in panel~c and the corresponding velocity Allan deviations in panel~d.
	
	The evolution of the Allan deviation cleanly discriminates between the even-soliton regime, the molecule-formation threshold, the Ising--Bloch region, and the oscillatory state.
	
	Within the drifting region, the Allan deviation rises smoothly toward long integration times, indicating that the repetition rate adiabatically follows the slow parameter drift—consistent with the experimentally observed behaviour, see Fig.~4.  
	This trend directly reflects the explicit dependence of the molecule’s group velocity on the system parameters.
	
	Interestingly, the even-soliton state below the molecule-formation threshold is significantly more sensitive to slow drifts.  
	This is particularly evident in Fig.~S7.2c, where the molecule frequently collapses under gain fluctuations, as indicated by the energy of the even solitons (red dots) dropping to zero.
	This analysis shows that long-term parameter noise affects the different dynamical regimes in distinct ways, but, most importantly, that gain drifts can induce an early collapse before molecule formation, whereas the fully developed M\"obius molecule remains robust even under sustained drifting conditions.

	\newpage

	
	\section*{S8. M\"obius Molecule Bifurcation Analysis with Group-Velocity Mismatch — Technical}
	
	This section provides a concise commentary on the effect of a soft symmetry breaking of the parity symmetry.  
	Terms that break the $\mathcal{P}$ symmetry — thus removing it completely from the governing equations — include the group-velocity mismatch~$\delta$ between cavities or asymmetric dispersion.
	
	Here we analyse the case of group-velocity mismatch, $\delta$, which is the most physically relevant representative of this class.  
	As shown in the previous sections, the M\"obius molecule is notably robust to both optical and parametric perturbations. Here we specifically analyse the effect of a non-zero group-velocity mismatch~$\delta$.
	Figure~S8.1 summarises the behaviour of molecule formation for $\delta = \pm0.0001$ and $\delta = \pm0.001$.  
	The analysis follows the same approach as in Section~S6.2, where the system is initialised from the even soliton and the odd soliton emerges from noise, allowing direct observation of the bias introduced in the group velocity.  
	A key observation is that the presence of a finite~$\delta$ biases but does not suppress the transition features.
	
	Even when the transition becomes strongly biased due to a large~$\delta$, the molecule’s velocity still undergoes a distinct shift at the bifurcation.  
	Regarding the dependence on the $(g,\,\Delta)$ parameters, we verified that the key dynamical features — molecule formation, drift–symmetry breaking, and subsequent Andronov--Hopf-type bifurcation — are robust and consistently observed throughout the existence region.
	
	The cases at higher group-velocity mismatches ($\delta = \pm0.001$) clearly show the parametric dependence of the velocity on the gain.  
	This confirms that the M\"obius molecule preserves its bifurcation diagram even under symmetry-breaking conditions.
	
	\begin{figure}[H]
		\centering
		\includegraphics[width=0.9\linewidth]{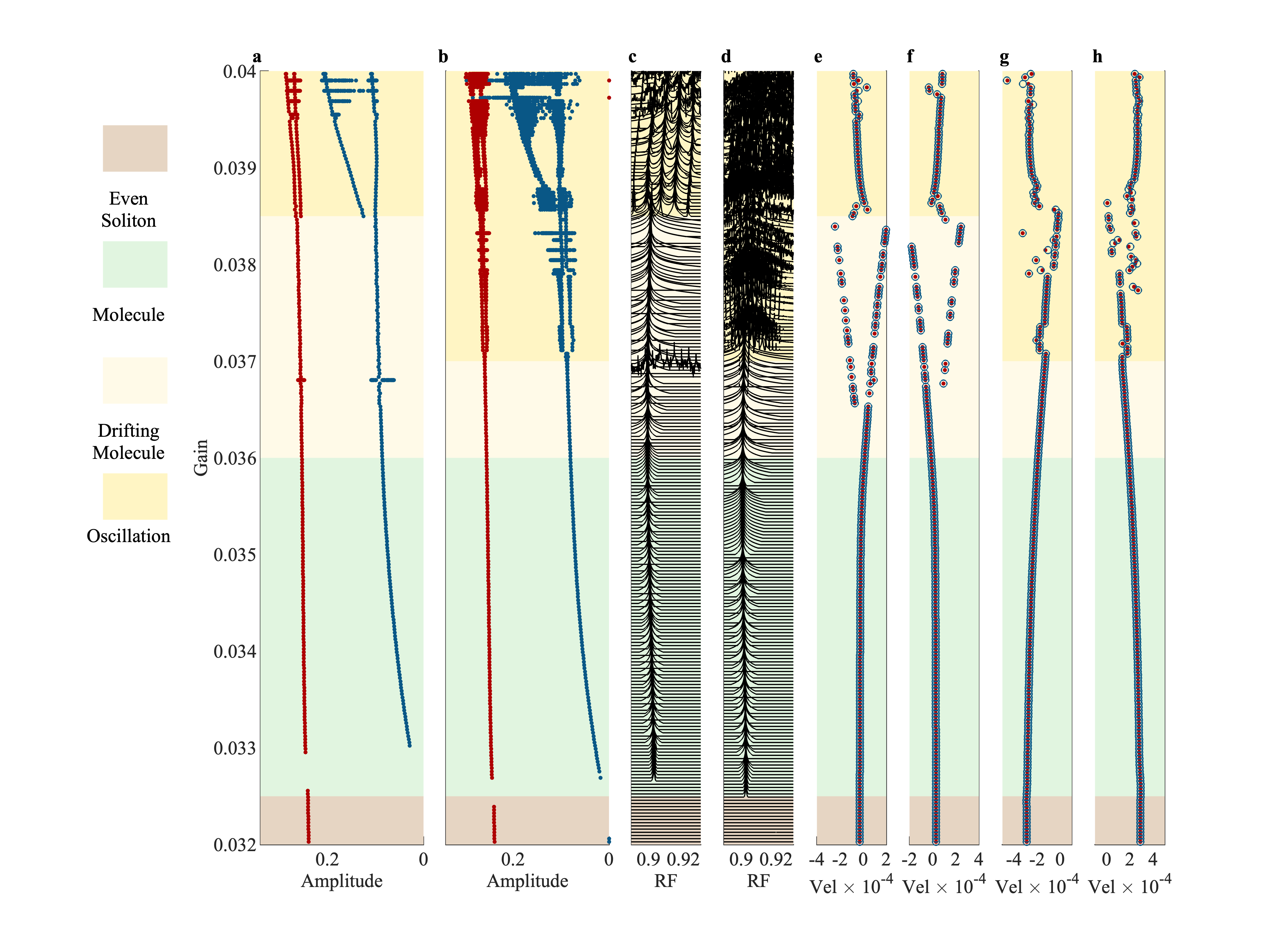}
		\caption{\small\textbf{Fig.~S8.1.} Theoretical analysis of the M\"obius molecule with varying group-velocity mismatch~$\delta$.  
			Results are shown as a function of gain~$g$ along the reference path in detuning, for different values of~$\delta$.
			Background colours indicate the dynamical regimes as defined in the legend.  
			The system is initialised from a high-energy even-soliton state; low-energy states emerge from noise.  
			\textbf{a}, Bifurcation diagram of average power for even (red) and odd (blue) solitons at $\delta = 0.0001$; the case $\delta = -0.0001$ is qualitatively equivalent.
			\textbf{b}, Same as panel~a, but for $\delta = 0.001$.  
			\textbf{c},\textbf{d}, Radio-frequency spectra around $f \sim F_b$ corresponding to panels~a and~b.  
			\textbf{e},\textbf{f}, Group-velocity maps for $\delta = 0.0001$ and $\delta = -0.0001$, respectively.  
			\textbf{g},\textbf{h}, Group-velocity maps for $\delta = 0.001$ and $\delta = -0.001$, respectively.}
		\label{fig:S8.1}
	\end{figure}
	\bibliographystyle{naturemag}

\begin{thebibliography}{10}
\expandafter\ifx\csname url\endcsname\relax
  \def\url#1{\texttt{#1}}\fi
\expandafter\ifx\csname urlprefix\endcsname\relax\def\urlprefix{URL }\fi
\providecommand{\bibinfo}[2]{#2}
\providecommand{\eprint}[2][]{\url{#2}}

\bibitem{Kivshar2003Optical}
\bibinfo{author}{Kivshar, Y.~S.} \& \bibinfo{author}{Agrawal, G.~P.}
\newblock \emph{\bibinfo{title}{Optical solitons: from fibers to photonic
  crystals}} (\bibinfo{publisher}{Academic Press}, \bibinfo{address}{Amsterdam;
  Boston}, \bibinfo{year}{2003}).

\bibitem{Lugiato2015Nonlinear}
\bibinfo{author}{Lugiato, L.~A.}, \bibinfo{author}{Prati, F.} \&
  \bibinfo{author}{Brambilla, M.}
\newblock \emph{\bibinfo{title}{Nonlinear optical systems}}
  (\bibinfo{publisher}{Cambridge University Press}, \bibinfo{year}{2015}).

\bibitem{Grelu2016Nonlinear}
\bibinfo{author}{Grelu, P.}
\newblock \emph{\bibinfo{title}{Nonlinear {Optical} {Cavity} {Dynamics}}}
  (\bibinfo{publisher}{John Wiley \& Sons, Incorporated},
  \bibinfo{year}{2016}).

\bibitem{Kippenberg2018Dissipative}
\bibinfo{author}{Kippenberg, T.~J.}, \bibinfo{author}{Gaeta, A.~L.},
  \bibinfo{author}{Lipson, M.} \& \bibinfo{author}{Gorodetsky, M.~L.}
\newblock \bibinfo{title}{Dissipative {Kerr} solitons in optical
  microresonators}.
\newblock \emph{\bibinfo{journal}{Science}} \textbf{\bibinfo{volume}{361}},
  \bibinfo{pages}{eaan8083} (\bibinfo{year}{2018}).

\bibitem{Pasquazi2018Micro}
\bibinfo{author}{Pasquazi, A.} \emph{et~al.}
\newblock \bibinfo{title}{Micro-combs: A novel generation of optical sources}.
\newblock \emph{\bibinfo{journal}{Physics Reports}}
  \textbf{\bibinfo{volume}{729}}, \bibinfo{pages}{1--81}
  (\bibinfo{year}{2018}).

\bibitem{Herr2014Temporal}
\bibinfo{author}{Herr, T.} \emph{et~al.}
\newblock \bibinfo{title}{Temporal solitons in optical microresonators}.
\newblock \emph{\bibinfo{journal}{Nature Photonics}}
  \textbf{\bibinfo{volume}{8}}, \bibinfo{pages}{145--152}
  (\bibinfo{year}{2014}).

\bibitem{Brasch2016Photonic}
\bibinfo{author}{Brasch, V.} \emph{et~al.}
\newblock \bibinfo{title}{Photonic chip--based optical frequency comb using
  soliton {Cherenkov} radiation}.
\newblock \emph{\bibinfo{journal}{Science}} \textbf{\bibinfo{volume}{351}},
  \bibinfo{pages}{357--360} (\bibinfo{year}{2016}).

\bibitem{Huang2016broadband}
\bibinfo{author}{Huang, S.-W.} \emph{et~al.}
\newblock \bibinfo{title}{A broadband chip-scale optical frequency synthesizer
  at 2.7 x 10-16 relative uncertainty}.
\newblock \emph{\bibinfo{journal}{Science Advances}}
  \textbf{\bibinfo{volume}{2}}, \bibinfo{pages}{e1501489}
  (\bibinfo{year}{2016}).

\bibitem{Obrzud2017Temporal}
\bibinfo{author}{Obrzud, E.}, \bibinfo{author}{Lecomte, S.} \&
  \bibinfo{author}{Herr, T.}
\newblock \bibinfo{title}{Temporal solitons in microresonators driven by
  optical pulses}.
\newblock \emph{\bibinfo{journal}{Nature Photonics}}
  \textbf{\bibinfo{volume}{11}}, \bibinfo{pages}{600--607}
  (\bibinfo{year}{2017}).

\bibitem{Stern2018Battery}
\bibinfo{author}{Stern, B.}, \bibinfo{author}{Ji, X.},
  \bibinfo{author}{Okawachi, Y.}, \bibinfo{author}{Gaeta, A.~L.} \&
  \bibinfo{author}{Lipson, M.}
\newblock \bibinfo{title}{Battery-operated integrated frequency comb
  generator}.
\newblock \emph{\bibinfo{journal}{Nature}} \textbf{\bibinfo{volume}{562}},
  \bibinfo{pages}{401--408} (\bibinfo{year}{2018}).

\bibitem{Pavlov2018Narrow}
\bibinfo{author}{Pavlov, N.~G.} \emph{et~al.}
\newblock \bibinfo{title}{Narrow-linewidth lasing and soliton {Kerr} microcombs
  with ordinary laser diodes}.
\newblock \emph{\bibinfo{journal}{Nature Photonics}}
  \textbf{\bibinfo{volume}{12}}, \bibinfo{pages}{694--699}
  (\bibinfo{year}{2018}).

\bibitem{Bao2019Laser}
\bibinfo{author}{Bao, H.} \emph{et~al.}
\newblock \bibinfo{title}{Laser cavity-soliton microcombs}.
\newblock \emph{\bibinfo{journal}{Nature Photonics}}
  \textbf{\bibinfo{volume}{13}}, \bibinfo{pages}{384--389}
  (\bibinfo{year}{2019}).

\bibitem{Huang2019Temporal}
\bibinfo{author}{Huang, Y.} \emph{et~al.}
\newblock \bibinfo{title}{Temporal soliton and optical frequency comb
  generation in a {Brillouin} laser cavity}.
\newblock \emph{\bibinfo{journal}{Optica}} \textbf{\bibinfo{volume}{6}},
  \bibinfo{pages}{1491--1497} (\bibinfo{year}{2019}).

\bibitem{Piccardo2020Frequency}
\bibinfo{author}{Piccardo, M.} \emph{et~al.}
\newblock \bibinfo{title}{Frequency combs induced by phase turbulence}.
\newblock \emph{\bibinfo{journal}{Nature}} \textbf{\bibinfo{volume}{582}},
  \bibinfo{pages}{360--364} (\bibinfo{year}{2020}).

\bibitem{Tikan2021Emergent}
\bibinfo{author}{Tikan, A.} \emph{et~al.}
\newblock \bibinfo{title}{Emergent nonlinear phenomena in a driven dissipative
  photonic dimer}.
\newblock \emph{\bibinfo{journal}{Nature Physics}}
  \textbf{\bibinfo{volume}{17}}, \bibinfo{pages}{604--610}
  (\bibinfo{year}{2021}).

\bibitem{Helgason2021Dissipative}
\bibinfo{author}{Helgason, {\' O}.~B.} \emph{et~al.}
\newblock \bibinfo{title}{Dissipative solitons in photonic molecules}.
\newblock \emph{\bibinfo{journal}{Nature Photonics}}
  \textbf{\bibinfo{volume}{15}}, \bibinfo{pages}{305--310}
  (\bibinfo{year}{2021}).

\bibitem{Liang2015High}
\bibinfo{author}{Liang, W.} \emph{et~al.}
\newblock \bibinfo{title}{High spectral purity {Kerr} frequency comb radio
  frequency photonic oscillator}.
\newblock \emph{\bibinfo{journal}{Nature Communications}}
  \textbf{\bibinfo{volume}{6}} (\bibinfo{year}{2015}).

\bibitem{Marin2017Microresonator}
\bibinfo{author}{Marin-Palomo, P.} \emph{et~al.}
\newblock \bibinfo{title}{Microresonator-based solitons for massively parallel
  coherent optical communications}.
\newblock \emph{\bibinfo{journal}{Nature}} \textbf{\bibinfo{volume}{546}},
  \bibinfo{pages}{274--279} (\bibinfo{year}{2017}).

\bibitem{Suh2018Soliton}
\bibinfo{author}{Suh, M.-G.} \& \bibinfo{author}{Vahala, K.~J.}
\newblock \bibinfo{title}{Soliton microcomb range measurement}.
\newblock \emph{\bibinfo{journal}{Science}} \textbf{\bibinfo{volume}{359}},
  \bibinfo{pages}{884--887} (\bibinfo{year}{2018}).

\bibitem{Spencer2018optical}
\bibinfo{author}{Spencer, D.~T.} \emph{et~al.}
\newblock \bibinfo{title}{An optical-frequency synthesizer using integrated
  photonics}.
\newblock \emph{\bibinfo{journal}{Nature}} \textbf{\bibinfo{volume}{557}},
  \bibinfo{pages}{81--88} (\bibinfo{year}{2018}).

\bibitem{Hu2018Single}
\bibinfo{author}{Hu, H.} \emph{et~al.}
\newblock \bibinfo{title}{Single-source chip-based frequency comb enabling
  extreme parallel data transmission}.
\newblock \emph{\bibinfo{journal}{Nature Photonics}}
  \textbf{\bibinfo{volume}{12}}, \bibinfo{pages}{469--473}
  (\bibinfo{year}{2018}).

\bibitem{Fulop2018High}
\bibinfo{author}{F{\" u}l{\" o}p, A.} \emph{et~al.}
\newblock \bibinfo{title}{High-order coherent communications using mode-locked
  dark-pulse {Kerr} combs from microresonators}.
\newblock \emph{\bibinfo{journal}{Nature Communications}}
  \textbf{\bibinfo{volume}{9}}, \bibinfo{pages}{1598} (\bibinfo{year}{2018}).

\bibitem{Corcoran2020Ultra}
\bibinfo{author}{Corcoran, B.} \emph{et~al.}
\newblock \bibinfo{title}{Ultra-dense optical data transmission over standard
  fibre with a single chip source}.
\newblock \emph{\bibinfo{journal}{Nature Communications}}
  \textbf{\bibinfo{volume}{11}}, \bibinfo{pages}{2568} (\bibinfo{year}{2020}).

\bibitem{Xu202111}
\bibinfo{author}{Xu, X.} \emph{et~al.}
\newblock \bibinfo{title}{11 {TOPS} photonic convolutional accelerator for
  optical neural networks}.
\newblock \emph{\bibinfo{journal}{Nature}} \textbf{\bibinfo{volume}{589}},
  \bibinfo{pages}{44--51} (\bibinfo{year}{2021}).

\bibitem{Feldmann2021Parallel}
\bibinfo{author}{Feldmann, J.} \emph{et~al.}
\newblock \bibinfo{title}{Parallel convolutional processing using an integrated
  photonic tensor core}.
\newblock \emph{\bibinfo{journal}{Nature}} \textbf{\bibinfo{volume}{589}},
  \bibinfo{pages}{52--58} (\bibinfo{year}{2021}).

\bibitem{Riemensberger2020Massively}
\bibinfo{author}{Riemensberger, J.} \emph{et~al.}
\newblock \bibinfo{title}{Massively parallel coherent laser ranging using a
  soliton microcomb}.
\newblock \emph{\bibinfo{journal}{Nature}} \textbf{\bibinfo{volume}{581}},
  \bibinfo{pages}{164--170} (\bibinfo{year}{2020}).

\bibitem{Liu2020Monolithic}
\bibinfo{author}{Liu, J.} \emph{et~al.}
\newblock \bibinfo{title}{Monolithic piezoelectric control of soliton
  microcombs}.
\newblock \emph{\bibinfo{journal}{Nature}} \textbf{\bibinfo{volume}{583}},
  \bibinfo{pages}{385--390} (\bibinfo{year}{2020}).

\bibitem{Meng2020Mid}
\bibinfo{author}{Meng, B.} \emph{et~al.}
\newblock \bibinfo{title}{Mid-infrared frequency comb from a ring quantum
  cascade laser}.
\newblock \emph{\bibinfo{journal}{Optica}} \textbf{\bibinfo{volume}{7}},
  \bibinfo{pages}{162--167} (\bibinfo{year}{2020}).

\bibitem{Boggio2022Efficient}
\bibinfo{author}{Boggio, J. M.~C.} \emph{et~al.}
\newblock \bibinfo{title}{Efficient {Kerr} soliton comb generation in
  micro-resonator with interferometric back-coupling}.
\newblock \emph{\bibinfo{journal}{Nature Communications}}
  \textbf{\bibinfo{volume}{13}}, \bibinfo{pages}{1292} (\bibinfo{year}{2022}).

\bibitem{Wu2025Vernier}
\bibinfo{author}{Wu, K.} \emph{et~al.}
\newblock \bibinfo{title}{Vernier microcombs for integrated optical atomic
  clocks}.
\newblock \emph{\bibinfo{journal}{Nature Photonics}}
  \textbf{\bibinfo{volume}{19}}, \bibinfo{pages}{400--406}
  (\bibinfo{year}{2025}).

\bibitem{Zhao2024All}
\bibinfo{author}{Zhao, Y.} \emph{et~al.}
\newblock \bibinfo{title}{All-optical frequency division on-chip using a single
  laser}.
\newblock \emph{\bibinfo{journal}{Nature}} \textbf{\bibinfo{volume}{627}},
  \bibinfo{pages}{546--552} (\bibinfo{year}{2024}).

\bibitem{Jang2018Synchronization}
\bibinfo{author}{Jang, J.~K.} \emph{et~al.}
\newblock \bibinfo{title}{Synchronization of coupled optical microresonators}.
\newblock \emph{\bibinfo{journal}{Nature Photonics}}
  \textbf{\bibinfo{volume}{12}}, \bibinfo{pages}{688} (\bibinfo{year}{2018}).

\bibitem{Kim2021Synchronization}
\bibinfo{author}{Kim, B.~Y.} \emph{et~al.}
\newblock \bibinfo{title}{Synchronization of nonsolitonic {Kerr} combs}.
\newblock \emph{\bibinfo{journal}{Science Advances}}
  \textbf{\bibinfo{volume}{7}}, \bibinfo{pages}{eabi4362}
  (\bibinfo{year}{2021}).

\bibitem{Moille2023Kerr}
\bibinfo{author}{Moille, G.} \emph{et~al.}
\newblock \bibinfo{title}{Kerr-induced synchronization of a cavity soliton to
  an optical reference}.
\newblock \emph{\bibinfo{journal}{Nature}} \textbf{\bibinfo{volume}{624}},
  \bibinfo{pages}{267--274} (\bibinfo{year}{2023}).

\bibitem{Lei2024Self}
\bibinfo{author}{Lei, F.} \emph{et~al.}
\newblock \bibinfo{title}{Self-injection-locked optical parametric oscillator
  based on microcombs}.
\newblock \emph{\bibinfo{journal}{Optica}} \textbf{\bibinfo{volume}{11}},
  \bibinfo{pages}{420} (\bibinfo{year}{2024}).

\bibitem{Wu2023Vernier}
\bibinfo{author}{Wu, K.} \emph{et~al.}
\newblock \bibinfo{title}{Vernier microcombs for high-frequency carrier
  envelope offset and repetition rate detection}.
\newblock \emph{\bibinfo{journal}{Optica}} \textbf{\bibinfo{volume}{10}},
  \bibinfo{pages}{626} (\bibinfo{year}{2023}).

\bibitem{Kudelin2024Photonic}
\bibinfo{author}{Kudelin, I.} \emph{et~al.}
\newblock \bibinfo{title}{Photonic chip-based low-noise microwave oscillator}.
\newblock \emph{\bibinfo{journal}{Nature}} \textbf{\bibinfo{volume}{627}},
  \bibinfo{pages}{534--539} (\bibinfo{year}{2024}).

\bibitem{Sun2024Integrated}
\bibinfo{author}{Sun, S.} \emph{et~al.}
\newblock \bibinfo{title}{Integrated optical frequency division for microwave
  and {mmWave} generation}.
\newblock \emph{\bibinfo{journal}{Nature}} \textbf{\bibinfo{volume}{627}},
  \bibinfo{pages}{540--545} (\bibinfo{year}{2024}).

\bibitem{Wildi2024Phase}
\bibinfo{author}{Wildi, T.}, \bibinfo{author}{Ulanov, A.~E.},
  \bibinfo{author}{Voumard, T.}, \bibinfo{author}{Ruhnke, B.} \&
  \bibinfo{author}{Herr, T.}
\newblock \bibinfo{title}{Phase-stabilised self-injection-locked microcomb}.
\newblock \emph{\bibinfo{journal}{Nature Communications}}
  \textbf{\bibinfo{volume}{15}}, \bibinfo{pages}{7030} (\bibinfo{year}{2024}).

\bibitem{sun_microcavity_2025}
\bibinfo{author}{Sun, S.} \emph{et~al.}
\newblock \bibinfo{title}{Microcavity {Kerr} optical frequency division with
  integrated {SiN} photonics}.
\newblock \emph{\bibinfo{journal}{Nature Photonics}}  (\bibinfo{year}{2025}).

\bibitem{jin_microresonator-referenced_2025}
\bibinfo{author}{Jin, X.} \emph{et~al.}
\newblock \bibinfo{title}{Microresonator-referenced soliton microcombs with
  zeptosecond-level timing noise}.
\newblock \emph{\bibinfo{journal}{Nature Photonics}}  (\bibinfo{year}{2025}).

\bibitem{Yi2017Single}
\bibinfo{author}{Yi, X.} \emph{et~al.}
\newblock \bibinfo{title}{Single-mode dispersive waves and soliton microcomb
  dynamics}.
\newblock \emph{\bibinfo{journal}{Nature Communications}}
  \textbf{\bibinfo{volume}{8}}, \bibinfo{pages}{14869} (\bibinfo{year}{2017}).

\bibitem{Lei2022Optical}
\bibinfo{author}{Lei, F.} \emph{et~al.}
\newblock \bibinfo{title}{Optical linewidth of soliton microcombs}.
\newblock \emph{\bibinfo{journal}{Nature Communications}}
  \textbf{\bibinfo{volume}{13}}, \bibinfo{pages}{3161} (\bibinfo{year}{2022}).

\bibitem{Liu2020XKBandMicrocombs}
\bibinfo{author}{Liu, J.}, \bibinfo{author}{Lucas, E.}, \bibinfo{author}{Raja,
  A.~S.} \emph{et~al.}
\newblock \bibinfo{title}{Photonic microwave generation in the x- and k-band
  using integrated soliton microcombs}.
\newblock \emph{\bibinfo{journal}{Nature Photonics}}
  \textbf{\bibinfo{volume}{14}}, \bibinfo{pages}{486--491}
  (\bibinfo{year}{2020}).

\bibitem{Lucas2020Ultralow}
\bibinfo{author}{Lucas, E.} \emph{et~al.}
\newblock \bibinfo{title}{Ultralow-noise photonic microwave synthesis using a
  soliton microcomb-based transfer oscillator}.
\newblock \emph{\bibinfo{journal}{Nature Communications}}
  \textbf{\bibinfo{volume}{11}}, \bibinfo{pages}{374} (\bibinfo{year}{2020}).

\bibitem{Yang2021DispersiveNoise}
\bibinfo{author}{Yang, Q.-F.}, \bibinfo{author}{Ji, Q.-X.},
  \bibinfo{author}{Wu, L.} \emph{et~al.}
\newblock \bibinfo{title}{Dispersive-wave induced noise limits in miniature
  soliton microwave sources}.
\newblock \emph{\bibinfo{journal}{Nature Communications}}
  \textbf{\bibinfo{volume}{12}}, \bibinfo{pages}{1442} (\bibinfo{year}{2021}).

\bibitem{Triscari2023Quiet}
\bibinfo{author}{Triscari, A.~C.}, \bibinfo{author}{Tusnin, A.},
  \bibinfo{author}{Tikan, A.} \& \bibinfo{author}{Kippenberg, T.~J.}
\newblock \bibinfo{title}{Quiet point engineering for low-noise microwave
  generation with soliton microcombs}.
\newblock \emph{\bibinfo{journal}{Communications Physics}}
  \textbf{\bibinfo{volume}{6}}, \bibinfo{pages}{318} (\bibinfo{year}{2023}).

\bibitem{Shen2020Integrated}
\bibinfo{author}{Shen, B.} \emph{et~al.}
\newblock \bibinfo{title}{Integrated turnkey soliton microcombs}.
\newblock \emph{\bibinfo{journal}{Nature}} \textbf{\bibinfo{volume}{582}},
  \bibinfo{pages}{365--369} (\bibinfo{year}{2020}).

\bibitem{Rowley2022Self}
\bibinfo{author}{Rowley, M.} \emph{et~al.}
\newblock \bibinfo{title}{Self-emergence of robust solitons in a microcavity}.
\newblock \emph{\bibinfo{journal}{Nature}} \textbf{\bibinfo{volume}{608}},
  \bibinfo{pages}{303--309} (\bibinfo{year}{2022}).

\bibitem{Nie2022Dissipative}
\bibinfo{author}{Nie, M.} \emph{et~al.}
\newblock \bibinfo{title}{Dissipative soliton generation and real-time dynamics
  in microresonator-filtered fiber lasers}.
\newblock \emph{\bibinfo{journal}{Light: Science \& Applications}}
  \textbf{\bibinfo{volume}{11}}, \bibinfo{pages}{296} (\bibinfo{year}{2022}).

\bibitem{Cutrona2023Nonlocal}
\bibinfo{author}{Cutrona, A.} \emph{et~al.}
\newblock \bibinfo{title}{Nonlocal bonding of a soliton and a blue-detuned
  state in a microcomb laser}.
\newblock \emph{\bibinfo{journal}{Communications Physics}}
  \textbf{\bibinfo{volume}{6}}, \bibinfo{pages}{259} (\bibinfo{year}{2023}).

\bibitem{Opacak2024NozakiBekki}
\bibinfo{author}{Opa{\v c}ak, N.} \emph{et~al.}
\newblock \bibinfo{title}{Nozaki--{Bekki} solitons in semiconductor lasers}.
\newblock \emph{\bibinfo{journal}{Nature}} \textbf{\bibinfo{volume}{625}},
  \bibinfo{pages}{685--690} (\bibinfo{year}{2024}).

\bibitem{Ulanov2024Synthetic}
\bibinfo{author}{Ulanov, A.~E.} \emph{et~al.}
\newblock \bibinfo{title}{Synthetic reflection self-injection-locked
  microcombs}.
\newblock \emph{\bibinfo{journal}{Nature Photonics}}
  \textbf{\bibinfo{volume}{18}}, \bibinfo{pages}{294--299}
  (\bibinfo{year}{2024}).

\bibitem{MkrtchyanMicroring}
\bibinfo{author}{Mkrtchyan, A.~A.} \emph{et~al.}
\newblock \bibinfo{title}{Microring resonator as a rayleigh mirror for
  broadband laser cavity comb generation}  (\bibinfo{year}{2025}).
\newblock \bibinfo{note}{ArXiv:2503.09166 [physics]}.

\bibitem{Lu2014Topological}
\bibinfo{author}{Lu, L.}, \bibinfo{author}{Joannopoulos, J.~D.} \&
  \bibinfo{author}{Solja\v{c}i\'{c}, M.}
\newblock \bibinfo{title}{Topological photonics}.
\newblock \emph{\bibinfo{journal}{Nature Photonics}}
  \textbf{\bibinfo{volume}{8}}, \bibinfo{pages}{821--829}
  (\bibinfo{year}{2014}).

\bibitem{Ozawa2019Topological}
\bibinfo{author}{Ozawa, T.} \emph{et~al.}
\newblock \bibinfo{title}{Topological photonics}.
\newblock \emph{\bibinfo{journal}{Reviews of Modern Physics}}
  \textbf{\bibinfo{volume}{91}}, \bibinfo{pages}{015006}
  (\bibinfo{year}{2019}).

\bibitem{Smirnova2020Nonlinear}
\bibinfo{author}{Smirnova, D.}, \bibinfo{author}{Leykam, D.},
  \bibinfo{author}{Chong, Y.} \& \bibinfo{author}{Kivshar, Y.}
\newblock \bibinfo{title}{Nonlinear topological photonics}.
\newblock \emph{\bibinfo{journal}{Applied Physics Reviews}}
  \textbf{\bibinfo{volume}{7}}, \bibinfo{pages}{021306} (\bibinfo{year}{2020}).

\bibitem{Nasari2023NonHermitian}
\bibinfo{author}{Nasari, H.}, \bibinfo{author}{Pyrialakos, G.~G.},
  \bibinfo{author}{Christodoulides, D.~N.} \& \bibinfo{author}{Khajavikhan, M.}
\newblock \bibinfo{title}{Non-hermitian topological photonics}.
\newblock \emph{\bibinfo{journal}{Optical Materials Express}}
  \textbf{\bibinfo{volume}{13}}, \bibinfo{pages}{870--885}
  (\bibinfo{year}{2023}).

\bibitem{Khanikaev2024Beyond}
\bibinfo{author}{Khanikaev, A.~B.} \& \bibinfo{author}{Al\`{u}, A.}
\newblock \bibinfo{title}{Topological photonics: robustness and beyond}.
\newblock \emph{\bibinfo{journal}{Nature Communications}}
  \textbf{\bibinfo{volume}{15}}, \bibinfo{pages}{931} (\bibinfo{year}{2024}).

\bibitem{Wong2025Nonlinear}
\bibinfo{author}{Wong, S.} \emph{et~al.}
\newblock \bibinfo{title}{Nonlinear {Topological} {Photonics}: Capturing
  {Nonlinear} {Dynamics} and {Optical} {Thermodynamics}}.
\newblock \emph{\bibinfo{journal}{ACS Photonics}}
  \bibinfo{pages}{acsphotonics.4c02430} (\bibinfo{year}{2025}).

\bibitem{SzameitRechtsman2024NonlinearReview}
\bibinfo{author}{Szameit, A.} \& \bibinfo{author}{Rechtsman, M.~C.}
\newblock \bibinfo{title}{Discrete nonlinear topological photonics}.
\newblock \emph{\bibinfo{journal}{Nature Physics}}
  \textbf{\bibinfo{volume}{20}}, \bibinfo{pages}{905--912}
  (\bibinfo{year}{2024}).

\bibitem{HasanKane2010TopIns}
\bibinfo{author}{Hasan, M.~Z.} \& \bibinfo{author}{Kane, C.~L.}
\newblock \bibinfo{title}{Colloquium: Topological insulators}.
\newblock \emph{\bibinfo{journal}{Reviews of Modern Physics}}
  \textbf{\bibinfo{volume}{82}}, \bibinfo{pages}{3045--3067}
  (\bibinfo{year}{2010}).

\bibitem{Wang2009TopoEM}
\bibinfo{author}{Wang, Z.}, \bibinfo{author}{Chong, Y.},
  \bibinfo{author}{Joannopoulos, J.~D.} \& \bibinfo{author}{Solja\v{c}i\'{c},
  M.}
\newblock \bibinfo{title}{Observation of unidirectional backscattering-immune
  topological electromagnetic states}.
\newblock \emph{\bibinfo{journal}{Nature}} \textbf{\bibinfo{volume}{461}},
  \bibinfo{pages}{772--775} (\bibinfo{year}{2009}).

\bibitem{Rechtsman2013FloquetTI}
\bibinfo{author}{Rechtsman, M.~C.} \emph{et~al.}
\newblock \bibinfo{title}{Photonic floquet topological insulators}.
\newblock \emph{\bibinfo{journal}{Nature}} \textbf{\bibinfo{volume}{496}},
  \bibinfo{pages}{196--200} (\bibinfo{year}{2013}).

\bibitem{Zak1989}
\bibinfo{author}{Zak, J.}
\newblock \bibinfo{title}{Berry’s phase for energy bands in solids}.
\newblock \emph{\bibinfo{journal}{Physical Review Letters}}
  \textbf{\bibinfo{volume}{62}}, \bibinfo{pages}{2747--2750}
  (\bibinfo{year}{1989}).

\bibitem{Berry1984}
\bibinfo{author}{Berry, M.~V.}
\newblock \bibinfo{title}{Quantal phase factors accompanying adiabatic
  changes}.
\newblock \emph{\bibinfo{journal}{Proc. R. Soc. Lond. A}}
  \textbf{\bibinfo{volume}{392}}, \bibinfo{pages}{45--57}
  (\bibinfo{year}{1984}).

\bibitem{Resta1994}
\bibinfo{author}{Resta, R.}
\newblock \bibinfo{title}{Macroscopic polarization in crystalline dielectrics:
  the geometric phase approach}.
\newblock \emph{\bibinfo{journal}{Rev. Mod. Phys.}}
  \textbf{\bibinfo{volume}{66}}, \bibinfo{pages}{899--915}
  (\bibinfo{year}{1994}).

\bibitem{Trillo1997Stable}
\bibinfo{author}{Trillo, S.}, \bibinfo{author}{Haelterman, M.} \&
  \bibinfo{author}{Sheppard, A.}
\newblock \bibinfo{title}{Stable topological spatial solitons in optical
  parametric oscillators}.
\newblock \emph{\bibinfo{journal}{Optics Letters}}
  \textbf{\bibinfo{volume}{22}}, \bibinfo{pages}{970} (\bibinfo{year}{1997}).

\bibitem{Oppo1999From}
\bibinfo{author}{Oppo, G.-L.}, \bibinfo{author}{Scroggie, A.~J.} \&
  \bibinfo{author}{Firth, W.~J.}
\newblock \bibinfo{title}{From domain walls to localized structures in
  degenerate optical parametric oscillators}.
\newblock \emph{\bibinfo{journal}{Journal of Optics B: Quantum and
  Semiclassical Optics}} \textbf{\bibinfo{volume}{1}},
  \bibinfo{pages}{133--138} (\bibinfo{year}{1999}).

\bibitem{Englebert2024Topological}
\bibinfo{author}{Englebert, N.} \emph{et~al.}
\newblock \bibinfo{title}{Topological soliton frequency comb in nanophotonic
  lithium niobate} (\bibinfo{year}{2025}).
\newblock \urlprefix\url{https://arxiv.org/abs/2511.01856}.
\newblock \eprint{2511.01856}.

\bibitem{aceves_optical_2000}
\bibinfo{author}{Aceves, A.~B.}
\newblock \bibinfo{title}{Optical gap solitons: {Past}, present, and future;
  theory and experiments}.
\newblock \emph{\bibinfo{journal}{Chaos: An Interdisciplinary Journal of
  Nonlinear Science}} \textbf{\bibinfo{volume}{10}}, \bibinfo{pages}{584--589}
  (\bibinfo{year}{2000}).

\bibitem{Lumer2013SelfLocalized}
\bibinfo{author}{Lumer, Y.}, \bibinfo{author}{Plotnik, Y.},
  \bibinfo{author}{Rechtsman, M.~C.} \& \bibinfo{author}{Segev, M.}
\newblock \bibinfo{title}{Self-localized states in photonic topological
  insulators}.
\newblock \emph{\bibinfo{journal}{Physical Review Letters}}
  \textbf{\bibinfo{volume}{111}}, \bibinfo{pages}{243905}
  (\bibinfo{year}{2013}).

\bibitem{LeykamChong2016EdgeSolitons}
\bibinfo{author}{Leykam, D.} \& \bibinfo{author}{Chong, Y.~D.}
\newblock \bibinfo{title}{Edge solitons in nonlinear-photonic topological
  insulators}.
\newblock \emph{\bibinfo{journal}{Physical Review Letters}}
  \textbf{\bibinfo{volume}{117}}, \bibinfo{pages}{143901}
  (\bibinfo{year}{2016}).

\bibitem{MukherjeeRechtsman2020FloquetSolitons}
\bibinfo{author}{Mukherjee, S.} \& \bibinfo{author}{Rechtsman, M.~C.}
\newblock \bibinfo{title}{Observation of floquet solitons in a topological
  bandgap}.
\newblock \emph{\bibinfo{journal}{Science}} \textbf{\bibinfo{volume}{368}},
  \bibinfo{pages}{856--859} (\bibinfo{year}{2020}).

\bibitem{Pernet2022GapSolitons}
\bibinfo{author}{Pernet, N.}, \bibinfo{author}{Zhong, J.},
  \bibinfo{author}{Benisty, H.} \& \bibinfo{author}{Liew, T. C.~H.}
\newblock \bibinfo{title}{Gap solitons in a one-dimensional driven-dissipative
  topological lattice}.
\newblock \emph{\bibinfo{journal}{Nature Physics}}
  \textbf{\bibinfo{volume}{18}}, \bibinfo{pages}{678--684}
  (\bibinfo{year}{2022}).

\bibitem{Bongiovanni2021DynamicalTopo}
\bibinfo{author}{Bongiovanni, D.}, \bibinfo{author}{Juki\'{c}, D.},
  \bibinfo{author}{Hu, Z.} \emph{et~al.}
\newblock \bibinfo{title}{Dynamically emerging topological phase transitions in
  nonlinear interacting soliton lattices}.
\newblock \emph{\bibinfo{journal}{Physical Review Letters}}
  \textbf{\bibinfo{volume}{127}}, \bibinfo{pages}{184101}
  (\bibinfo{year}{2021}).

\bibitem{Hu2021HOTIBIC}
\bibinfo{author}{Hu, Z.}, \bibinfo{author}{Bongiovanni, D.},
  \bibinfo{author}{Juki\'{c}, D.} \emph{et~al.}
\newblock \bibinfo{title}{Nonlinear control of photonic higher-order
  topological bound states in the continuum}.
\newblock \emph{\bibinfo{journal}{Light: Science \& Applications}}
  \textbf{\bibinfo{volume}{10}}, \bibinfo{pages}{164} (\bibinfo{year}{2021}).

\bibitem{Wang2023SubSymmetryTopo}
\bibinfo{author}{Wang, Z.}, \bibinfo{author}{Wang, X.}, \bibinfo{author}{Hu,
  Z.} \emph{et~al.}
\newblock \bibinfo{title}{Sub-symmetry-protected topological states}.
\newblock \emph{\bibinfo{journal}{Nature Physics}}
  \textbf{\bibinfo{volume}{19}}, \bibinfo{pages}{992--998}
  (\bibinfo{year}{2023}).

\bibitem{Fan2022Topological}
\bibinfo{author}{Fan, Z.}, \bibinfo{author}{Puzyrev, D.~N.} \&
  \bibinfo{author}{Skryabin, D.~V.}
\newblock \bibinfo{title}{Topological soliton metacrystals}.
\newblock \emph{\bibinfo{journal}{Communications Physics}}
  \textbf{\bibinfo{volume}{5}}, \bibinfo{pages}{248} (\bibinfo{year}{2022}).

\bibitem{Mittal2021Topological}
\bibinfo{author}{Mittal, S.}, \bibinfo{author}{Moille, G.},
  \bibinfo{author}{Srinivasan, K.}, \bibinfo{author}{Chembo, Y.~K.} \&
  \bibinfo{author}{Hafezi, M.}
\newblock \bibinfo{title}{Topological frequency combs and nested temporal
  solitons}.
\newblock \emph{\bibinfo{journal}{Nature Physics}}
  \textbf{\bibinfo{volume}{17}}, \bibinfo{pages}{1169--1176}
  (\bibinfo{year}{2021}).

\bibitem{Flower2024Observation}
\bibinfo{author}{Flower, C.~J.} \emph{et~al.}
\newblock \bibinfo{title}{Observation of topological frequency combs}.
\newblock \emph{\bibinfo{journal}{Science}} \textbf{\bibinfo{volume}{384}},
  \bibinfo{pages}{1356--1361} (\bibinfo{year}{2024}).

\bibitem{Coen2024Nonlinear}
\bibinfo{author}{Coen, S.} \emph{et~al.}
\newblock \bibinfo{title}{Nonlinear topological symmetry protection in a
  dissipative system}.
\newblock \emph{\bibinfo{journal}{Nature Communications}}
  \textbf{\bibinfo{volume}{15}}, \bibinfo{pages}{1398} (\bibinfo{year}{2024}).

\bibitem{Wang2023Experimental}
\bibinfo{author}{Wang, J.} \emph{et~al.}
\newblock \bibinfo{title}{Experimental observation of {Berry} phases in optical
  {M}{\" o}bius-strip microcavities}.
\newblock \emph{\bibinfo{journal}{Nature Photonics}}
  \textbf{\bibinfo{volume}{17}}, \bibinfo{pages}{120--125}
  (\bibinfo{year}{2023}).

\bibitem{Maitland2020Stationary}
\bibinfo{author}{Maitland, C.}, \bibinfo{author}{Conforti, M.},
  \bibinfo{author}{Mussot, A.} \& \bibinfo{author}{Biancalana, F.}
\newblock \bibinfo{title}{Stationary states and instabilities of a {M}{\"
  o}bius fiber resonator}.
\newblock \emph{\bibinfo{journal}{Physical Review Research}}
  \textbf{\bibinfo{volume}{2}}, \bibinfo{pages}{043195} (\bibinfo{year}{2020}).

\bibitem{Chen2023Topological}
\bibinfo{author}{Chen, Y.}, \bibinfo{author}{Hou, J.}, \bibinfo{author}{Zhao,
  G.}, \bibinfo{author}{Chen, X.} \& \bibinfo{author}{Wan, W.}
\newblock \bibinfo{title}{Topological resonances in a {M}{\" o}bius ring
  resonator}.
\newblock \emph{\bibinfo{journal}{Communications Physics}}
  \textbf{\bibinfo{volume}{6}}, \bibinfo{pages}{84} (\bibinfo{year}{2023}).

\bibitem{NonlinearDynamics2007}
\bibinfo{author}{Anishchenko, V.~S.}, \bibinfo{author}{Astakhov, V.},
  \bibinfo{author}{Neiman, A.}, \bibinfo{author}{Vadivasova, T.} \&
  \bibinfo{author}{Schimansky-Geier, L.}
\newblock \emph{\bibinfo{title}{Nonlinear Dynamics of Chaotic and Stochastic
  Systems}} (\bibinfo{publisher}{Springer}, \bibinfo{address}{Berlin ; New
  York}, \bibinfo{year}{2007}).

\bibitem{Heckelmann2023Quantum}
\bibinfo{author}{Heckelmann, I.} \emph{et~al.}
\newblock \bibinfo{title}{Quantum walk comb in a fast gain laser}.
\newblock \emph{\bibinfo{journal}{Science}} \textbf{\bibinfo{volume}{382}},
  \bibinfo{pages}{434--438} (\bibinfo{year}{2023}).

\bibitem{Bourgon2025Mode}
\bibinfo{author}{Bourgon, E.} \emph{et~al.}
\newblock \bibinfo{title}{Mode-locking in a semiconductor photonic bandgap
  laser}.
\newblock \emph{\bibinfo{journal}{Communications Physics}}
  \textbf{\bibinfo{volume}{8}}, \bibinfo{pages}{458} (\bibinfo{year}{2025}).

\bibitem{Cutrona2023Stability}
\bibinfo{author}{Cutrona, A.} \emph{et~al.}
\newblock \bibinfo{title}{Stability of laser cavity-solitons for metrological
  applications}.
\newblock \emph{\bibinfo{journal}{Applied Physics Letters}}
  \textbf{\bibinfo{volume}{122}}, \bibinfo{pages}{121104}
  (\bibinfo{year}{2023}).

\bibitem{Coullet1990Breaking}
\bibinfo{author}{Coullet, P.}, \bibinfo{author}{Lega, J.},
  \bibinfo{author}{Houchmanzadeh, B.} \& \bibinfo{author}{Lajzerowicz, J.}
\newblock \bibinfo{title}{Breaking chirality in nonequilibrium systems}.
\newblock \emph{\bibinfo{journal}{Physical Review Letters}}
  \textbf{\bibinfo{volume}{65}}, \bibinfo{pages}{1352--1355}
  (\bibinfo{year}{1990}).

\bibitem{Michaelis2001Universal}
\bibinfo{author}{Michaelis, D.}, \bibinfo{author}{Peschel, U.},
  \bibinfo{author}{Lederer, F.}, \bibinfo{author}{Skryabin, D.~V.} \&
  \bibinfo{author}{Firth, W.~J.}
\newblock \bibinfo{title}{Universal criterion and amplitude equation for a
  nonequilibrium {Ising}-{Bloch} transition}.
\newblock \emph{\bibinfo{journal}{Physical Review E}}
  \textbf{\bibinfo{volume}{63}}, \bibinfo{pages}{066602}
  (\bibinfo{year}{2001}).

\bibitem{De2002Domain}
\bibinfo{author}{De~Valc{\' a}rcel, G.~J.}, \bibinfo{author}{P{\' e}rez-Arjona,
  I.} \& \bibinfo{author}{Rold{\' a}n, E.}
\newblock \bibinfo{title}{Domain {Walls} and {Ising}-{Bloch} {Transitions} in
  {Parametrically} {Driven} {Systems}}.
\newblock \emph{\bibinfo{journal}{Physical Review Letters}}
  \textbf{\bibinfo{volume}{89}}, \bibinfo{pages}{164101}
  (\bibinfo{year}{2002}).

\bibitem{Esteban2005Controlled}
\bibinfo{author}{Esteban-Mart{\' i}n, A.}, \bibinfo{author}{Taranenko, V.~B.},
  \bibinfo{author}{Garc{\' i}a, J.}, \bibinfo{author}{De~Valc{\' a}rcel, G.~J.}
  \& \bibinfo{author}{Rold{\' a}n, E.}
\newblock \bibinfo{title}{Controlled {Observation} of a {Nonequilibrium}
  {Ising}-{Bloch} {Transition} in a {Nonlinear} {Optical} {Cavity}}.
\newblock \emph{\bibinfo{journal}{Physical Review Letters}}
  \textbf{\bibinfo{volume}{94}}, \bibinfo{pages}{223903}
  (\bibinfo{year}{2005}).

\bibitem{Li2023Zak}
\bibinfo{author}{Li, G.} \emph{et~al.}
\newblock \bibinfo{title}{Direct extraction of topological zak phase with the
  synthetic dimension}.
\newblock \emph{\bibinfo{journal}{Light: Science \& Applications}}
  \textbf{\bibinfo{volume}{12}}, \bibinfo{pages}{81} (\bibinfo{year}{2023}).

\bibitem{Gallego2000Self}
\bibinfo{author}{Gallego, R.}, \bibinfo{author}{San~Miguel, M.} \&
  \bibinfo{author}{Toral, R.}
\newblock \bibinfo{title}{Self-similar domain growth, localized structures, and
  labyrinthine patterns in vectorial {Kerr} resonators}.
\newblock \emph{\bibinfo{journal}{Physical Review E}}
  \textbf{\bibinfo{volume}{61}}, \bibinfo{pages}{2241--2244}
  (\bibinfo{year}{2000}).

\bibitem{Garbin2021Dissipative}
\bibinfo{author}{Garbin, B.} \emph{et~al.}
\newblock \bibinfo{title}{Dissipative {Polarization} {Domain} {Walls} in a
  {Passive} {Coherently} {Driven} {Kerr} {Resonator}}.
\newblock \emph{\bibinfo{journal}{Physical Review Letters}}
  \textbf{\bibinfo{volume}{126}}, \bibinfo{pages}{023904}
  (\bibinfo{year}{2021}).

\bibitem{Lucas2024Polarization}
\bibinfo{author}{Lucas, E.} \emph{et~al.}
\newblock \bibinfo{title}{Polarization faticons: Chiral localized structures in
  self-defocusing kerr resonators}.
\newblock \emph{\bibinfo{journal}{Phys. Rev. Lett.}}
  \textbf{\bibinfo{volume}{135}}, \bibinfo{pages}{063803}
  (\bibinfo{year}{2025}).

\bibitem{Letsou2025Hybridized}
\bibinfo{author}{Letsou, T.~P.} \emph{et~al.}
\newblock \bibinfo{title}{Hybridized {Soliton} {Lasing} in {Coupled}
  {Semiconductor} {Lasers}}.
\newblock \emph{\bibinfo{journal}{Physical Review Letters}}
  \textbf{\bibinfo{volume}{134}}, \bibinfo{pages}{023802}
  (\bibinfo{year}{2025}).

\end{thebibliography}

\begin{thebibliography}{10}
\expandafter\ifx\csname url\endcsname\relax
  \def\url#1{\texttt{#1}}\fi
\expandafter\ifx\csname urlprefix\endcsname\relax\def\urlprefix{URL }\fi
\providecommand{\bibinfo}[2]{#2}
\providecommand{\eprint}[2][]{\url{#2}}

\bibitem{Bao2019Laser}
\bibinfo{author}{Bao, H.} \emph{et~al.}
\newblock \bibinfo{title}{Laser cavity-soliton microcombs}.
\newblock \emph{\bibinfo{journal}{Nature Photonics}}
  \textbf{\bibinfo{volume}{13}}, \bibinfo{pages}{384--389}
  (\bibinfo{year}{2019}).

\bibitem{Rowley2022Self}
\bibinfo{author}{Rowley, M.} \emph{et~al.}
\newblock \bibinfo{title}{Self-emergence of robust solitons in a microcavity}.
\newblock \emph{\bibinfo{journal}{Nature}} \textbf{\bibinfo{volume}{608}},
  \bibinfo{pages}{303--309} (\bibinfo{year}{2022}).

\bibitem{NonlinearDynamics2007}
\bibinfo{author}{Anishchenko, V.~S.}, \bibinfo{author}{Astakhov, V.},
  \bibinfo{author}{Neiman, A.}, \bibinfo{author}{Vadivasova, T.} \&
  \bibinfo{author}{Schimansky-Geier, L.}
\newblock \emph{\bibinfo{title}{Nonlinear Dynamics of Chaotic and Stochastic
  Systems}} (\bibinfo{publisher}{Springer}, \bibinfo{address}{Berlin ; New
  York}, \bibinfo{year}{2007}).

\bibitem{Lucas2024Polarization}
\bibinfo{author}{Lucas, E.} \emph{et~al.}
\newblock \bibinfo{title}{Polarization faticons: Chiral localized structures in
  self-defocusing kerr resonators}.
\newblock \emph{\bibinfo{journal}{Phys. Rev. Lett.}}
  \textbf{\bibinfo{volume}{135}}, \bibinfo{pages}{063803}
  (\bibinfo{year}{2025}).

\bibitem{Bourgon2025Mode}
\bibinfo{author}{Bourgon, E.} \emph{et~al.}
\newblock \bibinfo{title}{Mode-locking in a semiconductor photonic bandgap
  laser}.
\newblock \emph{\bibinfo{journal}{Communications Physics}}
  \textbf{\bibinfo{volume}{8}}, \bibinfo{pages}{458} (\bibinfo{year}{2025}).

\bibitem{Heckelmann2023Quantum}
\bibinfo{author}{Heckelmann, I.} \emph{et~al.}
\newblock \bibinfo{title}{Quantum walk comb in a fast gain laser}.
\newblock \emph{\bibinfo{journal}{Science}} \textbf{\bibinfo{volume}{382}},
  \bibinfo{pages}{434--438} (\bibinfo{year}{2023}).

\bibitem{Michaelis2001Universal}
\bibinfo{author}{Michaelis, D.}, \bibinfo{author}{Peschel, U.},
  \bibinfo{author}{Lederer, F.}, \bibinfo{author}{Skryabin, D.~V.} \&
  \bibinfo{author}{Firth, W.~J.}
\newblock \bibinfo{title}{Universal criterion and amplitude equation for a
  nonequilibrium {Ising}-{Bloch} transition}.
\newblock \emph{\bibinfo{journal}{Physical Review E}}
  \textbf{\bibinfo{volume}{63}}, \bibinfo{pages}{066602}
  (\bibinfo{year}{2001}).

\bibitem{Coullet1990Breaking}
\bibinfo{author}{Coullet, P.}, \bibinfo{author}{Lega, J.},
  \bibinfo{author}{Houchmanzadeh, B.} \& \bibinfo{author}{Lajzerowicz, J.}
\newblock \bibinfo{title}{Breaking chirality in nonequilibrium systems}.
\newblock \emph{\bibinfo{journal}{Physical Review Letters}}
  \textbf{\bibinfo{volume}{65}}, \bibinfo{pages}{1352--1355}
  (\bibinfo{year}{1990}).

\bibitem{DeValcarcel2002}
\bibinfo{author}{De~Valc{\'a}rcel, G.~J.}, \bibinfo{author}{P{\'e}rez-Arjona,
  I.} \& \bibinfo{author}{Rold{\'a}n, E.}
\newblock \bibinfo{title}{Domain walls and ising-bloch transitions in
  parametrically driven systems}.
\newblock \emph{\bibinfo{journal}{Phys. Rev. Lett.}}
  \textbf{\bibinfo{volume}{89}}, \bibinfo{pages}{164101}
  (\bibinfo{year}{2002}).

\bibitem{Zak1989}
\bibinfo{author}{Zak, J.}
\newblock \bibinfo{title}{Berry’s phase for energy bands in solids}.
\newblock \emph{\bibinfo{journal}{Physical Review Letters}}
  \textbf{\bibinfo{volume}{62}}, \bibinfo{pages}{2747--2750}
  (\bibinfo{year}{1989}).

\bibitem{Fan2022Topological}
\bibinfo{author}{Fan, Z.}, \bibinfo{author}{Puzyrev, D.~N.} \&
  \bibinfo{author}{Skryabin, D.~V.}
\newblock \bibinfo{title}{Topological soliton metacrystals}.
\newblock \emph{\bibinfo{journal}{Communications Physics}}
  \textbf{\bibinfo{volume}{5}}, \bibinfo{pages}{248} (\bibinfo{year}{2022}).

\bibitem{Li2023Zak}
\bibinfo{author}{Li, G.} \emph{et~al.}
\newblock \bibinfo{title}{Direct extraction of topological zak phase with the
  synthetic dimension}.
\newblock \emph{\bibinfo{journal}{Light: Science \& Applications}}
  \textbf{\bibinfo{volume}{12}}, \bibinfo{pages}{81} (\bibinfo{year}{2023}).

\bibitem{Trillo1997Stable}
\bibinfo{author}{Trillo, S.}, \bibinfo{author}{Haelterman, M.} \&
  \bibinfo{author}{Sheppard, A.}
\newblock \bibinfo{title}{Stable topological spatial solitons in optical
  parametric oscillators}.
\newblock \emph{\bibinfo{journal}{Optics Letters}}
  \textbf{\bibinfo{volume}{22}}, \bibinfo{pages}{970} (\bibinfo{year}{1997}).

\bibitem{Oppo1999From}
\bibinfo{author}{Oppo, G.-L.}, \bibinfo{author}{Scroggie, A.~J.} \&
  \bibinfo{author}{Firth, W.~J.}
\newblock \bibinfo{title}{From domain walls to localized structures in
  degenerate optical parametric oscillators}.
\newblock \emph{\bibinfo{journal}{Journal of Optics B: Quantum and
  Semiclassical Optics}} \textbf{\bibinfo{volume}{1}},
  \bibinfo{pages}{133--138} (\bibinfo{year}{1999}).

\bibitem{Oppo2001}
\bibinfo{author}{Oppo, G.-L.}, \bibinfo{author}{Scroggie, A.~J.} \&
  \bibinfo{author}{Firth, W.~J.}
\newblock \bibinfo{title}{Characterization, dynamics and stabilization of
  diffractive domain walls and dark ring cavity solitons in parametric
  oscillators}.
\newblock \emph{\bibinfo{journal}{Phys. Rev. E}} \textbf{\bibinfo{volume}{63}},
  \bibinfo{pages}{066209} (\bibinfo{year}{2001}).

\bibitem{Englebert2024Topological}
\bibinfo{author}{Englebert, N.} \emph{et~al.}
\newblock \bibinfo{title}{Topological soliton frequency comb in nanophotonic
  lithium niobate} (\bibinfo{year}{2025}).
\newblock \urlprefix\url{https://arxiv.org/abs/2511.01856}.
\newblock \eprint{2511.01856}.

\bibitem{Gallego2000Self}
\bibinfo{author}{Gallego, R.}, \bibinfo{author}{San~Miguel, M.} \&
  \bibinfo{author}{Toral, R.}
\newblock \bibinfo{title}{Self-similar domain growth, localized structures, and
  labyrinthine patterns in vectorial {Kerr} resonators}.
\newblock \emph{\bibinfo{journal}{Physical Review E}}
  \textbf{\bibinfo{volume}{61}}, \bibinfo{pages}{2241--2244}
  (\bibinfo{year}{2000}).

\bibitem{Garbin2021Dissipative}
\bibinfo{author}{Garbin, B.} \emph{et~al.}
\newblock \bibinfo{title}{Dissipative {Polarization} {Domain} {Walls} in a
  {Passive} {Coherently} {Driven} {Kerr} {Resonator}}.
\newblock \emph{\bibinfo{journal}{Physical Review Letters}}
  \textbf{\bibinfo{volume}{126}}, \bibinfo{pages}{023904}
  (\bibinfo{year}{2021}).

\bibitem{Alberucci2007}
\bibinfo{author}{Alberucci, A.}, \bibinfo{author}{Peccianti, M.} \&
  \bibinfo{author}{Assanto, G.}
\newblock \bibinfo{title}{Nonlinear bouncing of nonlocal spatial solitons at
  the boundaries}.
\newblock \emph{\bibinfo{journal}{Opt. Lett.}} \textbf{\bibinfo{volume}{32}},
  \bibinfo{pages}{2795--2797} (\bibinfo{year}{2007}).

\bibitem{EstebanMartin2005}
\bibinfo{author}{Esteban-Mart{\'i}n, A.}, \bibinfo{author}{Taranenko, V.~B.},
  \bibinfo{author}{Garc{\'i}a, J.}, \bibinfo{author}{De~Valc{\'a}rcel, G.~J.}
  \& \bibinfo{author}{Rold{\'a}n, E.}
\newblock \bibinfo{title}{Controlled observation of a nonequilibrium
  ising-bloch transition in a nonlinear optical cavity}.
\newblock \emph{\bibinfo{journal}{Phys. Rev. Lett.}}
  \textbf{\bibinfo{volume}{94}}, \bibinfo{pages}{223903}
  (\bibinfo{year}{2005}).

\bibitem{Coullet89a}
\bibinfo{author}{Coullet, P.}, \bibinfo{author}{Gil, L.} \&
  \bibinfo{author}{Lega, J.}
\newblock \bibinfo{title}{Defect-mediated turbulence}.
\newblock \emph{\bibinfo{journal}{Physical Review Letters}}
  \textbf{\bibinfo{volume}{62}}, \bibinfo{pages}{1619--1622}
  (\bibinfo{year}{1989}).

\bibitem{Coullet89b}
\bibinfo{author}{Coullet, P.}, \bibinfo{author}{Gil, L.} \&
  \bibinfo{author}{Rocca, F.}
\newblock \bibinfo{title}{Optical vortices}.
\newblock \emph{\bibinfo{journal}{Optics Communications}}
  \textbf{\bibinfo{volume}{73}}, \bibinfo{pages}{403--408}
  (\bibinfo{year}{1989}).

\bibitem{Zhao21}
\bibinfo{author}{Zhao, L.-C.}, \bibinfo{author}{Qin, Y.-H.},
  \bibinfo{author}{Lee, C.} \& \bibinfo{author}{Liu, J.}
\newblock \bibinfo{title}{Classification of dark solitons via topological
  vector potentials}.
\newblock \emph{\bibinfo{journal}{Phys. Rev. E}}
  \textbf{\bibinfo{volume}{103}}, \bibinfo{pages}{L040204}
  (\bibinfo{year}{2021}).

\bibitem{Su1979}
\bibinfo{author}{Su, W.~P.}, \bibinfo{author}{Schrieffer, J.~R.} \&
  \bibinfo{author}{Heeger, A.~J.}
\newblock \bibinfo{title}{Solitons in polyacetylene}.
\newblock \emph{\bibinfo{journal}{Phys. Rev. Lett.}}
  \textbf{\bibinfo{volume}{42}}, \bibinfo{pages}{1698--1701}
  (\bibinfo{year}{1979}).

\bibitem{Berry1984}
\bibinfo{author}{Berry, M.~V.}
\newblock \bibinfo{title}{Quantal phase factors accompanying adiabatic
  changes}.
\newblock \emph{\bibinfo{journal}{Proc. R. Soc. Lond. A}}
  \textbf{\bibinfo{volume}{392}}, \bibinfo{pages}{45--57}
  (\bibinfo{year}{1984}).

\bibitem{Resta1994}
\bibinfo{author}{Resta, R.}
\newblock \bibinfo{title}{Macroscopic polarization in crystalline dielectrics:
  the geometric phase approach}.
\newblock \emph{\bibinfo{journal}{Rev. Mod. Phys.}}
  \textbf{\bibinfo{volume}{66}}, \bibinfo{pages}{899--915}
  (\bibinfo{year}{1994}).

\bibitem{Ozawa2019}
\bibinfo{author}{Ozawa, T.} \emph{et~al.}
\newblock \bibinfo{title}{Topological photonics}.
\newblock \emph{\bibinfo{journal}{Rev. Mod. Phys.}}
  \textbf{\bibinfo{volume}{91}}, \bibinfo{pages}{015006}
  (\bibinfo{year}{2019}).

\bibitem{Smirnova2020}
\bibinfo{author}{Smirnova, D.}, \bibinfo{author}{Leykam, D.},
  \bibinfo{author}{Chong, Y.} \& \bibinfo{author}{Kivshar, Y.}
\newblock \bibinfo{title}{Nonlinear topological photonics}.
\newblock \emph{\bibinfo{journal}{Appl. Phys. Rev.}}
  \textbf{\bibinfo{volume}{7}}, \bibinfo{pages}{021306} (\bibinfo{year}{2020}).

\bibitem{Hasan2010}
\bibinfo{author}{Hasan, M.~Z.} \& \bibinfo{author}{Kane, C.~L.}
\newblock \bibinfo{title}{Colloquium: Topological insulators}.
\newblock \emph{\bibinfo{journal}{Reviews of Modern Physics}}
  \textbf{\bibinfo{volume}{82}}, \bibinfo{pages}{3045--3067}
  (\bibinfo{year}{2010}).

\bibitem{Smirnova2020Nonlinear}
\bibinfo{author}{Smirnova, D.}, \bibinfo{author}{Leykam, D.},
  \bibinfo{author}{Chong, Y.} \& \bibinfo{author}{Kivshar, Y.}
\newblock \bibinfo{title}{Nonlinear topological photonics}.
\newblock \emph{\bibinfo{journal}{Applied Physics Reviews}}
  \textbf{\bibinfo{volume}{7}}, \bibinfo{pages}{021306} (\bibinfo{year}{2020}).

\bibitem{khanikaev_alù_2024}
\bibinfo{author}{Khanikaev, A.~B.} \& \bibinfo{author}{Alù, A.}
\newblock \bibinfo{title}{Topological photonics: robustness and beyond}.
\newblock \emph{\bibinfo{journal}{Nature Communications}}
  \textbf{\bibinfo{volume}{15}}, \bibinfo{pages}{931} (\bibinfo{year}{2024}).

\bibitem{GarciaOjalvo1999}
\bibinfo{author}{Garc{\'i}a-Ojalvo, J.} \& \bibinfo{author}{Sancho, J.~M.}
\newblock \emph{\bibinfo{title}{Noise in Spatially Extended Systems}}
  (\bibinfo{publisher}{Springer}, \bibinfo{address}{New York, NY},
  \bibinfo{year}{1999}).

\end{thebibliography}

	\end{document}